\documentclass{jaa}

\usepackage{graphicx}
\usepackage{natbib}
\usepackage[colorlinks=true,citecolor=blue,linkcolor=blue]{hyperref}%
\usepackage{url}
\usepackage{float}

\begin{document}\sloppy

\title{High-precision distance measurements with classical pulsating stars}


\author{Anupam Bhardwaj\textsuperscript{1}}
\affilOne{\textsuperscript{1}Kavli Institute for Astronomy and Astrophysics, Peking University, Yi He Yuan Lu 5, Hai Dian District, Beijing 100871, China.\\}


\twocolumn[{

\maketitle

\corres{anupam.bhardwajj@gmail.com}

\msinfo{21 Jun 2020}{29 Jun 2020. In original form 12 May 2020.}

\begin{abstract}

Classical Cepheid and RR Lyrae variables are radially pulsating stars that trace young and old-age stellar populations, respectively. These classical pulsating stars are the most sensitive probes for the precision stellar astrophysics and the extragalactic distance measurements. Despite their extensive use as standard candles thanks to their well-defined Period-Luminosity relations, distance measurements based on these objects suffer from their absolute primary calibrations, metallicity effects, and other systematic uncertainties. Here, I present a review of classical Cepheid, RR Lyrae, and type II Cepheid variables starting with a historical introduction and describing their basic evolutionary and pulsational properties. I will focus on recent theoretical and observational efforts to establish  absolute scale for these standard candles at multiple wavelengths. The application of these classical pulsating stars to high-precision cosmic distance scale will be discussed along with observational systematics. I will summarize with an outlook for further improvements in our understanding of these classical pulsators in the upcoming era of extremely large telescopes.

\end{abstract}


}]


\doinum{12.3456/s78910-011-012-3}
\artcitid{\#\#\#\#}
\volnum{000}
\year{0000}
\pgrange{1--}
\setcounter{page}{1}
\lp{1}

\section{Introduction}
Stars are primary engines of cosmic evolution and play a crucial role in our understanding of the Universe. Variable stars, in particular, provide information about the stellar properties including physical parameters, internal and external envelope structure and composition, and probe both the stellar evolution and cosmic distances. The first variable star was discovered more than four centuries back in 1596 by David Fabricius which was later named as {\it Omicron Ceti} or Mira and now represents one of the subclasses belonging to the long-period variables. The short-period, typically fainter, variable stars were not well-known until two British astronomers, Edward Pigott and John Goodricke started observations of $\beta$ Persei (Algol) in 1782 \citep{goodricke1783}. A few years later, Pigott detected the variability in $\eta$ Aquilae, the first known Cepheid variable. At the same time, Goodricke discovered $\delta$ Cephei \citep{goodricke1786}, which represents classical Cepheid variables as one of the most important classes of pulsating variables in the modern astronomy.
 
About a century later the first variable stars within a Galactic globular cluster (GGC) were discovered by Wilhelmina Flemming and reported in \cite{pickering1889}. Following this discovery, Solon Bailey initiated a search for variable stars in the GGCs from the Harvard College Observatory in 1893 and discovered hundreds of ``cluster variables''. Bailey later separated the cluster variables as RR Lyrae subtypes but the {\it RR Lyrae} itself was discovered by Wilhelmina Flemming \citep{pickering1901}. Historically, W Virginis was the prototype of Type II Cepheids (T2Cs) and it was discovered by 
\citet{schonfeld1866}. The short-period representative of T2Cs, BL Herculis was discovered by \citet{hoffmeister1929} and the variability of long-period RV Tauri was first observed by \citet{ceraski1905}{\footnote{\url{https://www.aavso.org}}}. A more detailed historical overview of classical Cepheids, RR Lyrae and T2Cs can be found in \cite{catelan2015} but this brief introduction demonstrates that the Cepheid and RR Lyrae stars represent two of the oldest and therefore well-studied subtypes of variable stars. 

The observations of Cepheids in the Magellanic Clouds \citep{leavitt1908} led to the discovery of a relation between their pulsation period and luminosity \citep{leavitt1912}. This relation is commonly known as ``Cepheid Period-Luminosity relation (PLR)'' or the {\it Leavitt Law} honouring the discoverer. Ever since, classical Cepheids have played a fundamental role in the extragalactic distance measurements. Edwin Hubble used Cepheid PLR to determine reliable distance to the M31 and discovered that Andromeda, assumed to be a gaseous nebula at that time, is another galaxy beyond our Milky Way \citep{hubble1926}. Cepheid-based distances to the galaxies as far as the Virgo cluster allowed Hubble to discover a linear correlation between the apparent distances to galaxies and their recessional velocities \citep{hubble1929} - the more distant the galaxy, the faster it moves away from us - now known as the {\it Hubble-Lema\^itre law}, providing the first evidence of the expanding universe. The slope of the velocity over distance is the Hubble constant ($H_0$), which parameterizes the current expansion rate of the Universe. The current $H_0$ values in the late evolutionary universe are in tension with early universe measurements \citep{riess2018, planck2018} and therefore understanding the systematics involved in standard candles is critical to resolve the $H_0$ tension, and improve the precision of cosmic distance scale. On the other hand, RR Lyrae, which are exclusively old and metal-poor stars, have been used as stellar tracers of the age, metallicity, extinction and structure of our Galaxy but their use as robust distance indicators gained importance more recently thanks to the boost of near-infrared (NIR) observations over the last two decades. 

The goal of this review is to focus on recent progress on absolute calibration of classical Cepheids, RR Lyrae and T2Cs, and their application to the extragalactic distance scale. I strongly emphasize here that a short review can not fully describe all the aspects of these classical pulsating stars as standard candles. The interested readers are referred to the books, for example, \cite{catelan2015} on pulsating variables and \cite{grijs2011} on introduction to the cosmic distance scale. Additionally, several excellent reviews are also available in the literature \citep[][and references within]{madore1991, feast1999, wallerstein2002, sandage2006, catelan2009, feast2013, subramanian2017, beaton2018}. \cite{mcwilliam2011} published an excellent set of online conference review articles on RR Lyrae stars focussed on different aspects beyond their use as distance indicators while a recent review of Cepheid and RR Lyrae as young and old stellar population tracers of the Galactic structure can be found in \citet{matsunaga2018} and \citet{kunder2018}, respectively. Note that while classical and T2Cs will be discussed extensively here, Anomalous Cepheids \citep[see,][and references therein]{wallerstein2002, fiorentino2006, groenewegen2017, jurkovic2018} are not included in this review.     

\begin{figure}[!t]
\begin{center}
\includegraphics[trim=4.2cm 0.0cm 4.2cm 0.0cm, clip=true, width=0.5\textwidth]{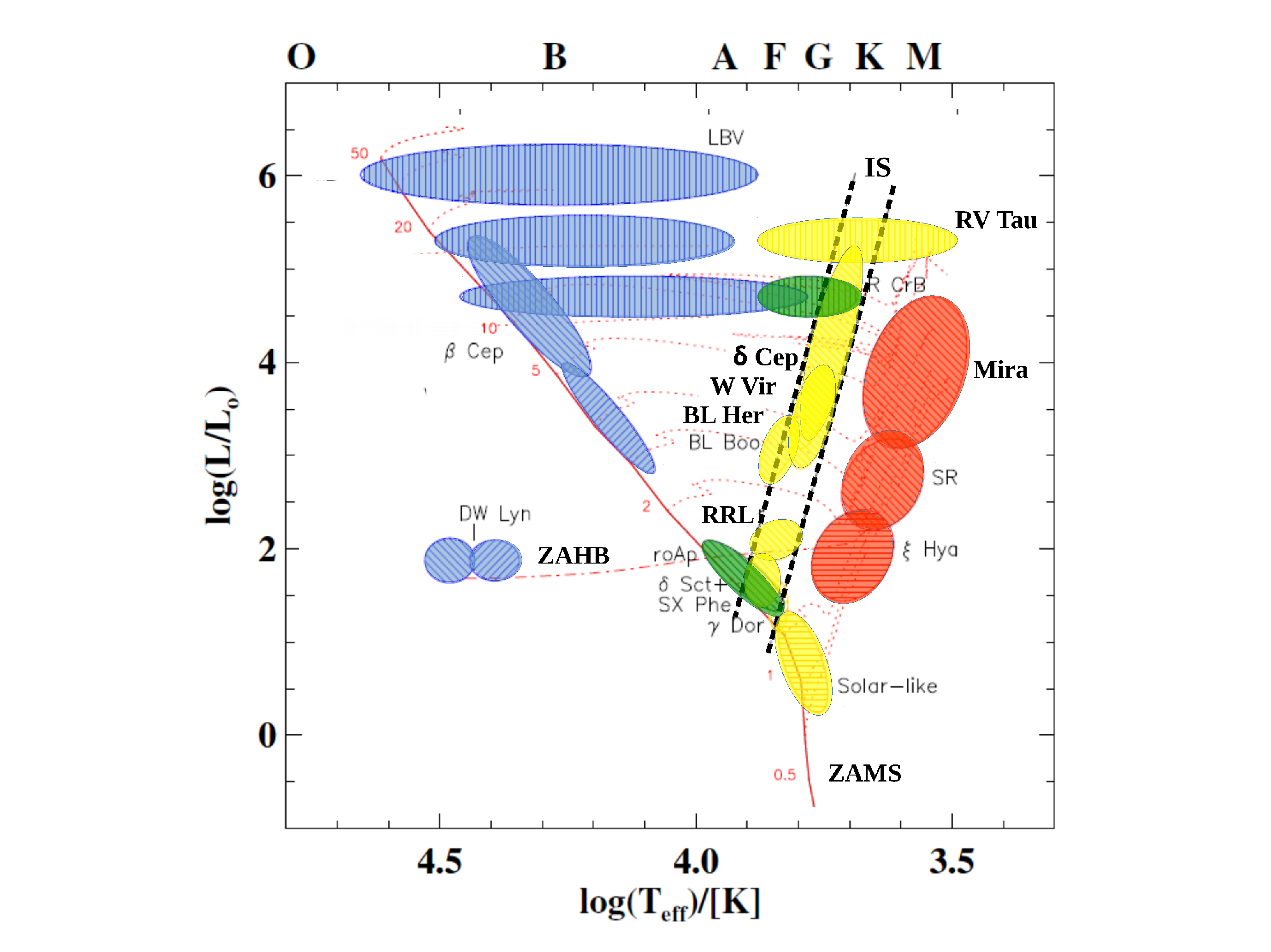}
\vspace{-22pt}
\caption{Hertzsprung-Russell diagram displaying schematic representation of classical pulsating variable stars. A modified version of figure taken from \cite{jeffery2016} is shown. The line-shaded regions represent approximate location of variables and the color represents approximate spectral class mentioned on the top. The zero-age main sequence (ZAMS) and the horizontal branch (ZAHB) are shown with solid and dashed red lines. Cepheid instability strip is shown with vertical black dashed lines. Dotted lines represent evolutionary tracks of stars with different masses. The label on the left of the ZAMS shows the stellar mass of each track.}
\label{fig:puls_hr}
\vspace{-18pt}
\end{center}
\end{figure}

This review is organised as follows: I describe briefly the description of evolutionary and pulsational scenario for classical pulsating stars in Section~2 and their light curve variations in Section~3. The Sections~4 to 6 focus on classical Cepheids, RR Lyrae and T2Cs as distance indicators both from the observational and theoretical perspectives at multiple wavelengths. The absolute scale for each standard candle and associated systematics is also addressed. Finally, summary with an outlook for the future will be briefly presented in Section~7.
\vspace{-3pt}

\section{Evolutionary and Pulsational Scenario}
\label{sec:sec2}

Cepheid and RR Lyrae represent radially pulsating class of variable stars. Classical Cepheids are young ($\sim$10-300 Myr), intermediate-mass ($\sim$3-10$M_\odot$), metal-rich stars while RR Lyrae are old ($\geq10$ Gyr), low-mass ($\sim$0.5-0.8$M_\odot$) metal-poor stars. T2Cs also belong to old, low-mass, metal-poor stellar populations.  Classical pulsating variables populate a well-defined narrow vertical region in temperature in the Hertzsprung-Russell (HR) diagram, known as the instability strip (IS). Fig.~\ref{fig:puls_hr} shows the location of classical pulsating stars including Cepheid and RR Lyrae within the IS in the HR diagram. Classical Cepheids, represented by the prototype $\delta$ Cep, are luminous yellow giant variables that pulsate in fundamental (FU), first-overtone (FO), second-overtone harmonics and multiperiodic (double/triple) modes \citep{soszynski2015}. RR Lyrae occupy the region between the cross-section of the Horizontal Branch (HB) and the IS. Although RR Lyrae stars also pulsate primarily in the fundamental-mode (RRab) and first-overtone modes (RRc), few variables pulsating in more than one mode simultaneously (RRd) have also been discovered \citep[for example,][]{soszynski2017}. 

\begin{figure}[!t]
\begin{center}
\includegraphics[width=1.0\columnwidth,angle =0]{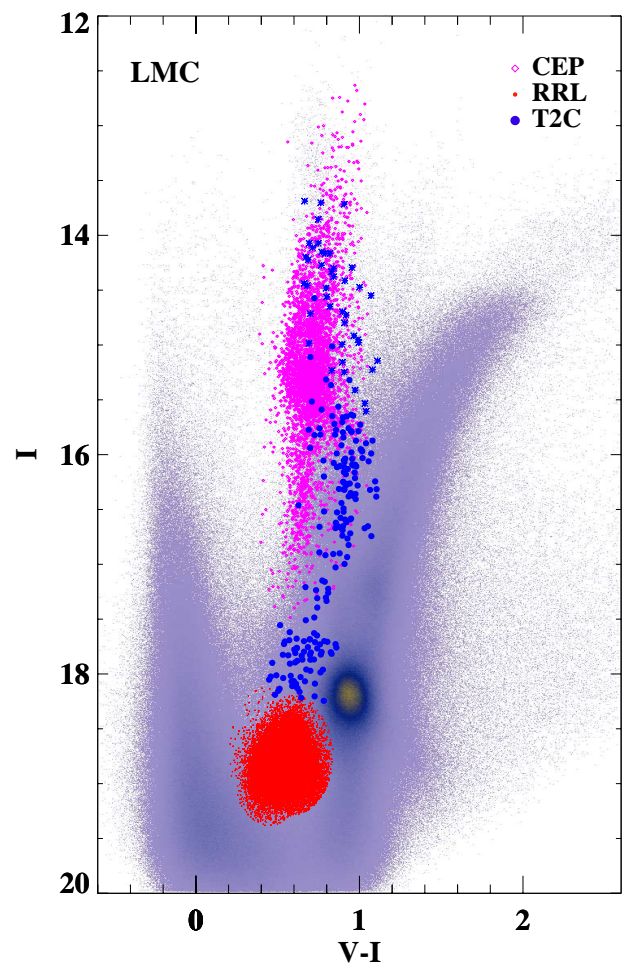}
\caption{Optical color-magnitude diagram for the LMC with data from the Magellanic Clouds Photometric Survey \citep{zaritsky2004} without any extinction corrections. Classical Cepheids, RR Lyrae and T2Cs are also overplotted using data from the OGLE survey \citep{soszynski2015, soszynski2016, soszynski2017}. Only the central clusters of each of these pulsating stars are shown for visualization purposes.}
\label{fig:puls_cmd}
\end{center} 
\end{figure}

The T2Cs represent different evolutionary states from post HB to the asymptotic giant branch (AGB) phase and a preliminary classification is done based on their pulsation periods: BL Herculis (BL Her, $1\lesssim\!P\!\lesssim4$~d), W Virginis (W Vir, $4\lesssim\!P\!\lesssim20$~d) and RV Tauri (RV Tau, $P\!\gtrsim 20$~d). \citet{soszynski2008a} suggested another subtype, peculiar W Virginis (pW Vir, $4\lesssim\!P\!\lesssim10$~d), with distinct light curves and these peculiar stars are mostly brighter and bluer than W Vir. T2Cs primarily pulsate in the fundamental mode but BL Hers pulsating in the first-overtone mode have also been discovered by \citet{soszynski2019}. Fig.~\ref{fig:puls_cmd} shows distribution of classical Cepheids, RR Lyrae, and T2Cs on the observed color-magnitude diagram in the Large Magellanic Cloud (LMC) from the optical gravitational lensing experiment \citep[OGLE,][]{udalski1993, soszynski2015, soszynski2016, soszynski2018}. The T2C population is located along the IS and have luminosities that are intermediate between classical cepheids and RR Lyrae. However, the RV Tau and some W Vir overlap the region of classical Cepheids but the T2Cs are typically significantly less abundant than classical Cepheids and RR Lyrae. The basic properties of Cepheids and RR Lyrae are given in Table~\ref{tab:basics}. Depending on the pulsation periods, classical Cepheids are systematically $\sim$2-3 magnitude brighter than T2Cs at a fixed period and up to $\sim 8$ mag brighter than RR Lyrae.

\begin{table*}[htb]
\tabularfont
\begin{center}
\caption{Basic Properties of Cepheid and RR Lyrae variables.}\label{tab:basics}
\begin{tabular}{lllcccccc}
\topline
\textbf{Star}&	\textbf{}		&\textbf{Subtype}	&\textbf{Mass}	&\textbf{Period range}	&\textbf{Period}	&\textbf{$M_V$}&	\textbf{$M_K$}	&\textbf{$\Delta I$}\\
		&			& 			& M$_\odot$	&	days		&	days		& mag		&	mag	& mag\\
\midline
Classical Cepheids	&	Pop I 	&	Fundamental mode (FU)	&  $\sim$3-10	&1 - 100	&1	&	$\sim$ -1.5	&	$\sim$ -2.5	& $\sim$ 0.45 \\
			&		&				&		&		&10	&	$\sim$ -4.0	&	$\sim$ -6.0	& $\sim$ 0.20 \\
			&		&				&		&		&50	&	$\sim$ -6.0	&	$\sim$ -8.0	& $\sim$ 0.65 \\
			&		&	First-overtone mode (FO)&$\sim$3-10	&0.5 - 6	&1	&	$\sim$ -1.5	&	$\sim$ -3.0	& $\sim$ 0.20 \\
			&		&				&		&		&5	&	$\sim$ -4.0	&	$\sim$ -5.5	& $\sim$ 0.20\\				
\hline
RR Lyrae		&	Pop II	&	Fundamental mode (RRab)	& $\sim$0.5-0.8	&0.3 - 1.0	&0.4	&	$\sim$ +0.8	&	$\sim$ -0.1	& $\sim$ 0.80 \\
			&		&				&		&		&0.6	&	$\sim$ +0.8	&	$\sim$ -0.5	& $\sim$ 0.35 \\
			&		&	First-overtone mode (RRc)&$\sim$0.5-0.8	&0.2 - 0.5	&0.3	&	$\sim$ +0.7	&	$\sim$ -0.1	& $\sim$ 0.25 \\
\hline
Type II Cepheids	&	Pop II	&	BL Herculis (BL Her)	&$\sim$0.5-0.6	&1 - 4		&1	&	$\sim$ +0.2	&	$\sim$ -1.0	& $\sim$ 0.50 \\
			&		&	W Virginis (W Vir)	&$\sim$$<$ 1	&4 - 20		&10	&	$\sim$ -1.3	&	$\sim$ -3.5	& $\sim$ 0.25 \\
			&		&	RV Tauri (RV Tau)	&$\sim$$<$ 1	&20 - 80 	&50	&	$\sim$ -4.0	&	$\sim$ -5.5	& $\sim$ 0.30 \\
\hline
\end{tabular}
\end{center} 
\tablenotes{ \footnotesize {\it Notes:} The reader should be cautious regarding numbers shown in this table which are only crude approximation and presented here for a relative comparison. Population I Cepheids are young (10-300 Myr) and Population II RR Lyrae are old ($\gtrsim$ 10 Gyr) stellar populations. The period-range and $I$-band amplitudes corresponding to the period listed in the column 4 are estimated within 90\% percentile range from the OGLE-LMC data \citep{soszynski2015, soszynski2016, soszynski2018}. Absolute $V$-band and $K$-band magnitudes for the given period in the column 4 are derived from the LMC PLRs discussed in the next sections. }
\end{table*}

\subsection{Stellar evolutionary states}

Let us first consider the evolution of Cepheid-like intermediate-mass ($\sim3\textrm{-}10$M$_{\odot}$) stars in their post main-sequence phase and going through the IS. Once a star has exhausted hydrogen in the core, it expands to become a red giant with a temporarily inert helium core that is surrounded by a hydrogen burning shell. The expansion of a star happens very rapidly, and therefore, it is difficult to observe it during this short evolutionary phase which reflects in the {\it Hertzsprung gap} between main sequence and red giant stars \citep{kippenhahn1991}. For a classical Cepheid-like star (say $\sim5$M$_{\odot}$) the expansion of stellar envelope moves star to cooler temperature in the HR diagram during the first crossing through the IS. The first crossing is usually very rapid ($10^3\textrm{-}10^4$ years) and the star exits the red edge of the IS while the hydrogen shell is still burning. Once the ignition of the helium starts, the star contracts and heats up, and makes a loop towards the hotter effective temperature in the HR diagram. During this phase, the star crosses the IS for the second time and undergoes a ``blue loop'' \citep[See Chapter 31, Figures 31.2 \& 31.4 in][]{kippenhahn1991}. Since the central helium burning evolutionary phase lasts for a longer time-scale, the star remains in the IS for a greater time than the first crossing. The star can undergo a third crossing through the IS during the blue loop or return without crossing the blue edge of the IS. The exact location of the blue loops is a function of stellar mass and of the chemical composition. For higher mass stars, the extent of blue loops increases while the low-mass stars can undergo only one crossing through the IS (see Fig.~\ref{fig:puls_hr}). At the late stage of evolution of high mass stars, the stellar core contains a degenerate mixture of carbon and oxygen which can ignite a supernova explosion if the mass limit reaches 1.4$M_{\odot}$. While the initial mass of Cepheids for this to happen is not well constrained, typically an intermediate-mass star evolves onto the AGB while the most massive Cepheid can become a supernova. Interested readers are referred to \citet[][and references therein]{kippenhahn1991, chiosi1992, bono2000, salaris2005, anderson2014, catelan2015} for more details regarding the evolution of intermediate-mass star in the central-helium burning phase. Similar to the stellar evolutionary timescale, the time spent in a Cepheid phase decreases dramatically as a function of mass. Note that the higher mass stars have longer pulsation periods. Therefore, short-period Cepheid variables are discovered in greater numbers than long-period ones if both are within the observational limits.

RR Lyrae, similar to classical Cepheids, are core helium burning stars and occupy a region in the HR diagram which is the intersection between the Cepheid IS and the HB. A low-mass ($\sim1$M$_{\odot}$) star evolves to become a red giant in its post main-sequence phase and enters the HB evolutionary phase with helium burning core. The morphology of the HB itself is quite complex and a broad spectrum of
HB-related topics are covered in the review by \citet{catelan2009}. The zero-age HB (ZAHB) star is characterized by the helium-burning in the core and the hydrogen shell burning surrounding the helium core. The location of ZAHB stars on an almost horizontal locus in the HR diagram for given helium core mass and envelope composition depends on the total mass (or the envelope mass). These stars have a wide range of effective temperatures such that massive envelopes lead to cooler temperatures. After the onset of degenerate central helium burning, only stars with initial main-sequence masses of $\lesssim 0.8M_\odot$ achieve the temperatures that place them within the IS. Such stars pulsate and become RR Lyrae variables either when they are close to the ZAHB or else when they evolve to the blue or red side in the HR diagram. The blue edge of the IS of RR Lyrae is located at an effective temperature of $\sim 7200$K at the ZAHB luminosity level which decreases with increasing luminosity. The red edge of the IS is located somewhere around 5900 K and is very sensitive to the efficiency of convection, and the topology of the IS is also dependent on the metal abundance \citep[see details in][]{bono1994, bono1996, bono1997a, salaris2005, catelan2015, marconi2015, marconi2018a}. 

T2Cs are in a post-HB evolutionary phase of low-mass stars evolving up the AGB. After the exhaustion of helium in the core, HB stars move towards brighter luminosities in the HR diagram evolving mainly into AGB. The post-HB evolution of star depends on its location on the HB or on the effective temperature. T2Cs represent the class of those pulsating stars that evolve from the blue tail of the HB and reach the IS at higher luminosities than those of RR Lyrae. These stars suffer shell flashes at the boundary between degenerate CO core and the helium region. Short-period BL Her stars evolve from the HB, bluer than the RR Lyrae gap, to AGB i.e., towards higher luminosity and larger radius in the process of depleting helium in their core. The intermediate period W Vir stars begin to undergo helium shell flashes as they reach AGB phase and make temporary excursions into the IS \citep{wallerstein2002}. However, \citet{groenewegen2017a} showed that the evolution of the W Vir subclass is not clear and they may have the binarity origin similar to pW Vir. The long-period RV Tau are thought to represent post-AGB evolution  \citep{wallerstein2002}. However, RV Tau may also evolve from the more massive and younger objects or represent binary evolution \citep{groenewegen2017,manick2018}. The evolutionary tracks of T2Cs were pioneered by \citet{gingold1976} and the updated theoretical calculations were presented by \citet{bono1997c, bono2016} and \citet{smolec2016}.

\subsection{Stellar pulsation mechanism}

I will briefly discuss the physical mechanism driving the pulsations in Cepheid and RR Lyrae variables. The classical relation between the pulsation period and the mean-density of a pulsating gaseous sphere was first developed by \cite{ritter1879} who demonstrated that for a homogeneous sphere experiencing adiabatic radial pulsation- 
\vspace{-2pt}
\begin{equation}
\label{eq:puleq}
P \propto \sqrt(R/g),
\end{equation}

\noindent where $P$ is the pulsation period, $R$ is the radius and $g$ is the surface gravity of gaseous sphere. Since, $g \propto M/R^2$ and using relation between mean density ($\rho$), mass and radius- 
\vspace{-2pt}
\begin{equation}
\label{eq:puls}
P \sqrt\rho = Q,
\end{equation}

\noindent where $Q$ is the pulsational constant and the equation is known as the pulsation equation or the period-mean-density equation. However, the hypothesis of radial pulsations in stars came much later when more detailed investigations showed that the above relation is also valid for real stars. 

Around early twentieth century, the periodic changes in the light and velocity curves of $\delta$ Cephei favoured the explanation that Cepheids were binary stars but the light variations of $\delta$ Cephei were significantly different from the confirmed spectroscopic binary {\it Algol}. Later, \cite{shapley1914} presented strong evidence against binary hypothesis noting that small parallaxes of Cepheids suggest the luminosities and radii of primary stars are on average $\sim 10^3 L_\odot$  and $5 R_\odot$, respectively. These results favoured stellar pulsation for causing light variations in Cepheid-like variables. The pulsation hypothesis for a single star was also used by \cite{martin1915} to explain the radial velocity variations of a RR Lyrae, then known as cluster variable. Finally, the most significant progress for the pulsating star hypothesis was made by \cite{eddington1918, eddington1919}, who developed a theory of adiabatic oscillations of a stellar atmosphere. He suggested that every star of intermediate mass will go through a Cepheid phase for a brief time during its life-cycle, and the physics of radial oscillations was presented in \cite{eddington1926}. Note that a PLR for pulsating stars follows directly from the Stefan-Boltzmann law and the pulsation equation (\ref{eq:puls}) such that the bolometric magnitudes can be written as:
\vspace{-2pt}
\begin{equation}
M_{bol} = a + b\log P + cT_{eff},
\end{equation}

\noindent where pulsation period ($P$) is used assuming its dependence on stellar mass and radius through equation (\ref{eq:puleq}). The observable color term can replace the $T_{eff}$ which results in a Period-Luminosity-Color (PLC) relation. In a two-dimensional plane, neglecting color-term, the PLRs in a given wavelength ($\lambda$) takes the form:
\vspace{-2pt}
\begin{equation}
M_{\lambda} = a + b\log P.
\label{eq:plr}
\end{equation}

The physical scenarios regarding the main driving mechanism behind Cepheid pulsation and stellar structure and evolution were explored by various authors \citep{christy1966, stobie1969a, cox1980a}.
The pulsation occurs in the stellar envelope for a specific range of surface effective temperatures i.e., within the IS, a region where stars are unstable to pulsation. For example, in a Cepheid-like star with temperature near 6000K, hydrogen ionization zone occurs close to the surface of the star. Further, helium becomes doubly ionized in another zone deeper in the stellar envelope. The increase in the opacity ($\kappa$) increases the ionization in both the hydrogen and helium ionization zones. Due to cyclic variations in the opacity, the energy is trapped during contraction, favouring instability. Since the ionization occurs deep inside the surface of the star, the pressure or excitation beneath drives stellar envelope expansion. The phenomena works as a mechanical valve and the expansion reduces the opacity and the energy is released. The temperature and pressure drop and the expansion occurs only due to momentum of the envelope structure. Finally, star starts contracting again and the temperature regains its initial value, thus re-starting the pulsation cycle. Since the mechanism responsible for pulsation is mainly the increase in the opacity of the ionization zones, it is known as the ``$\kappa$ mechanism'' \citep{kippenhahn1991, salaris2005, catelan2015}. In the case of radial pulsations, if all parts of a star move in and out together, the pulsation occurs in fundamental-mode but the star can have an infinite number of modes. Within the IS, classical Cepheid and RR Lyrae variables exhibit pulsations during their long-lasting central helium burning evolutionary phase and the pulsations in T2Cs occur during post-HB evolution. As a passing remark, non-radial pulsation and light curve modulations have also been discovered in classical pulsating stars \citep[for example see,][and reference within for more details]{dziembowski2004, netzel2015, moskalik2015, smolec2016a, anderson2016a}.

It is important to emphasize here that for classical Cepheids, evolutionary masses are systematically larger at the level of $10-20\%$ than the pulsation masses or masses derived from other independent methods \citep[][and references therein]{cox1980b, caputo2005d, prada2012, neilson2011, marconi2013a}. This {\it Cepheid mass discrepancy} originally proposed by \citet{christy1968, stobie1969a, stobie1969b} is an open problem. Discovery of classical Cepheid in the binary system \citep{piet2010} allowed precise dynamical mass estimates which were found to be consistent with masses derived from the pulsation models \citep[see also,][]{pilecki2018}. Therefore, non-standard phenomena like mass-loss, core overshooting and rotation have been explored in evolutionary models for consistency with pulsation masses \citep{prada2012, anderson2014}.

\section{Light curve morphology}
\label{sec:sec3}

\begin{figure*}[!t]
\centering
  \begin{tabular}{@{}c@{}}
    \includegraphics[width=1.0\textwidth]{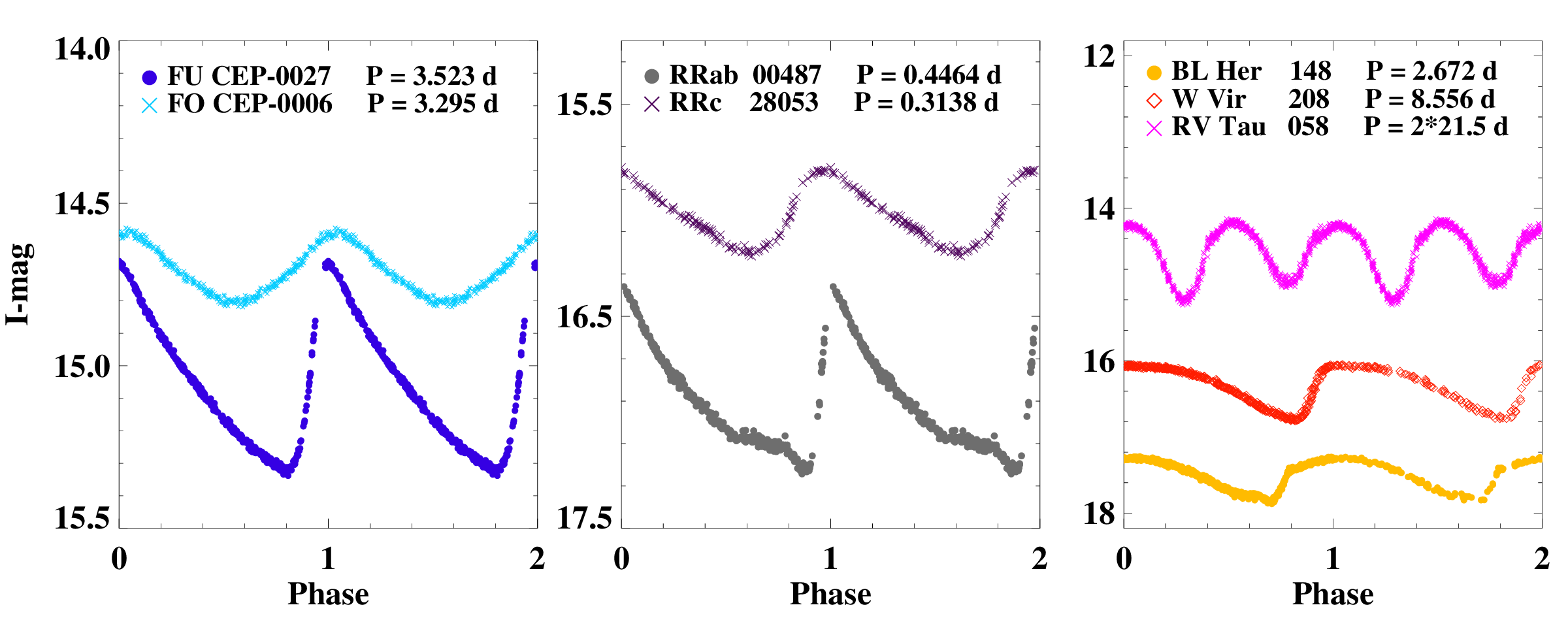} \\
    \includegraphics[width=1.0\textwidth]{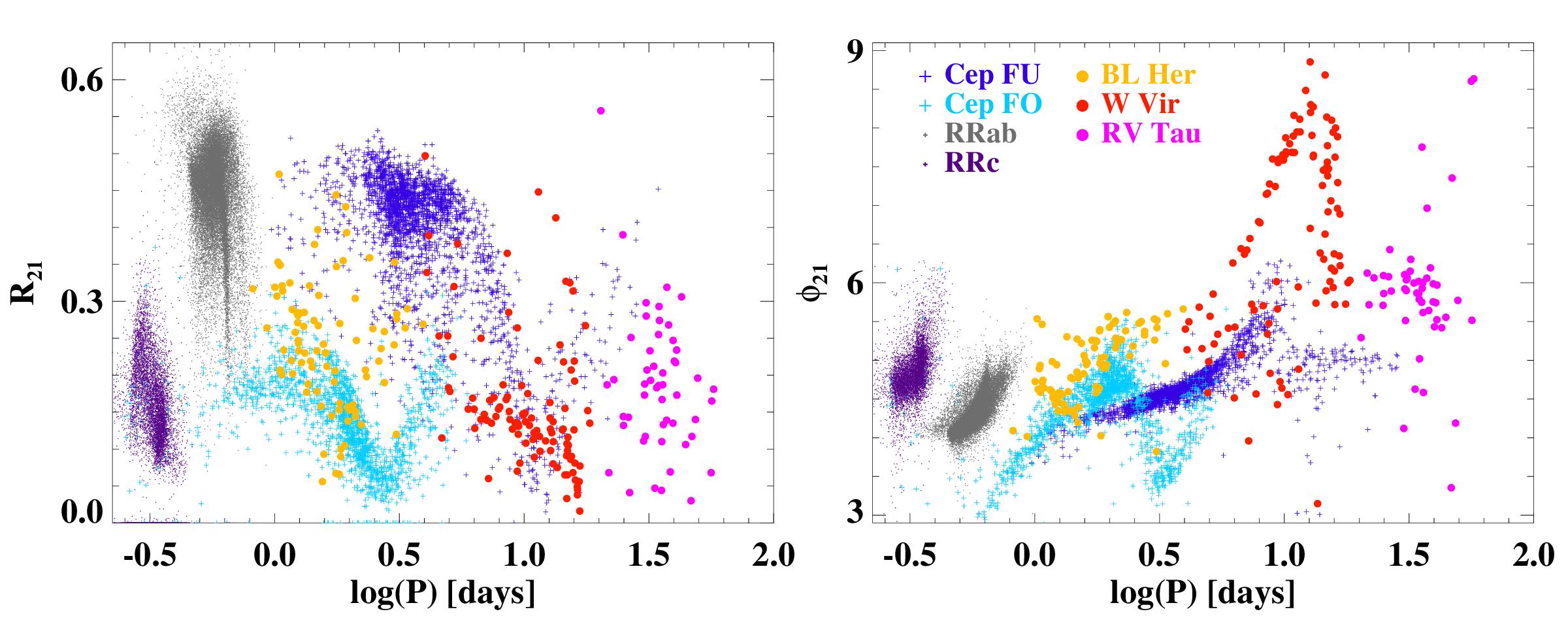} \\
  \end{tabular}
 \caption{{\it Top panels:} The representative light curves of classical Cepheids, RR Lyrae and T2Cs in the LMC in $I$-band taken from the OGLE survey \citep{soszynski2015, soszynski2016, soszynski2018}. The OGLE ID, subtype and periods are also listed on the top of each panel. {\it Bottom panles:} The $I$-band Fourier amplitude ($R_{21}$) and phase parameters ($\phi_{21}$) for classical pulsating stars plotted as a function of logarithm of pulsation period ($P$).}
\label{fig:lcs}
\end{figure*}

The analysis of the light curve structure of Cepheid and RR Lyrae variables is very useful for their identification and classification. At the same time, pulsation models can also be used to successfully predict the multiband light and radial velocity variations. Therefore, quantification of light curve structure can allow a rigorous comparison between observations and theory and provide constraints for the stellar pulsation models \citep{wood1997, marconi2013, marconi2017}. Top panels of Fig.~\ref{fig:lcs} show the $I$-band light curves of classical Cepheids, RR Lyrae and T2Cs in the LMC from the OGLE survey \citep{soszynski2015, soszynski2016, soszynski2018}. Typical optical light curves of fundamental mode Cepheids are symmetric with a saw-tooth feature while some Cepheids also exhibit ``bump'' along their light curves. \cite{hertzsprung1926} discovered that Galactic Cepheids present a relationship between the pulsation period and the location of the bump along the light curve - known as ``Hertzsprung Progression''. Classical Cepheids show a bump on the descending branch of both the light and velocity curves for periods between 6 and 16 days and it appears around the phases of maximum light for periods between 9 and 12 days. For longer period Cepheids, the bump feature appears on the rising branch. The central period of the Hertzsprung progression has been used to constrain models \citep{bono2000d}. It depends on the metal-abundance and wavelength such that it shifts to longer periods with decreasing metallicity or increasing wavelengths \citep{bhardwaj2017}. Note that ``bump Cepheids'' are single mode variables with strong regularity in their light curves while the so called ``Beat Cepheids' are mixed-mode variables that pulsate in two or more modes simultaneously.

The shape of fundamental-mode RR Lyrae optical light curves is more saw-toothed than that of classical Cepheids. The RRab light curves also exhibit a sharp rise from minima to maxima and a distinct bump near the minimum light. The first-overtone Cepheid and RR Lyrae variables display near-sinusoidal variations in the light curves even at optical wavelengths. T2Cs generally display complex light curve variations with BL Her showing variations similar to RRab while W Vir sometimes complement fundamental-mode classical Cepheids. RV Tau stars exhibit complex light curves with varying maxima and minima from cycle-to-cycle. At longer wavelengths, both amplitude and phase variations decrease significantly and the skewness and acuteness of Cepheid and RR Lyrae light curves attain a value close to unity implying a nearly symmetric sinusoidal variations as a function of pulsation phase. 

\cite{slee1981} used Fourier analysis method to study light curve of periodic variables and showed that the lower order Fourier coefficients can be used to describe the structure of Cepheid and RR Lyrae variables. In brief, a Fourier series can be fitted to the periodic light curves in the following form: 

\begin{equation}
m = m_{0}+\sum_{k=1}^{N}A_{k} \sin(2 \pi k x + \phi_{k}),
\label{eq:foufit}
\end{equation}

\noindent where, $m$ is the magnitude as a function of the pulsation phase ($x$). The Fourier-fit results in a mean-magnitude ($m_0$) and amplitude ($A_k$) and phase ($\phi_k$) coefficients which are used to construct Fourier amplitude ratios and phase differences: $R_{k1} = \frac{A_{k}}{A_{1}};~ \phi_{k1} = \phi_{k} - i\phi_{1},~\mathrm{for}~k>1$ \citep{bhardwaj2015}. Fourier analysis of classical Cepheid, RR Lyrae, and T2C light curves were first carried out by \citet{slee1981}, \citet{simon1982} and \citet{petersen1986}, respectively. A comparison of the observed light and velocity curves of classical Cepheids with theoretical models was followed in a number of studies \citep{simon1983, simon1985, stelling1986}. The phase lag obtained from Fourier decomposition of light curves was found to be the most useful parameter for comparison with observations. Later, \citet{jk1996} derived an empirical relation between period, Fourier phase parameter ($\phi_{31}$), and metallicity for fundamental mode RR Lyrae variables, which is used extensively in deriving photometric metallicities of the statistical samples of RR Lyrae with well-sampled light curves \citep[for example,][]{piet2015}. Fourier analysis of Cepheid and RR Lyrae have also been used for the classification of these variables \citep[for example,][]{deb2009, kains2019}. The lower-order Fourier parameters contain the most characteristic information about the light curve structure and occupy different regions in period and Fourier parameter planes. 

In the bottom panel of Fig.~\ref{fig:lcs}, Fourier amplitude and phase parameters are plotted against the pulsation period. Classical Cepheids display a distinct progression at 10 days in the case of fundamental mode Cepheids and at 2.5 days in the case of first-overtone mode Cepheids. The sharp changes in the Fourier plane at 10 days are attributed to the resonance $P_2/P_0 = 0.5$, in the normal mode spectrum \citep{simon1976, slee1981}. In case of multiwavelength light curves of Cepheids, the phase of maximum-light shifts to later phases as a function of wavelength \citep{madore1991}. Similarly, the Fourier amplitude parameters decrease while the Fourier phase parameters increase with wavelength at a given period for both Cepheid and RR Lyrae variables \citep{bhardwaj2017, das2018}. The Fourier parameters of RR Lyrae do not exhibit any significant structure within short-period range, as can be seen in Fig.~\ref{fig:lcs}. However, each subclass of T2Cs display a distinct structure on the Fourier parameter plane, and the amplitude and phase parameters also overlap with those of classical Cepheids. 

The modern stellar pulsation models are based on nonlinear, radial pulsation codes that account for nonlocal and time-dependent treatment of turbulent convection \citep{stellingwerf1982, bono1994, bono1999a}. These models accurately predict the observables, including the topology of the IS, pulsation modes, amplitudes, multiband light and radial velocity variations \citep{bono2000, marconi2013, marconi2015}. The model-fitting of observed light curves with pulsation models was first carried out by \citet{wood1997} resulting in a robust distance to the LMC. \citet{marconi2013a} performed model-fitting of Cepheids in an eclipsing binary system and predicted pulsation masses that are consistent with dynamical estimates, and later extended model-fitting to multiband light curves of Cepheids in the Small Magellanic Cloud \citep[SMC,][]{marconi2017}. \citet{bhardwaj2017} and \citet{das2018} performed a multiwavelength comparison of Cepheid and RR Lyrae light curve parameters and found that models are consistent with observations in most period bins. While the theoretical amplitudes are systematically larger than the observed amplitudes, this discrepancy can be remedied by increasing the convective efficiency in the models. Using a machine-learning approach, \citet{bellinger2020} compared observed and modelled Fourier light curve parameters of Cepheid and RR Lyrae and provided a preliminary estimates of physical parameters such as mass, luminosity, temperature, radius, and distances to the observed stars in the Galaxy and the Magellanic Clouds with a precision limited by a finer grid of models covering entire period range.  

At shorter wavelengths, ultravoilet (UV) and X-ray studies of classical pulsators are very limited, and aimed at exploring evolutionary, pulsational and atmospheric properties of these variables \citep[for example,][and references therein]{downes2004, engle2015, siegel2015, neilson2016, sachkov2018}. At UV wavelengths, the amplitudes of classical pulsators are significantly large \citep[up to 4 mag in RR Lyrae,][]{kinman2014, siegel2015}, which makes their identification and classification easier provided sufficient time coverage is available. Combining with the light curves at longer wavelengths, the large amplitudes of UV light curves can be used to constrain the impact of convective efficiency in the non-linear pulsation models. Furthermore, simultaneous model-fits to UV, optical and IR data can also provide insight into the physical parameters of these pulsating stars \citep{wheatley2012}.

\section{Classical Cepheids as distance indicators}
\label{sec:sec4}

\begin{figure*}[!t]
\begin{center}
\includegraphics[width=\textwidth]{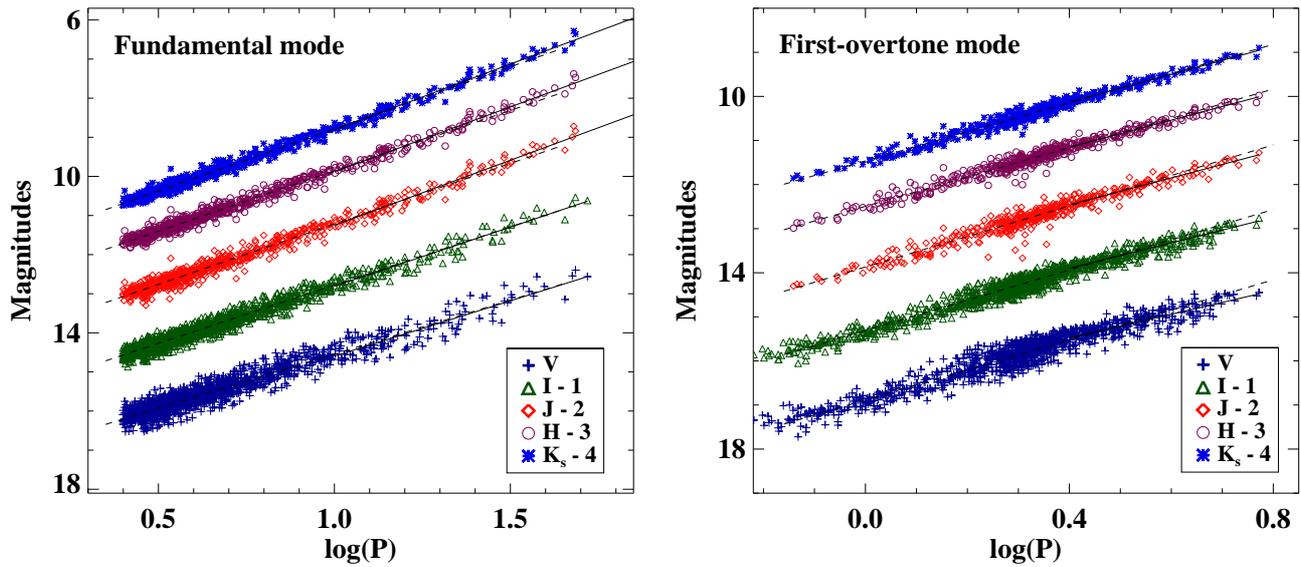}
\caption{Multiwavelength PLRs of fundamental and first-overtone mode classical Cepheids in the LMC. The optical data is taken from OGLE survey \citep{soszynski2015} and the near-infrared photometry is adopted from \citet{macri2015}. The dashed/solid lines represent linear regression over entire/long period range only with a break period at 10 days for fundamental-mode Cepheids and at 2.5 days for first-overtone mode Cepheids.}
\label{fig:plr_cep}
\end{center}
\end{figure*}

Over the past century, Cepheid variables have been used as standard candles with considerable interest in determining distances to star-forming galaxies out to $\sim 40$~Mpc. The Cepheid PLRs in the Galaxy and the LMC have played a vital role in calibrating the distant type Ia supernovae in the local universe, and connecting to the Hubble flow to determine a value of $H_0$ \citep[see the review by][]{freedman2010}. The {\it Hubble Space Telescope} (HST) key project on extragalactic distance scale utilized traditional Cepheid-Supernovae distance ladder to estimate a 10$\%$ precise $H_0$ \citep{freedman2001}, thus, settling a debate on the factor of two uncertainty in the expansion rate of the universe. In the past decade, {\it Supernovae and $H_0$ for the Equation of State} (SH0ES) project has made a significant progress in reducing the systematics in Cepheid-Supernovae distance ladder to $2\%$ \citep{riess2011, riess2016, riess2019}. However, improved precision of local $H_0$ measurements have resulted in a tension with cosmic microwave background based {\it Plank} mission results \citep{planck2018}. The current $\sim 9\%$ discord in the $H_0$ measurements between two extreme ends of the universe hints at possible new physics in the standard model and is one of the key ongoing problems in modern cosmology \citep{freedman2019, riess2019, verde2019}.

Although Cepheids have been used successfully for the cosmic distance scale, their PLRs suffer from several systematic uncertainties that limit achieving a sub-percent precision in distance determination. The primary source of uncertainty arises from the lack of precise absolute calibration of Cepheid PLRs in our own Galaxy. Apart from the statistical photometric uncertainties, metallicity effects on PLRs and extinction corrections also contribute to the scatter in the Leavitt law thus limiting the precision of distance estimates to individual Cepheids. I will now discuss the recent progress in the Cepheid PLRs and some possible sources of uncertainties both from theoretical and observational side, wherever possible, in the following sections.

\subsection{Multiband Period-Luminosity Relations}

\subsubsection{LMC calibrations:}

Classical Cepheids in the LMC have played a crucial role in providing the calibration of the first-rung of the cosmic distance ladder. More than a century after the discovery of Cepheid PLR, \citet{soszynski2017a} claimed to have concluded the work by Heneritta Leavitt on identifying Cepheid variables in the Magellanic Clouds. OGLE survey has discovered more than 9500 classical Cepheids in the Magellanic Clouds allowing an empirical derivation of precise PLRs at optical wavelengths. Fig.~\ref{fig:plr_cep} shows PLRs for classical Cepheids at optical and NIR wavelengths. The scatter in the optical band Cepheid PLRs is significant ($\sim 0.2$ mag in $V$ and $\sim 0.15$ mag in $I$) due to the finite width of the IS. Once a color term as a proxy for temperature is included (extinction corrections are already applied), the scatter in these relations is reduced to within the observational uncertainties. Optical data for classical Cepheids in the Magellanic Clouds from the OGLE survey have been used to derive PLRs independently in several studies \citep{ngeow2015, bhardwaj2016b, bhardwaj2016c, wielgorski2017, gieren2018}. We list the $I$-band PLRs for fundamental-mode Cepheids in the Magellanic Clouds for a relative comparison in the form of equation~(\ref{eq:plr}):

\begin{align}
\label{eq:plr_cep_opt_mc}
I_{~\textrm{LMC}}&= 16.892 - 2.997\log(P) ~~~~~(\sigma=0.15),\nonumber \\
I_{~\textrm{SMC}}&= 17.264 - 2.947\log(P) ~~~~~(\sigma=0.22). 
\end{align}

These relations are adopted from \citet{wielgorski2017} and the statistical uncertainties on the slopes and zero-points are $\lesssim 0.02$ mag. Generally, the zero-point of the PLRs is adopted at 10 days or at the mean of underlying period range to minimize the correlated errors due the derived slopes. The slopes of $I$-band Cepheid PLRs in the LMC and SMC are consistent within uncertainties. The optical PLRs of Cepheids have been used extensively for the distance determination \citep[see reviews by][]{madore1991, feast1999, sandage2006, freedman2010}. However, significant scatter ($\sim 0.2$ mag) in the optical PLRs due to the temperature variations, extinction, and metallicity limits their use in the era of precision cosmology. 

In the past two decades, significant progress has been made in deriving precise PLRs for classical Cepheids at NIR wavelengths. The pioneering work of \citet{mcgonegal1982} showed that the scatter of Cepheid PLRs even with random phase observations at NIR wavelengths is almost $2.5$ times smaller than at bluer wavelengths. It is well known that temperature variations are significantly smaller at longer wavelengths and the impact of extinction is about ten times less in $K$-band compared to optical wavelenghts \citep{madore1991}. Therefore, both the impact of differential extinction and the measurement uncertainties on reddening are reduced significantly. Further, the pulsation amplitudes are smaller than in the optical bands allowing accurate mean-magnitude determination with sparsely sampled light curves. Also, the light curves in the infrared are typically sinusoidal and thus easier to model generating excellent templates. This allows for more precise measurement of period and mean magnitudes from fewer epochs which is particularly important for more distant systems where deep observations are very limited. While all these advantages and less sensitivity to metallicity effects makes infrared PLRs excellent tools for distance determination, smaller amplitude variations also create difficulty in their identification and classification. 

One of the earliest statistically significant sample of 92 LMC Cepheids with NIR light curves was provided by \citet{persson2004}. The authors also derived Cepheid PLRs and PLC relations with a scatter of $\sim 0.13$ mag but their sample predominantly included long-period Cepheids. The increasingly larger sample of Cepheids with NIR time-series are available with time-domain surveys such as VISTA NIR survey of the Magellanic Clouds \citep[VMC,][]{cioni2011} which is targetting almost all OGLE fields in the $JK_s$-bands. Preliminary results on Cepheid PLRs in the LMC in $JK_s$-bands from the VMC survey were provided by \citet{ripepi2012} and \citet{morretti2014}. Although, the VMC survey does not cover $H$-band, it is expected to provide near-complete complementary sample of $JK_s$ observations to OGLE Cepheids in the Magellanic Clouds \citep{ripepi2017}.

Another excellent sample of NIR light curves of $\sim1500$ Cepheids in the central bar of the LMC was provided by the LMC NIR synoptic survey \citep{macri2015}. Fig.~\ref{fig:plr_cep} shows the Cepheid PLRs from the survey of \citet{macri2015} where the scatter in $K_s$ band PLR is only $\sim0.08$~mag. While this survey provided homogeneous time-series of Cepheids in the LMC, single-epoch NIR observations of larger samples of Cepheids have also been used extensively in deriving PLRs and Cepheid-based distance determinations \citep[for example,][]{ita2004a, ita2004b, inno2013}. \citet{ripepi2017} also provided time-series NIR photometry for Cepheids in the SMC from the VMC survey. The $K_s$-band PLRs for Cepheids in the LMC \citep{macri2015} and SMC \citep{ripepi2017} in the form of equation~(\ref{eq:plr}) are listed below.

\begin{align}
\label{eq:plr_cep_nir_mc}
K_{s~{\textrm{LMC}}} &= 16.023 - 3.247\log(P) ~~~~~(\sigma=0.09), \nonumber\\
K_{s~{\textrm{SMC}}} &= 16.530 - 3.224\log(P) ~~~~~(\sigma=0.17).
\end{align}

In the $K_s$-band, the slopes of the PLRs are similar within the uncertainties ($\lesssim 0.02$ mag) for LMC and SMC Cepheids. The scatter in the $K_s$-band PLRs has reduced significantly ($\sim 45\%$ for LMC Cepheids and $\sim 25\%$ for SMC Cepheids) as compared to $I$-band (equation~\ref{eq:plr_cep_opt_mc}). The difference in the zero-points gives a relative distance between the Clouds and a precise calibration of LMC Cepheid PLRs can be used to estimate robust distance to the SMC. At present, the most precise primary calibration of Cepheid PLRs for distance scale studies is based on LMC anchored using its $\sim1\%$ accurate late-type eclipsing binary distance \citep[$\mu_{LMC}=18.477 \pm 0.004$ (statistical) $\pm 0.026$ (systematic) mag][]{piet2019}.  

\subsubsection{Galactic calibrations:}

Despite the significant use of Cepheids for extragalactic distance determinations, the calibrations of Galactic Cepheid PLRs are not as precise as their LMC counterparts.  The main reason is that the precise geometric distances to Galactic Cepheids were available only for a small sample with parallaxes from {\it Hipparcos} \citep{van2007} and {\it HST} \citep{benedict2007, riess2014}. This is changing with increasingly accurate parallaxes from progressive {\it Gaia} data releases providing unprecedently precise astrometry \citep{lindegren2016, clementini2017, ripepi2018}. In the pre-{\it Gaia} era, the most accurate parallaxes for Cepheids were limited to nearby objects \citep[$D\lesssim 4$~kpc with {\it HST,}][]{benedict2007, riess2014, riess2018}. Cepheid distances have also been measured to relatively high precision by a number of independent-methods such as the Infrared Surface Brightness technique and Baade-Wesselink methods, cluster main-sequence fitting, and SpectroPhoto-Interferometry \citep[see, ][and references therein for more details]{gieren1998, kervella2004, fouque2007, turner2010, storm2011, merand2015, gieren2018}.

The uncertainties in the available Galactic calibrations of Cepheid PLRs are evident from the fact that their application results in a Cepheid-based LMC distance having systematics typically more than $3\%$ using most empirical calibrations, while a geometric distance to the LMC is now known to $1\%$ precision \citep{piet2019}. The different calibrations of Galactic Cepheid PLRs lead to an active debate regarding the universality of Cepheid PLRs between the Galaxy and the LMC as the metallicity and extinction effects may change the slope as well as the intercept of the PLRs \citep{sandage2006}. For example, a multiwavelength calibration of Galactic Cepheid PLRs was carried out by \citet{fouque2007} using distances to Cepheids based on several independent methods mentioned previously, including trigonometric parallaxes. The authors did not find any significant variation in the Cepheid PLRs between the Galaxy and the LMC. \citet{storm2011} calibrated PLRs using distances derived from infrared surface brightness method and found no variation in the slope and a marginal change in the zero-point between Galactic and LMC Cepheid PLRs in the NIR bands. Several other studies also provided calibration of Galactic Cepheid PLRs \citep{ngeow2012, gmat2013, bhardwaj2016a} but they all used nearly the same sample of distances to nearby Cepheids. The Galactic Cepheid PLRs based on Baade-Wesselink distances from \citet{gieren2018} differ from their Magellanic Cloud counterparts at all wavelengths. The $I$ and $K_s$-band PLRs from \citet{gieren2018} are given here in the form of equation~(\ref{eq:plr}):

\begin{align}
\label{eq:plr_cep_mw}
\textrm{M}_{I_{~{\textrm{MW}}}}  &= -2.149 - 2.664\log(P) ~~~~~(\sigma=0.21),\nonumber\\
\textrm{M}_{K_{s~{\textrm{MW}}}} &= -2.424 - 3.258\log(P) ~~~~~(\sigma=0.23),
\end{align}

\noindent where the uncertainties on the slopes and zero-points are $\sim 0.1$ mag and $\sim 0.03$ mag, respectively. Comparing with the equations (\ref{eq:plr_cep_opt_mc}) and (\ref{eq:plr_cep_nir_mc}), it is evident that the slopes of $K_s$-band PLRs are similar between the Galaxy and the Magellanic Clouds while the slopes of $I$-band PLRs in the Milky Way differs from the ones in the Magellanic Clouds but still consistent within $3\sigma$ uncertainty. \citet{bhardwaj2016a} also provided absolute calibration of the Galactic relations based on several distance determination methods accounting for the
intrinsic scatter of each technique. The authors derived a $K_s$-band PLR similar to the equation (\ref{eq:plr_cep_mw}) and determined an independent distance to the LMC of $\mu_{LMC} = 18.47 \pm 0.07$ (statistical) mag based on NIR photometry of Cepheid from \citet{macri2015} in concordance with the geometric distance.

\subsubsection{Theoretical calibrations:}

Multiwavelength calibrations of Cepheid PLRs based on stellar pulsation models have been used to provide comparison with the empirical relations and explore possible systematics in the predicted distance scale. The nonlinear modelling of Cepheids incorporating coupling between hydrodynamical equations and time-dependent convection by \citet{stellingwerf1982, stellingwerf1984, bono1994} formed a solid basis for such comparisons. \citet{bono1999b} derived theoretical PLR and PLC relations for models representative of Cepheids in the Galaxy and the LMC and showed that theoretical $VK_s$-band relations are consistent with empirical investigations. \citet{caputo2000} extended model computations to multiple wavelengths and their PLRs were also fairly consistent with observations but also displayed some dependence on metallicity. \citet{bono2002a} also presented first-overtone Cepheid models in the Magellanic Clouds and suggested that a mild overshooting in pulsation models is needed for the consistency between empirical and theoretical PLRs. They did not find any metallicity dependence and estimated distance to the Magellanic Clouds that agree at the 2\% level with empirical results.

\begin{figure}[!t]
\begin{center}
\includegraphics[width=0.5\textwidth]{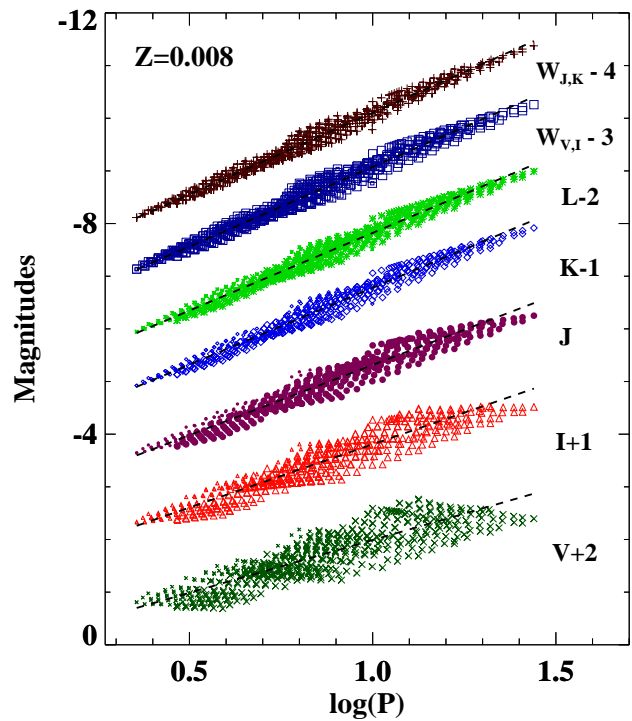}
\caption{Multiwavelength theoretical PLRs of models representative of fundamental and first-overtone mode Cepheids in the LMC with metal abundance Z=0.008 \citep{marconi2013a}. Small symbol size represents first-overtone mode Cepheids. The dashed lines represent linear regression over the entire period range.}
\label{fig:model_cep_plr}
\end{center}
\end{figure}

\begin{figure*}[!t]
\begin{center}
\includegraphics[width=\textwidth]{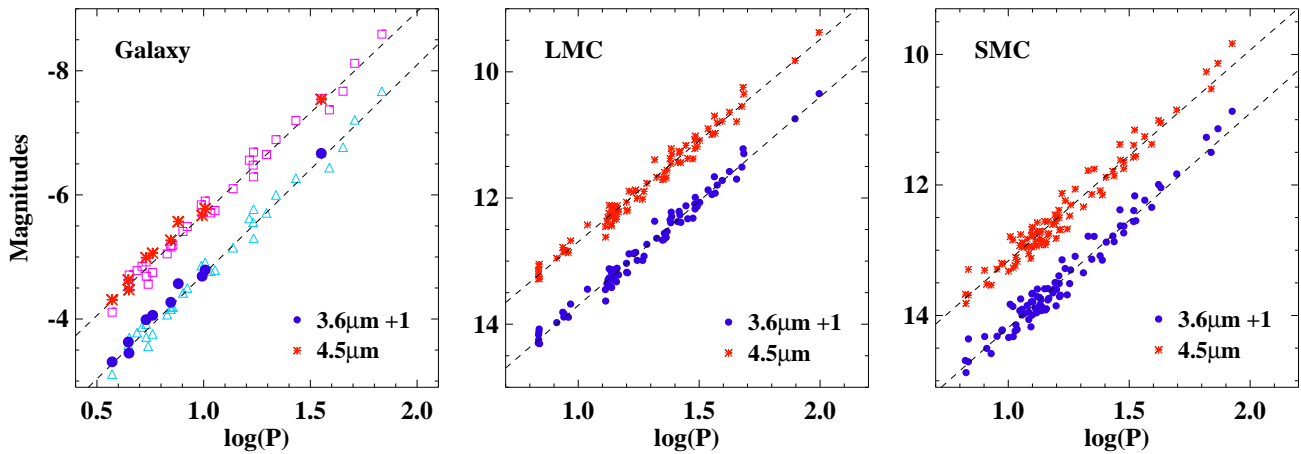}
\caption{The PLRs of classical Cepheids at mid-infrared wavelengths in the Galaxy, LMC and SMC adopted from \citet{monson2012}, \citet{scowcroft2011} and \citet{scowcroft2016}, respectively. The solid lines represent linear regression over entire period range.}
\label{fig:mir_cep}
\end{center}
\end{figure*}

In stellar pulsation models, for a given chemical composition, the major systematics in the absolute calibration of Cepheid PLRs arises due to poorly understood phenomenon like mass-loss, core overshooting and rotation. It is very difficult to disentangle the effects of these phenomenon on the mass-luminosity relation of classical Cepheids adopted as input to the pulsation models. Note that the canonical mass-luminosity relations are those that come from stellar evolutionary calculations and the non-canonical mass-luminosity relations 
typically have brighter luminosity levels by 0.25 dex to account for non-standard phenomenon \citep{marconi2013}. However, the zero-point of the adopted mass-luminosity also affects the zero-point of PLRs. For example, an increase in the luminosity level by 0.25 dex at fixed mass, due to one or more of the above mentioned non-standard phenomena, implies a decrease of 0.2 mag ($10\%$ on distance) in the estimated distance moduli from the PLRs \citep{marconi2005, fiorentino2007}. Furthermore, the zero-point of theoretical PLRs is also dependent on the treatment of convective efficiency through the variation in the mixing-length parameter in the pulsation models \citep{fiorentino2007}. I will also discuss the theoretical predictions of chemical composition on Cepheid PLRs later when comparing to empirical investigations. 

Using stellar evolutionary models, \citet{anderson2014} investigated the effect of rotation on Cepheids and found that it affects the mass-luminosity relations particularly during the blue loop phase. The authors showed that the difference in Cepheid luminosities between different crossings of the IS also increases with faster rotation. Furthermore, rotation also contributes to the dispersion in Cepheid PLRs \citep{anderson2014, anderson2016}, and more importantly, can also resolve the {\it Cepheid mass discrepancy} problem \citep{stobie1969a, stobie1969b}. 

Fig.~\ref{fig:model_cep_plr} displays PLRs at multiple wavelengths for metal-abundance (Z=0.008, Y=0.25) representative of Cepheids in the LMC. The first-overtone mode Cepheids are fundamentalized using the equation: $\log(P_{FU}) = \log(P_{FO}) + 0.127$. The Cepheid models are adopted from \citet{marconi2013a} and used in \citet{bhardwaj2017}. These models include Cepheid masses from $4.5-9 M_\odot$ adopting both canonical and non-canonical mass-luminosity relations, and both the standard ($\alpha=1.5$) and increased convective efficiency ($\alpha=1.8$). The PLRs for fundamental mode Cepheids in the period range, $0.45 < \log(P) < 1.45$ days, are listed below: 

\begin{align}
\label{eq:plr_cep_th}
\textrm{M}_{I_{~{\textrm{Theory}}}}  &= -2.179 - 2.626\log(P) ~~~~~(\sigma=0.19),\nonumber\\
\textrm{M}_{K_{s~{\textrm{Theory}}}} &= -2.716 - 3.062\log(P) ~~~~~(\sigma=0.11).
\end{align}

While the theoretical $I$ \& $K_s$-band Cepheid PLRs in the LMC are shallower than the empirical calibrations in the Magellanic Clouds, $I$-band PLR is  consistent with the empirical calibration in the Galaxy. Note that the slopes of $I$ \& $K_s$-band theoretical PLRs listed in \citet[Table~2,][]{bono2010} are in excellent agreement with empirical relations but vary significantly between short ($\log(P) \lesssim 1$ day) and long-period  ($\log(P) > 1$ day) Cepheids. The slopes of PLRs in the equation~(\ref{eq:plr_cep_th}) are also in agreement with those of long-period Cepheids from \citet{bono2010}. Apart from the period range under consideration, theoretical PLRs also depend on the composition, adopted mass-luminosity relation and the efficiency of convection in the pulsation models. 

\subsubsection{Mid-infrared calibrations:}

The mid-infrared (MIR) observations of Cepheids hold a significant advantage with respect to shorter wavelengths because the extinction is more than an order of magnitude smaller ($A_V\sim 15A_{3.6\mu m}$) at 3.6$\mu$m band. Furthermore, the luminosity variations due to pulsations are mostly insensitive to effective temperature. Therefore, amplitude variations, which are smaller than $K$-band, predominantly occur from small radius fluctuations. The infrared Cepheid spectra are also mostly free from line blanketing thus reducing the dependence of the PLRs on metallicity, although CO band-head at $4.5\mu m$ is very sensitive to temperature variations \citep[see][for details]{scowcroft2016a}. Given increasing MIR observations in the past decade, several investigations were aimed at providing empirical calibrations of MIR PLRs for Cepheid variables, in particular, with InfraRed Array Camera \citep[IRAC,][]{fazio2004} onboard {\it Spitzer Space Telescope}.   

High-precision MIR photometry for Cepheids in the Galaxy and the LMC have been used to derive empirical PLRs at these wavelengths \citep{freedman2008, ngeow2008, madore2009, marengo2010}. Most of these studies utilized single-epoch photometry at 3.6, 4.5, 5.8 and 8.0 $\mu$m for Cepheids and the resulting PLRs exhibited a dispersion of $\sim 0.15$~mag, better than the optical counterparts with mean-magnitudes from well-sampled light curves. \citet{marengo2010} used two random epochs of photometry and provided Cepheid MIR PLRs including first-time ever at 24 and 70 $\mu$m wavelengths. The zero-points of their Galactic calibrations were primarily anchored using the HST parallaxes from \citet{benedict2007}. The MIR PLRs of Cepheids were extended to NGC 6822 \citep{madore2009}, IC 1613 \citet{freedman2009}, and for the OGLE sample of fundamental-mode \citep{ngeow2009, ngeow2015} and first-overtone mode Cepheids in the Magellanic Clouds \citep{bhardwaj2016c}.

\begin{figure*}[!t]
\begin{center}
\includegraphics[width=\textwidth]{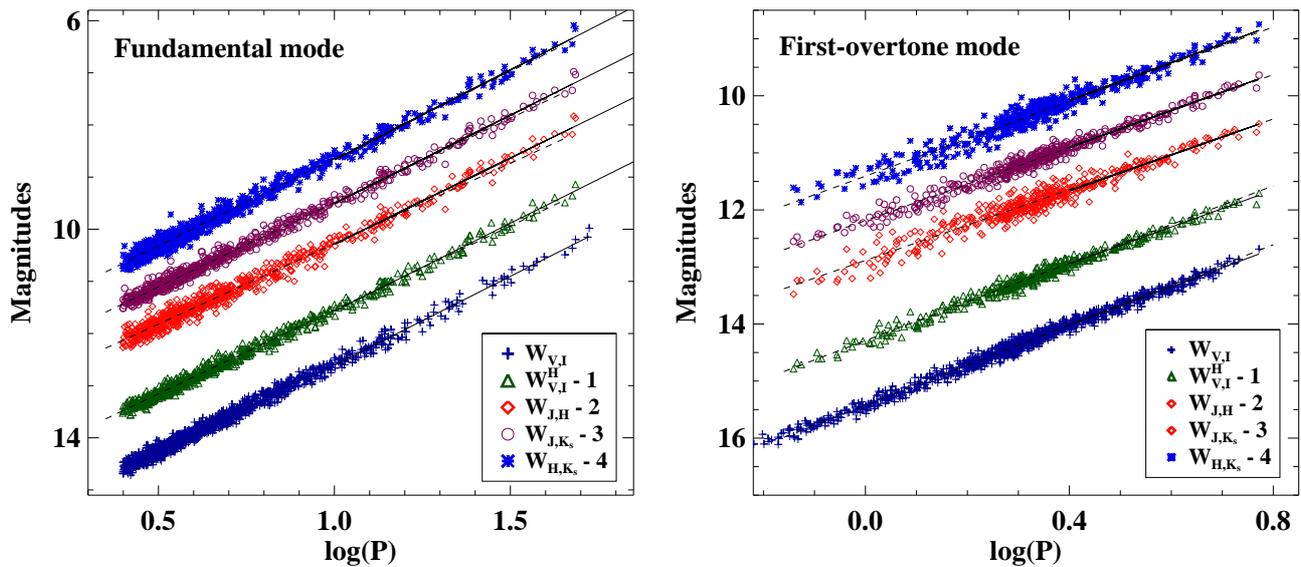}
\caption{Multiwavelength PWRs of fundamental and first-overtone mode classical Cepheids in the LMC. The optical data is taken from OGLE survey \citep{soszynski2015} and the near-infrared photometry is adopted from \citet{macri2015}. The dashed/solid lines represent linear regression over entire/long period range only with a break period at 10 days for fundamental-mode Cepheids and at 2.5 days for first-overtone mode Cepheids.}
\label{fig:pwr_cep}
\end{center}
\end{figure*}

Due to significant advantages of MIR observations, {\it Carnegie Hubble Program} was aimed at measuring a $H_0$ with a precision of $\sim2\%$ using the absolute calibration of Cepheid PLRs at $3.6$ and $4.5\mu m$ \citep{freedman2011}. Time-series observations of Galactic and Magellanic Clouds Cepheids spanning over 24 epochs were obtained as part of this program. Fig.~\ref{fig:mir_cep} shows MIR Cepheid PLRs in the Galaxy, LMC and the SMC from \citet{monson2012}, \citet{scowcroft2011} and \citet{scowcroft2016}. The $3.6\mu m$-band PLRs in these three galaxies are listed below:

\begin{align}
\label{eq:plr_cep_mir}
\textrm{M}_{3.6\mu m~{\textrm{MW}}}  &= -2.49 - 3.33\log(P) ~~~~~(\sigma=0.09),\nonumber\\
\textrm{m}_{3.6\mu m~{\textrm{LMC}}}  &= 16.01 - 3.31\log(P) ~~~~~(\sigma=0.11),\nonumber\\
\textrm{m}_{3.6\mu m~{\textrm{SMC}}}  &= 16.50 -3 .31\log(P) ~~~~~(\sigma=0.16).
\end{align}

Note that the Galactic calibration was still based on the {\it HST} parallaxes and other independent methods discussed previously but the scatter in MIR Cepheid PLRs was reduced to 0.1~mag with a zero-point uncertainty of only $\sim 3\%$. Equation~\ref{eq:plr_cep_mir} suggests that $3.6\mu m$-band PLR in the Galaxy and Magellanic Clouds is universal. The Galactic calibration leads to a precise distance to the LMC ($\mu_{LMC} = 18.48 \pm 0.04$ mag) and SMC ($\mu_{LMC} = 18.96 \pm 0.04$ mag).  \citet{scowcroft2016a} also found that  ([3.6]-[4.5]) colour is a reliable metallicity indicator for Cepheids. The Galactic (zero-point) and LMC (slope) calibrations of Cepheid MIR PLRs led to a factor of three decrease in the systematic uncertainties resulting in a $2.8\%$ precise $H_0$ measurement \citep{freedman2012}. The absolute calibrations of Cepheid PLRs at MIR wavelengths will be critical in the era of {\it James Webb Space Telescope} (JWST) thanks to the higher resolution and higher sensitivity enabling access to crowded and extincted regions of more distant supernovae host galaxies. 

\subsection{Period-Wesenheit relations}

Multiwavelength observations of Cepheids (or RR Lyrae) allow us to obtain distances and color excess simultaneously. Given a reddening law and photometry in at least two filters, PLRs can be used to solve for two unknowns - distance modulus ($\mu$) and extinction ($A_\lambda$). Similar to this approach, to circumvent the problem of extinction, \citet{vanden1975, madore1982} constructed reddening free Wesenheit magnitudes that are used in deriving Period-Wesenheit relations (PWRs). At given wavelengths, say $\lambda_1$, $\lambda_2$, $\lambda_3$, the Wesenheit functions can be written in the following form:

\begin{align}
\label{eq:pw_all}
W^{\lambda_{3}}_{\lambda_{2},\lambda_{1}}~ & = ~~m_{\lambda_{3}} - R^{\lambda_{2},\lambda_{1}}_{\lambda_3} (m_{\lambda_{2}}-m_{\lambda_{1}}), \nonumber \\
R^{\lambda_{2},\lambda_{1}}_{\lambda_3} ~& = ~~\left[\frac{A_{\lambda_{3}}}{E(m_{\lambda_{2}}-m_{\lambda_{1}})}\right], 
\end{align}

\noindent where $m_{\lambda_i}$ represents the mean magnitude at wavelength $\lambda_i$ and $\lambda_{1}>\lambda_{2}$.
Generally, the superscript $\lambda_3$ is dropped from $W^{\lambda_{3}}_{\lambda_{2},\lambda_{1}}$ for simplicity when $\lambda_1 = \lambda_3$. The total-to-selective absorption ratios are adopted based on a reddening law \citep[for example,][]{card1989} assuming a value of $R^{B,V}_V$ \citep{fouque2007, inno2013}. The Wesenheit relations are a proxy for PLC relations such that the effects of the width of the IS are reduced due to the additional color term. Fig.~\ref{fig:pwr_cep} displays optical and NIR PWRs for classical Cepheids in the LMC from \citet{bhardwaj2016a}. The optical PWRs for Cepheids in the Magellanic Clouds from the OGLE survey are derived as $W_{V,I} = I - 1.55(V - I)$, and the empirical relations are listed as follows:  

\begin{align}
\label{eq:pwr_cep_opt_mc}
W_{{V,I}~{\textrm{LMC}}} &= 15.904 - 3.332\log(P) ~~~~~(\sigma=0.083), \nonumber\\
W_{{V,I}~{\textrm{SMC}}} &= 16.385 - 3.330\log(P) ~~~~~(\sigma=0.146).
\end{align}


The dispersion in the optical Wesenheit ($W_{V,I}$) is significantly smaller when compared to optical LMC Cepheid PLRs in the $V$ and $I$-bands ($\sim 60\%$ and $\sim 30\%$, respectively, see Fig.~\ref{fig:plr_cep}). Theoretically, NIR and optical-NIR PWRs have additional advantage because these relations are independent of metal-abundance and linear over the entire period range \citep{bono2010}. The most commonly used NIR PWR is defined as $W_{J,K_s} = K_s - 0.69(J-K_s)$, and these relations in the Galaxy and the Magellanic Clouds are:

\begin{align}
\label{eq:pwr_cep_nir_mc}
W_{{J,K_s}~{\textrm{MW~}}}  &= -2.63 - 3.36\log(P) ~~~~~(\sigma=0.24),\nonumber\\ 
W_{{J,K_s}~{\textrm{LMC}}} &=  15.76 - 3.28\log(P) ~~~~~(\sigma=0.08),\nonumber\\ 
W_{{J,K_s}~{\textrm{SMC}}} &=  16.36 - 3.33\log(P) ~~~~~(\sigma=0.16),
\end{align}

\noindent which are adopted from \citet[MW,][]{gieren2018}, \citet[LMC,][]{bhardwaj2016a} and \citet[SMC,][]{ripepi2017}, respectively. The slopes of the PWRs are consistent within their uncertainties ($0.1$ mag for the MW and $\sim 0.02$ mag for the Magellanic Clouds). Several theoretical and empirical studies have employed different combinations of filters to derive PLRs and subsequently estimate Cepheid-based distances \citep[][and references therein]{fiorentino2007, bono2010, ngeow2012, inno2013, bhardwaj2016a}. 

It is important to emphasize that SH0ES project utilizes W$^H_{V,I}$ Wesenheit magnitudes in deriving PWRs (see Fig.~\ref{fig:pwr_cep})  for $H_0$ determination. The use of three band PWRs leads to smaller dispersion possibly due to lower correlated systematics in photometry used in the color term. The total-to-selective absorption ratio ($R^H_{V,I} = 0.41$) is small and any possible variations in this parameter, due to the choice of adopted reddening law, do not lead to large systematics in PWRs. Based on stellar evolutionary models, \citet{anderson2016} also suggested that W$^H_{V,I}$ Wesenheit leads to smallest scatter in the PWRs.

\begin{figure}[!t]
\begin{center}
\includegraphics[width=0.48\textwidth]{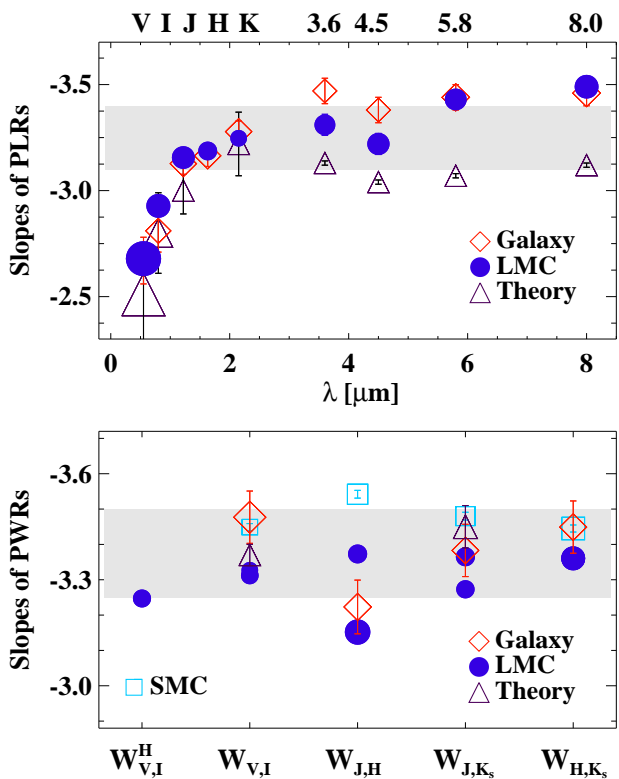}
\caption{{\it Top:} Multiwavelength slopes of Cepheid PLRs as a function of wavelength. {\it Bottom:} A comparison of the slopes of Cepheid PWRs. Shaded regions give a crude approximation of the range of most commonly derived slopes of Cepheid PLRs and PWRs for distance measurements. The bigger symbol size represents larger dispersion in the underlying PLRs or PWRs.}
\label{fig:cep_slope}
\end{center}
\end{figure}

\subsubsection{Comparison of multiband slopes:}

The slopes of fundamental-mode Cepheid PLRs as a function of wavelength are shown in Fig.~\ref{fig:cep_slope}. The slopes of LMC Cepheid PLRs are adopted from \citet[$VI$,][]{bhardwaj2016c}, \citet[$JHK_s$,][]{macri2015}, \citet[$3.6~\&~4.5\mu m$,][]{scowcroft2011} and \citet[$5.8~\&~8.0\mu m$,][]{madore2009a}. The Galactic calibrations are adopted from \citet[$VI$,][]{storm2011}, \citet[$JHK_s$,][]{bhardwaj2016a}, and \citet[$3.6~-~8.0\mu m$,][]{marengo2010}. The slopes of the empirical PLRs for Cepheids in the Galaxy and LMC from different studies are consistent within uncertainties. The theoretical calibrations are adopted from \citet[$VIJK_s$,][]{bono2010} and \citet[$3.6~-~8.0\mu m$,][]{marengo2010} for metal-abundance (Z=0.02) representative of Cepheids in the Galaxy. The theoretical slopes of the $VIJK_s$-band PLRs also agree well with empirical relations but difference in the slopes is relatively larger at wavelengths longer than $K_s$-band. However, \citet{marengo2010} also found that the slopes of Galactic MIR PLRs calibrated based on the astrometric distances are in excellent agreement with the theoretical predictions.

The bottom panel of Fig.~\ref{fig:cep_slope} displays the slopes of PWRs from \citet{inno2013} and \citet{bhardwaj2016a} for LMC Cepheids, \citet{storm2011} and \citet{bhardwaj2016b} for Galactic calibrations, and \citet{inno2013} for the SMC Cepheids. The theoretical calibrations are adopted from \citet{bono2010}. The slopes of PWRs are in good agreement among different studies except in the case of $W_{J,H}$ Wesenheit. The inconsistency in slopes of PLRs and PWRs may be due to, for example,  different sample sizes, different photometric systems, the uncertainty on the reddening correction, and single-epoch versus time-domain data in different studies. Regardless, the range of slopes of the optical and NIR PWRs is significantly smaller than the multiwavelength Cepheid PLRs suggesting that the PWRs are indeed excellent tools for Cepheid-based distance measurements. An example of application of different Cepheid PLRs and PWRs is the Araucaria Project \citep{piet2006} that has utilized variable stars as standard candles to measure distances to several Local Group galaxies \citep[for example,][]{piet2007, gieren2013, zgirski2017}, Sculptor Group galaxies \citep{gieren2005, gieren2009}, and improve the calibration of extragalactic distance scale.

\subsection{Systematic uncertainties in the Cepheid-based distance scale}

\subsubsection{Photometric mean-magnitudes:}

The photometric uncertainties in individual measurements for Cepheid variables contribute to the observed dispersion in the PLRs through the estimates of mean-magnitudes. Despite the increase in NIR observational facilities in the past decade, infrared time-series is limited and the light curves are typically sparsely sampled. Since Cepheids cover a wide period range, optimizing a cadence to obtain well-sampled light curves without large phase gaps is difficult when having only a few epochs of measurements. In the case of HST observations of Cepheids in the supernovae host galaxies at a distance of 20-40 Mpc, photometric uncertainties due to blending alone can be a few tenths of magnitudes and the random phase corrections can also amount to $\sim 0.15$ mag of additional errors \citep{riess2016}. The photometric uncertainties in the nearby galaxies are typically smaller ($\lesssim 0.1$ mag) on individual measurements.

The templates for Cepheid light curves are useful to estimate precise mean-magnitudes from sparsely sampled light curves. \citet{soszynski2005} provided NIR templates for classical Cepheids based on a small sample of calibrating Cepheids in the Galaxy and LMC. The new NIR templates for Cepheids were provided by \citet{inno2015} based on a very large set of $\sim 800$ Galactic and Magellanic Cloud Cepheids. These templates are divided in ten period bins to account for a wide range of Cepheid periods and allow mean-magnitude estimates with a precision ($\sim 0.02$ mag) only limited by the intrinsic accuracy of the templates. One new addition to these templates was the use of the phase of mean-magnitude along the rising branch as an anchor of phase zero-point which allows proper sampling of the light-curves of bump Cepheids. 

\subsubsection{Linear versus non-linear period-luminosity relations:}

The application of Cepheid PLRs to the distance scale follows a basic assumption that these relations are linear over the entire period range. The non-linearity of the PLRs has been a subject of many studies in the past decade \citep{tammann2003, sandage2004, ngeow2005, ngeow2006a, ngeow2008, varela2013, bhardwaj2016b}. The Cepheid PLRs in the LMC exhibit a change in the slope at 10 days for fundamental-mode Cepheids and at 2.5 days for first-overtone mode Cepheids at optical wavelengths \citep{bhardwaj2016b}. The short-period break at 2.5 days has  been noted for both fundamental and first-overtone mode Cepheids in the SMC \citep{bauer1999, ngeow2015, bhardwaj2016c}. The break in the PLRs at 10 days has also been observed for Cepheids in M31 \citep{kodric2015, kodric2018}. Furthermore, possible non-linearities in Cepheid PLRs have been investigated using a number of independent methods including both parametric and non-parametric statistical tests \citep{kanbur2007, varela2013, bhardwaj2016b}. \citet{bhardwaj2016b} found evidence of a break at 10 days in optical Cepheid PLRs and around 18 days in the NIR PLRs in the LMC. However, the authors did not find any significant bias between distance estimates using linear and non-linear models of PLRs when combining the LMC sample with Cepheids in the supernovae host galaxies. 

\citet{ngeow2006} estimated distances to the type Ia supernovae using calibrated linear and non-linear Cepheid PLRs and found marginal difference in the $H_0$ values and corresponding systematic uncertainties. In the traditional distance ladder, only long-period Cepheids in the LMC were used for the calibration of zero-point since distant Cepheids observed in the supernovae host galaxies predominantly have periods greater than 10 days \citep[for example in the SH0ES project,][]{riess2011}. However, a two-slope model for the calibrated Cepheid PLRs can provide a stronger constraint on the global slope of the PLR and also reduce corresponding systematic uncertainty \citep{bhardwaj2016b}. \citet{riess2016} included several variants of non-linear $W^H_{V,I}$ PWR in their analysis for the determination of the $H_0$ including two-slope model with possible break periods at 10 days or 60 days. The authors found negligible contribution to the systematic uncertainties on the $H_0$ estimates between the linear and non-linear model of Cepheid PLRs. However, considering that Cepheid PLRs in the supernovae host galaxies presently have a typical dispersion more than three times the scatter in the calibrator LMC PLRs, any possible changes in the slope of PLRs need careful investigation when precise relations become available with JWST and the extremely large ground-based telescopes.

The theoretical explanation for the cause of possible non-linearities in the Cepheid PLRs is not well-understood. \citet{kanbur2005a, ngeow2006a, kanbur2010} argued that the changes in the slope of LMC Cepheid period-color relation (and subsequently PLR) as a function of pulsation phase contribute to the observed non-linearities. The period-color and amplitude-color relations of long-period ($> 10$ days) classical Cepheids in the LMC exhibit a nearly flat slope at maximum light but a non-zero slope at minimum-light \citep{bhardwaj2014}. \citet[][and references therein]{kanbur2010} related these variations in the period-color relations with the interaction of hydrogen ionization front and the stellar photosphere and the properties of the Saha ionization equation, and suggested that the changes in the period-color relations affect the PLRs through PLC relations. However, the changes in the slope of Cepheid PLRs are also strongly correlated with the sharp structural changes in the Fourier parameters at the break periods \citep{bhardwaj2016b, bhardwaj2016c}. At the same time, metallicity is also expected to play a crucial role as metal-poor Cepheids are brighter than their metal-rich counterparts at fixed period \citep[][see next subsection]{romaniello2008}. The observed non-linearity at the long-period end can be an observational bias as including brightest LMC Cepheids from the OGLE shallow survey \citep{ulac2013} masks the evidence of non-linearity in optical Cepheid PLRs at 10 days \citep{bhardwaj2016b}.

\subsubsection{Metallicity effects:}

One of the most crucial issues in the Cepheid distance scale is the dependence on metallicity of both the slope and zero-point of the PLRs and PWRs. The validity of the basic assumption regarding universality of the Cepheid PLRs in different stellar environment critically depends on negligible metallicity effects. Theoretical studies by \citet{bono1999b, caputo2000b} based on non-linear convective models showed that both the zero point and the slope of the predicted PLRs are significantly dependent on metallicity with the amplitude of the metallicity effect decreasing at the longer wavelengths. At a given wavelength, the slope becomes steeper for lower metal-abundances. These models predicted that at a fixed period, metal-rich Cepheids should be fainter than the metal-poor ones \citep{bono1999b}. Interestingly, the slope of the optical and NIR PWRs is independent of the metal-content \citep{fiorentino2007, bono2010}. However, the metallicity dependence of the zero-point of the PWRs depends on the adopted filters and needs to taken into account. Theoretical models also predict a dependence on helium of Cepheid PLRs \citep{fiorentino2002, marconi2005} which was further investigated by \citet{carini2017}. The latter found negligible effect on PLRs based distance estimates and a systematic uncertainty of up to $7\%$ on PWRs based distances. Metallicity and helium variations simultaneously affect Cepheid (and RR Lyrae) pulsation properties, light curves and the PLRs thus it is difficult to disentangle the two contributions.

Empirically, several independent observations have suggested a wide range of estimates for the metallicity sensitivity on Cepheid distance scale \citep[see Table 1,][]{romaniello2008} that vary from $\sim$ -0.9 mag/dex to negligible dependence on metallicity at optical wavelengths. The indirect measurements of the metallicity in external galaxies mostly based on oxygen nebular abundances of H~II regions showed that the metal-rich Cepheids are brighter than metal-poor ones \citep{kennicutt1998, macri2006}, inconsistent with the predictions of nonlinear convective models \citep{bono2010}.  Several other investigations based on empirical Cepheid PLRs also found similar results \citep{tammann2003, sandage2004, storm2004, groenewegen2004} or negligible metallicity effects \citet{fouque2007}. Based on direct measurements of iron abundances for individual Cepheids, \citet{romaniello2005, romaniello2008} found that Cepheids become fainter as metallicity increases. They found significant metallicity effects on $V$-band PLRs such that metal-rich stars are fainter, a result consistent with theoretical predictions. However, no firm conclusion concerning the metallicity dependence on the $K_s$-band PLR has been achieved \citep{romaniello2008, bono2010}. 

In the last few years, \citet{wielgorski2017} utilized precise Cepheid PLRs in the Magellanic Clouds and found metallicity effects compatible with zero in all bands on PLRs and PWRs. \citet{gieren2018} employed Baade-Wesselink method to determine distances to Cepheids in the Galaxy and the Magellanic Clouds and quantified the strictly differential effect of metallicity on Cepheid PLRs by minimizing systematic zero-point uncertainties. The authors found a metallicity dependence in all bands ($\sim-0.23\pm0.06$ mag/dex in $K$-band) such that the more metal-poor Cepheids are intrinsically fainter than their metal-rich counterparts with similar pulsation periods. \citet{groenewegen2018} used parallaxes for Galactic Cepheids from {\it Gaia} second data release to investigate period-luminosity-metallicity relations and found no significant metallicity term. The author argued that the significant parallax zero-point offset present in {\it Gaia} data leads to systematic uncertainties of the order of 0.15 mag on the distance scale \citep[see also][]{riess2018a}. 

In more distant supernovae host galaxies, it is impossible to measure directly Cepheid metallicity from individual stars. Hence, the mean-metallicity of the host (and target) galaxy is adopted to constrain systematics due to metallicity effect on the $H_0$ estimates. \citet{riess2016} found a metallicity dependence ($\sim -0.24\pm0.06$ mag/dex) similar to \citet[][]{kennicutt1998} which ultimately contributes to 0.5$\%$ systematics in $H_0$ determinations. Even after decades of effort the metallicity effects on Cepheid PLRs are not well-understood and even the sign of metallicity sensitivity is debated. The precise parallaxes from the future {\it Gaia} data releases for Galactic Cepheids with high-resolution spectra \citep[for example,][]{andrievsky2002, lemasle2013, genovali2013, genovali2014, genovali2015, proxauf2018} and spectroscopic abundances for Magellanic Cloud Cepheids \citep{lemasle2017, mancino2020} are essential to resolve metallicity systematics in Galactic and LMC calibration on the Cepheid distance scale.

\subsubsection{Other systematic uncertainties:}

The impact of extinction on Cepheid-based distance measurements has been mitigated by using either Wesenheit functions or PLRs at the infrared wavelengths. However, the choice of adopted reddening law also contributes to the possible systematics due to extinction, specially in the regions with differential reddening where the reddening law may not be universal \citep{nishiyama2006, nishiyama2009, nataf2016}. For example, \citet{dekany2015} identified 35 classical Cepheids in the inner part of the Galactic disc but \citet{matsunaga2016} showed that there is lack of young population in the inner 2.5 kpc region of the Galactic disc except the nuclear stellar disk \citep{matsunaga2011a}. \citet{matsunaga2016} estimated a large impact of the reddening correction based on different reddening laws even at NIR wavelengths leading to an overestimate of distances to Cepheids in \citet{dekany2015} thus locating those in the inner part of the Galactic disc. 

Cepheids in the wide binaries and in open clusters can also contribute to a possible bias in distance estimates with additional light contribution to photometric measurements of extragalactic Cepheids due to blending and changing spatial resolution along the distance ladder \citep{anderson2018}. The authors found a negligible effect due to stellar companions and a relatively larger effect due to cluster populations which amounts to an overestimate of $0.23\%$ in $H_0$ determinations. 

\citet{anderson2019} investigated the impact of time-dilation on Cepheid light curves because redshift dilates the periods of variables in distant supernova-host galaxies relative to periods of those in the calibrator galaxies. He estimated a bias of 0.27\% in the $H_0$ values and argued that this effect will become increasingly relevant for Cepheids in more distant galaxies in the near-future.

\section{RR Lyrae variables as distance indicators}
\label{sec:sec5}

RR Lyrae, being fainter than classical Cepheids, have been used less for distance determinations. This is changing thanks to larger telescopes used for the time-domain surveys and increasing use of infrared observations. RR Lyrae are population II distance indicators and provide an independent primary calibration, and an alternate distance ladder to the traditional Cepheid-Supernovae distance scale. {\it Carnegie-Chicago Hubble Program} aims to use population II RR Lyraes and the tip of the red giant branch stars, and estimate distances to the supernovae host galaxies determining $H_0$ with a precision comparable to current Cepheid-based estimates \citep{beaton2016, freedman2019}. Recently, \citet{freedman2019} determined the tip of the red giant branch and supernovae based value of  $H_0$  with a precision of $2.4\%$ that sits midway the Cepheid-based and {\it Planck} measurements. Considering ongoing Hubble tension, it is important to independently test or complement tip of the red giant branch based distance estimates using independent population II distance indicators such as RR Lyrae variables. I will discuss basic properties of RR Lyrae that are relevant for distance scale studies and focus on NIR PLRs as useful tools to determine robust individual distances in the following sections.

\begin{figure*}[!t]
\begin{center}
\includegraphics[width=\textwidth]{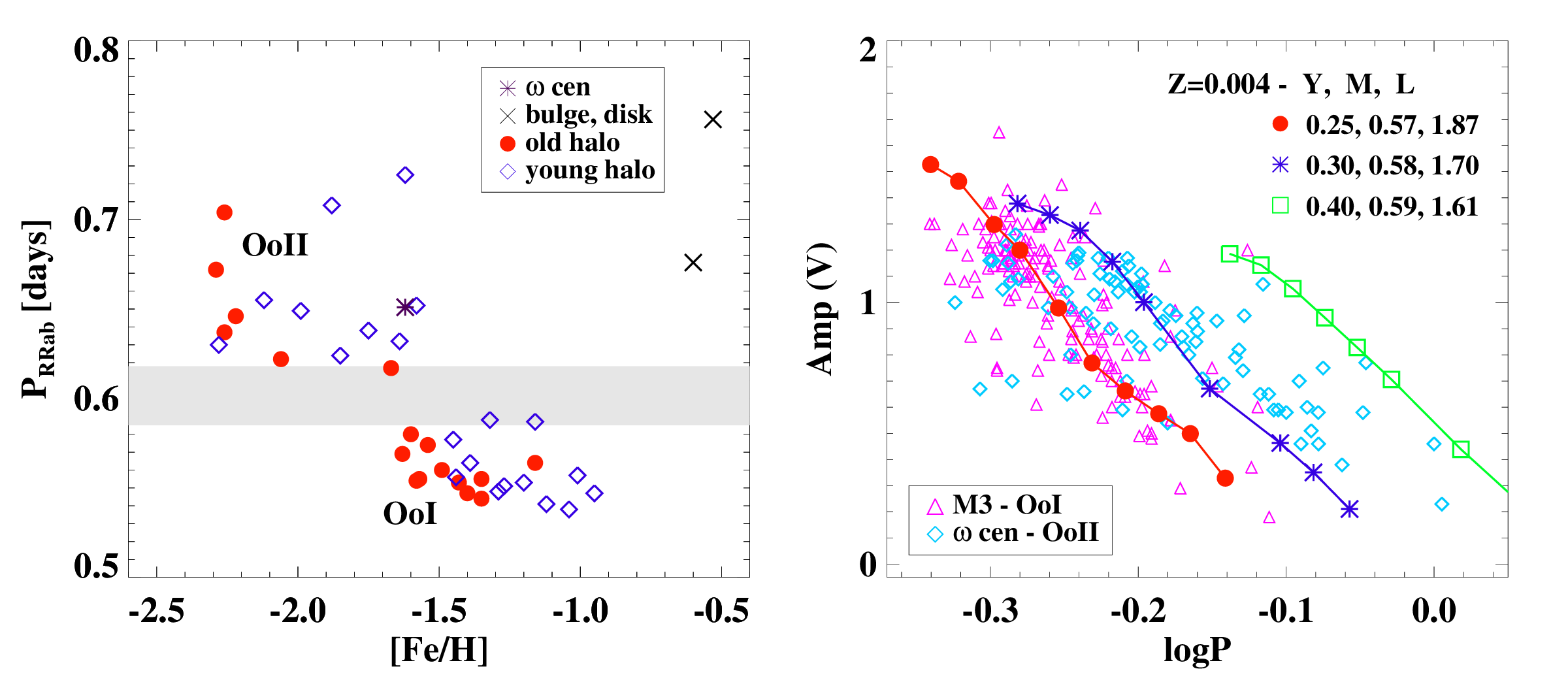}
\caption{{\it Left:} The Oosterhoff dichotomy in the Galactic globular clusters is shown plotting mean period of RR Lyrae against the metallicity using data compiled by \citet{catelan2009}. {\it Right:} Period-amplitude diagram for RRab in Oosterhoff I (M3) \& II ($\omega$ Cen) clusters from the catalog of \citet{clement2001}. For a fixed composition (Z=0.004), RR Lyrae models from \citet{marconi2015, marconi2018} with different helium abundance, mass and luminosities are also overplotted.}
\label{fig:rrl_amp_per}
\end{center}
\end{figure*}

\subsection{Period-amplitude diagrams}

At the beginning of the nineteenth century, Solon Bailey discovered hundreds of variable stars in the globular clusters and introduced RR Lyrae variables of a, b, and c Bailey types \citep{bailey1902}. These types are now typically separated in two classes based on their pulsation mode: RRab (or RR0) are pulsating in the fundamental radial mode while RRc (or RR1) are pulsating in the first-overtone radial mode. Bailey constructed period-amplitude diagrams (or `Bailey' diagrams) for RR Lyrae in the globular clusters and found that these diagrams differ from cluster to cluster. These differences  in the Bailey diagrams can be associated with the Oosterhoff type \citep{oosterhoff1939} of the globular cluster. RRab in Oosterhoff I (OoI) clusters have an average period of 0.55 days and [Fe/H]$\gtrsim -1.5$~dex while RRab in Oosterhoff II (OoII) have average period of 0.65 days and [Fe/H]$\lesssim -1.5$~dex \citep{oosterhoff1939, smith1995, catelan2009, catelan2015}. 

Fig.~\ref{fig:rrl_amp_per} displays Oosterhoff dichotomy in the GGCs. There is a distinct gap between OoI and OoII clusters in the periods versus metallicity plot. However, some metal-rich bulge clusters (for example, NGC 6441, NGC 6388, [Fe/H] $\sim$-0.6 dex) have a larger value of mean-period of RRab than OoII clusters. While GGCs display an Oosterhoff gap, globular clusters and dwarf galaxies in the Milky Way satellite systems do not show such dichotomy \citep[see][for more details]{catelan2009}. \citet{sandage1958a}, using equation (\ref{eq:puls}), showed that the absolute magnitude of the horizontal branches differ by 0.2 mag in $V$-band between OoII and OoI clusters, former being the brighter cluster. Oosterhoff dichotomy can be explained as the difference in the intrinsic luminosity for the RR Lyrae in two clusters, with the higher metallicity OoI clusters being fainter. 

The right panel of Fig.~\ref{fig:rrl_amp_per} shows period-amplitude diagram of RRab variables in a OoI and OoII cluster respectively. It is evident that RR Lyrae in the OoII type cluster have longer periods for a given amplitude. Note that several RR Lyrae stars display modulations in their amplitudes and phases from cycle-to-cycle, a phenomenon known as the Blazhko effect \citep{blazhko1907}, but the origin of these effects is still unexplained despite a number of investigations including those with unprecedently high-precision photometry from {\it Kepler} \citep[][]{jurcsik2009, kolenberg2010, szabo2010, buchler2011, skara2020}. The Blazhko effect in RR Lyrae is one of the main sources of scatter in the observational period-amplitude diagram shown in Fig.~\ref{fig:rrl_amp_per}. The RR Lyrae models from \citet{marconi2015, marconi2018, das2018}, computed at fixed metal content (Z=0.004) and primordial helium contents ranging from Y=0.25 to Y=0.40, are also shown in Fig.~\ref{fig:rrl_amp_per}. Theoretically, Bailey diagrams can also be used to constrain the helium content of RR Lyrae stars. The helium-enhancement leads to a systematic shift in periods which primarily occurs due to increased luminosity levels for similar masses \citep[see,][and references therein]{rood1973, sweigart1998, marconi2018a}. \citet{marconi2018} recently derived helium-abundance (Y=0.245) of RR Lyrae population in the Galactic bulge by comparing their minimum period with pulsation models. Bailey diagrams for RR Lyrae have also been constructed at near UV wavelengths \citep{siegel2015}. The large amplitudes in UV can be useful to constrain the composition effects on RR Lyrae pulsation properties. Further investigations are needed to examine the dependence of UV pulsation properties on metallicity and Oosterhoff classification.

\subsection{The visual magnitude-metallicity relation}

The Oosterhoff dichotomy was later extended to investigate empirical relations between the location of RR Lyrae stars in the period-amplitude diagram and both absolute magnitude and [Fe/H]. \citet{sandage1982} derived an empirical relation between the period shift of a star with a given amplitude from the mean period-amplitude relation and the metallicity. Later, period-amplitude-[Fe/H] relations were used to determine metallicities for RRab stars \citep[see,][]{kinemuchi2006, kunder2009}. However, the correlation between Bailey diagram and [Fe/H] is debated, for example, \citet{bono2007} showed that the Oosterhoff dichotomy plays a key role in determination of period-amplitude diagram rather than the [Fe/H].

The period-amplitude-[Fe/H] relations suggest a continuous correlation between period and both the luminosity and metallicity for RR Lyrae. An empirical relation between RR Lyrae $V$-band absolute magnitude ($M_V$) and stellar metallicity is usually written in the following form:

\begin{equation}
M_V = \alpha + \beta\mathrm{[Fe/H]},\\
\end{equation}

\noindent where, the slope ($\beta$) and the zero-point ($\alpha$) have been determined through several calibrations in the literature \citep[][ and references therein]{fernley1998, caputo2000, clementini2003, bono2003, muraveva2018}. Several investigations have also suggested deviations from the linear form of $M_V$-[Fe/H] relation \citep[see, ][]{caputo2000, bono2003, catelan2004, bono2007}, and also proposed a quadratic form of RR Lyrae $M_V$-[Fe/H] relation \citep{catelan2004, sandage2006, bono2007, muraveva2018}. With {\it Gaia} second data release, \citet{muraveva2018} suggested that the coefficients of metallicity on luminosity is much higher than previous studies in the literature. Although {\it Gaia} parallaxes suffer from systematic zero-point offset which varies with magnitudes, colors and position in the sky \citep{muraveva2018, riess2018a}, the improvement in the precision of parallaxes is significant. For interested readers, linear and quadratic form of $M_V$-[Fe/H] relation from \citet{muraveva2018} are provided here - 

\begin{align}
\nonumber
M_V &= 1.17(\pm0.04) + 0.34(\pm0.03)\mathrm{[Fe/H]},\nonumber\\ 
M_V &= 1.19(\pm0.06) + 0.39(\pm0.10)\mathrm{[Fe/H]} \nonumber \\ 
& 	~~~~+ 0.02(\pm0.04)\mathrm{[Fe/H]^2}. 
\end{align}

The coefficient of quadratic metallicity term in the  $M_V$-[Fe/H] relation is not significant and the zero-points are consistent between both linear and quadratic versions. While this empirical relation is very simple and useful tool to determine distances, several sources of uncertainties affect the precision of distance measurements based on this method. Firstly, the reddening effects are significant at optical wavelengths due to a large total-to-selective absorption ratio in $V$-band, $R_V = 3.1$ \citep{card1989}. Even in moderately extincted regions, the effect of reddening on optical luminosities is typically larger than the metallicity effects. In regions with heavy and differential extinction, reddening effects are a major drawback in using $M_V$-[Fe/H] relation for distance diagnostics. 
Another important concern is the evolutionary effects on RR Lyrae population. Typically the evolved-RR Lyrae have higher luminosities than those of ZAHB RR Lyrae for a given metallicity. However, there is significant overlap in the color-space for RR Lyrae evolving off the ZAHB and the stars on the ZAHB. This evolutionary effect results in the broadening of the HB and the distribution of optical magnitudes \citep{bono1995}. Further, systematic uncertainties in the metallicity measurements due to different metallicity scales and methodologies add another source of uncertainty in distance measurements with the visual magnitude metallicity relation. Note that the extinction and metallicity effects lead to large scatter in the PLRs at wavelengths shorter than $V$-band if the dependence on period is significant. For example, \citet{siegel2015} found a significant dependence of NUV PLRs on metallicity with difference of up to half a magnitude between coolest RRab stars in M3 and M15 clusters.  

\subsection{Multiband Period-Luminosity relations}

\begin{figure*}[!t]
\begin{center}
\includegraphics[width=\textwidth]{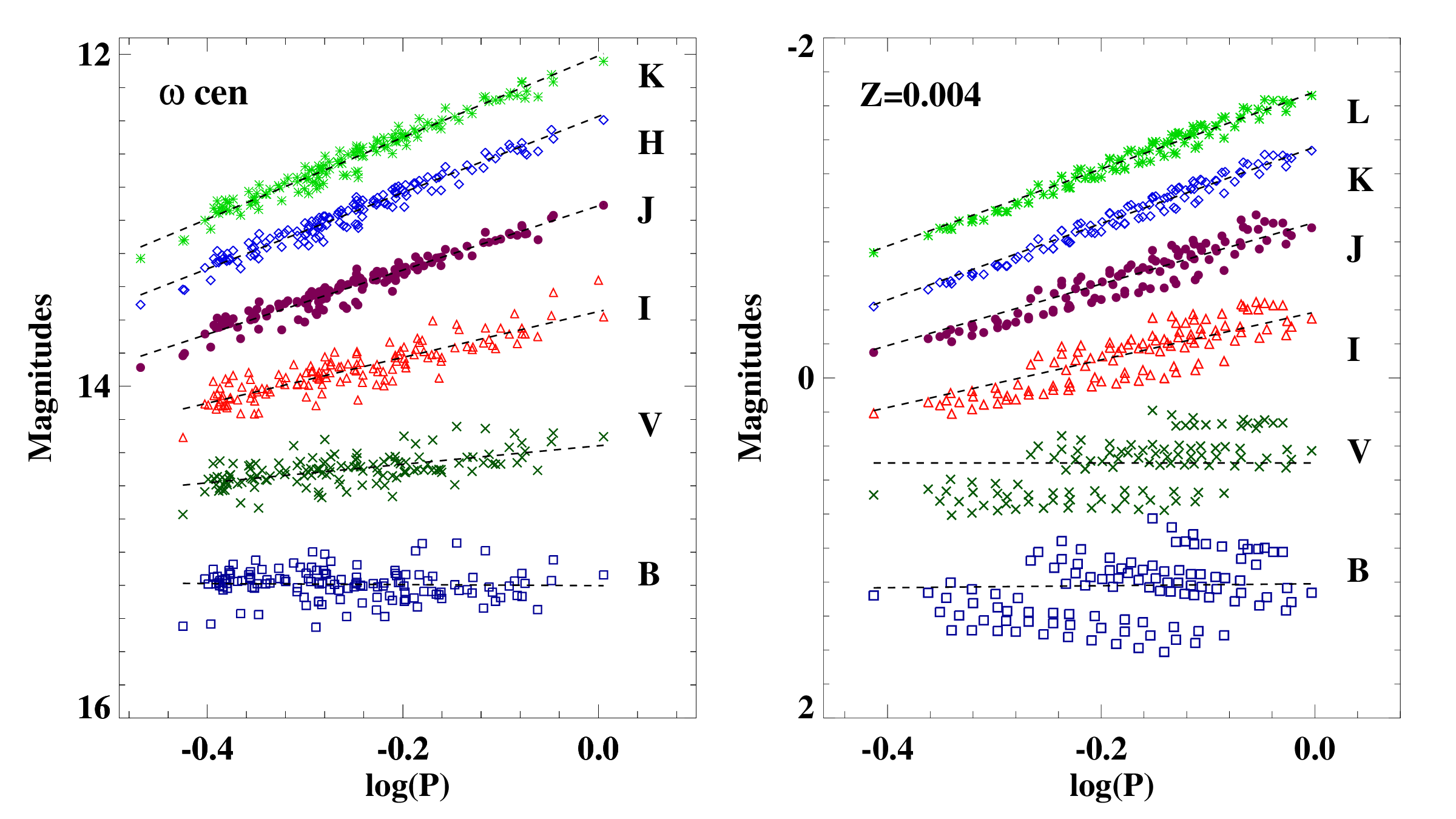}
\caption{{\it Left:} Multiwavelength PLRs of RR Lyrae in $\omega$ Cen cluster using data from \citet{braga2016} and \citet{braga2018}. {\it Right:} Multiband theoretical PLRs for RR Lyrae using models adopted from \citet{marconi2015}. The solid lines represent linear regression over entire period range. The periods of RRc stars are fundamentalize to include those in the PLR fits. The magnitudes in different bands are offset by some arbitrary amount for visualization purposes and do not exactly correspond to the scale on the $y$-axis.}
\label{fig:plr_wcen}
\end{center}
\end{figure*}

RR Lyrae are known to exhibit a very tight PLR at infrared wavelengths which makes them excellent standard candles. \citet{longmore1986} were the first to derive an empirical RR Lyrae PLR in the $K$-band. The pulsation equation implies a Period-Luminosity-color relation for RR Lyrae but the use of such relation suffers from uncertainties due to evolutionary effects,
effective temperature and metallicity predominantly at optical wavelengths. \citet{longmore1986} showed that a PLR in $K$-band comes naturally from pulsation equation because bolometric corrections increase with effective temperature such that redder RR Lyrae are brighter in $K$-band. This results in an empirical period-magnitude relation in $K$-band. Later, \citet{bono2001} derived theoretical $K$-band Period-Luminosity-Metallicity (PLZ) relation and showed that the uncertainties on the mass and luminosity also do not effect the PLRs significantly at this wavelength.

The empirical NIR PLRs of RR Lyrae have been a subject of several investigations, particularly in the globular clusters and the Magellanic Clouds \citep[][and references therein]{nemec1994, butler2003, dallora2004, sollima2006, borissova2009, coppola2011, braga2015, muraveva2015, muraveva2018a}. Similar to Cepheids, NIR observations of RR Lyrae in most of these studies are limited to few epochs or sparsely sampled light curves. Therefore, NIR templates for RR Lyrae are crucial to determine accurate mean-magnitudes and derive precise PLRs. Using well-sampled $K_s$-band light curves of RR Lyrae from the VISTA VVV survey \citep{minnitivvv2010}, \citet{hajdu2018} used principal component analysis to generate $J$ and $H$-band mean-magnitudes from single-epoch measurements. Recently, \citet{braga2019} derived NIR templates of RR Lyrae and showed that 2\% precise mean-magnitudes can be estimated even from single-epoch NIR observations.

Figure~\ref{fig:plr_wcen} shows empirical PLRs in $\omega$ Centauri and theoretical PLRs for RR Lyrae variables at multiple wavelengths. The $\omega$ Cen is a well-studied cluster in terms of RR Lyrae populations \citep[][]{navarrete2015, braga2016, braga2018} which exhibits a spread in metallicity. Regardless of the metallicity contribution, the infrared PLRs do not exhibit large intrinsic dispersion. The apparent magnitudes in $V$-band for RR Lyrae luminosities are nearly constant as a function of period. Since the bolometric correction sensitivity to effective temperature starts playing a role $R$-band onwards, a true PLR is observed at longer wavelengths \citep{catelan2004}. The empirical PLRs for the global sample of RRab and RRc variables in $IJHK_s$-bands from \citet{braga2016} and \citet{braga2018} are presented:

\begin{align}
\label{eq:rrl_plr_wcen}
I_{\omega~{\textrm{Cen}}} &= 13.56 - 1.34\log(P) ~~~~~(\sigma=0.06), \nonumber \\
J_{\omega~{\textrm{Cen}}} &= 13.02 - 1.88\log(P) ~~~~~(\sigma=0.04), \nonumber \\
H_{\omega~{\textrm{Cen}}} &= 12.69 - 2.22\log(P) ~~~~~(\sigma=0.04),\nonumber \\  
K_{s~\omega~{\textrm{Cen}}} &= 12.63 - 2.38\log(P) ~~~~~(\sigma=0.05),
\end{align}

\noindent where the uncertainties in the slopes and zero-points are $\lesssim 0.02$ and $\lesssim 0.05$ mag, respectively. For a reference zero-point, a distance modulus to $\omega$ Cen is $13.67\pm0.04$~mag \citep{braga2018}. The spectroscopic metallicities for RR Lyrae in $\omega$ Cen \citep{sollima2006a} were used by \citet{braga2018} to investigate metallicity dependence on NIR PLRs. The authors found a correlation between PLR residuals with [Fe/H]. In GGCs with marginal metallicity spread, the dispersion in NIR PLRs of RR Lyrae is typically $\sim 0.05$~mag implying an uncertainty of 2.5\% in individual distance determination. The right panel of Figure~\ref{fig:plr_wcen} displays theoretical PLRs based on RR Lyrae pulsation models of \citet{marconi2015} for a fixed metal-abundance. It is evident that the $V$-band absolute magnitude is nearly constant as a function of period for different mass-luminosity levels. The theoretical and empirical investigations on metallicity effects on RR Lyrae PLRs will be discussed in the next subsection. 

Mid-infrared observations of RR Lyrae, similar to classical Cepheids, have indisputable advantages as discussed previously. \citet{klein2011} utilized Wide Field Infrared Survey Explorer (WISE) catalog of RR Lyrae to derive PLRs at MIR wavelengths. The calibrations of RR Lyrae PLRs in MIR bands using WISE data were further improved by \citet{madore2013, klein2014} and the latter found the dispersion in these relations to be $\lesssim 0.05$ mag. Using theoretical approach, \citet{neeley2017} used {\it Spitzer} observations of RR Lyrae in M4 to derive PLRs with a dispersion of $\sim 0.05$ mag in 3.6$\mu$m and 4.5$\mu$m bands. Recently, \citet{muraveva2018} also used {\it Spitzer} data in LMC old cluster Reticulum to derive PLRs and estimate a distance to the LMC as part of the {\it Carnegie RR Lyrae Program}.

\subsubsection{Metallicity effects:}

Although extinction and metallicity effects are expected to be smaller at longer wavelengths, the impact of metal and helium abundance on RR Lyrae PLRs is actively debated. Theoretically, \citet{bono2001} found that the dependence on the metallicity is quantitatively smaller ($\sim$0.17 mag/dex) in $K$-band than that in the optical bands ($>0.2$ mag/dex). \citet{catelan2004} derived metal-dependent PLRs for RR Lyrae based on the calculations of synthetic horizontal branch models and found a significant metallicity term (0.21-0.17 mag/dex) in $IJHK_s$-bands. Using a new theoretical framework of RR Lyrae, \citet{marconi2015} generated pulsation models covering a broad range of metal-abundance (Z=0.02 to 0.0001) and derived PLRs. They found a metallicity dependence ($\sim$0. 18 mag/dex) similar to \citet{bono2001}, on RR Lyrae NIR PLRs. The PLZ relations for RR Lyrae are written in the following form:

\begin{equation}
\label{eq:rrl_plz}
M = \alpha + \beta\log(P) + \gamma[Fe/H].\\
\end{equation}

For a relative comparison, $IJK$-band PLZ relations are provided in the form of equation (\ref{eq:rrl_plz}):

\begin{align}
\label{eq:rrl_plr_model}
\textrm{M}_{I_{TH}} &= -0.07 -1.53\log(P) + 0.17\textrm{[Fe/H]} ~~~(\sigma=0.09), \nonumber\\ 
\textrm{M}_{J_{TH}} &= -0.50 -1.90\log(P) + 0.18\textrm{[Fe/H]} ~~~(\sigma=0.06),  \nonumber\\ 
\textrm{M}_{K_{TH}} &= -0.82 -2.25\log(P) + 0.18\textrm{[Fe/H]} ~~~(\sigma=0.04),
\end{align}

\noindent where $TH$ represents theory and the uncertainties in the coefficients of PLZ relations are negligible. While the theoretical studies consistently predict an appreciable metallicity dependence on RR Lyrae PLRs, empirical investigations have most often resulted in a marginal dependence on metallicity.

\citet{sollima2006} used RR Lyrae in several GGCs to constrain the metallicity dependence and quantified a relatively small dependence of 0.08 mag/dex on [Fe/H]. In the case of LMC RR Lyrae, \citet{borissova2009} also found a very mild dependence on metallicity in $K_s$-band by combining NIR photometry and spectroscopic metallicities for a homogeneous sample of 50 RR stars in the inner regions. 
Similarly, \citet{muraveva2015} utilized low-dispersion spectroscopic metallicities of 70 RRLs in the bar of the LMC with NIR photometry from VISTA VMC survey \citep{cioni2011} and found a marginal dependence ($\sim0.03\pm0.07$ mag/dex) on [Fe/H] in RR Lyrae PLRs. More recently, \citet{neeley2019} used {\it Gaia} parallaxes for Galactic field RR Lyrae to derive multiband PLZ relations and found that the dispersion in these relations is dominated by the uncertainties in the parallaxes despite reproducing the metallicity dependence predicted from models. The high-resolution spectroscopy of RR Lyrae has been obtained mostly for the field \citep[][and references therein]{clementini1995, for2011, nemec2013, pancino2015} and globular cluster variables \citep[for example,][]{sollima2006a, magurno2018, magurno2019}. In the case of Magellanic Clouds, low-resolution spectroscopy has been limited to small samples of RR Lyrae stars \citep[e.g.][]{gratton2004, borissova2004, borissova2006, haschke2012}.

\subsubsection{Period-Wesenheit relations:}

The PWRs for RR Lyrae have also been used for distance determinations to negate the issues related with reddening corrections. \citet{marconi2015} presented new optical and NIR PWRs adopting \citet{card1989} reddening law and a total-to-selective absorption ratio, $R_V = 3.06$. The dual-band PWRs from \citet{marconi2015} are listed in the form of equation (\ref{eq:rrl_plz}):

\begin{align}
\label{eq:pwr_rrl_model}
\textrm{M}_V~_{TH} - 3.06(\textrm{M}_B-\textrm{M}_V)~_{TH} &= -1.07 - 2.49\log(P) \nonumber \\ 
		&+ 0.01\textrm{[Fe/H]} ~~(\sigma=0.08),\nonumber \\
\textrm{M}_K~_{TH} - 0.69(\textrm{M}_J-\textrm{M}_K)~_{TH} &= -1.05 - 2.50\log(P) \nonumber \\ 
			&+ 0.18\textrm{[Fe/H]} ~~(\sigma=0.04),
\end{align}

\noindent where the NIR Wesenheits display metallicity dependence similar to PLRs. Interestingly, optical Wesenheit using combination of $B$ and $V$-band is nearly metallicity independent and this is not true for other combination of filters used to construct Wesenheit functions. Using this Wesenheit function, it is possible to estimate precise distance independent of the uncertainties on the metallicity measurements \citep[see equation~(\ref{eq:pwr_rrl_model}) and][for an application to $\omega$ Cen]{braga2016}. However, the dispersion in this optical Wesenheit function is twice as large compared to NIR PLRs and PWRs. \citet{marconi2015} derived several combinations of triple-band PWRs relations, similar to $W^H_{V,I}$ function used in SH0ES project, but those require mean-magnitudes in three independent filters. 

As a passing remark, the theoretically predicted first-overtone blue edge (FOBE) on the $M_V-\log(P)$ plane is also a useful distance indicator for stellar systems that host a statistically significant number of RRc stars \citep[][]{caputo1997}. The FOBE is independent of the metallicity.  If the blue part of the IS is well-populated and the metallicity is known, assuming the mass, a period-luminosity-metallicity relation can be derived for the evolutionary FOBE pulsators \citep[see,][for details]{caputo1997, caputo2000, bono2003, beaton2018}. 

\subsubsection{Absolute calibrations:}

The lack of accurate parallax measurements for RR Lyrae limits the precision of the absolute calibration of PLRs at infrared wavelengths. Unlike classical Cepheids, the calibration based on the LMC exhibits large dispersion due to spread in metallicity distribution of RR Lyrae and its effect on the PLRs. \citet{feast2008} utilized {\it Hipparcos} and {\it HST} parallaxes of RR Lyrae itself to provide a zero-point calibration. \citet{benedict2011} presented HST parallaxes for 5 RR Lyrae variables and provided absolute calibrations in $K_s$-band for a PLZ relation. Recently, \citet{muraveva2018} provided absolute magnitudes for RR Lyrae in several bands using {\it Gaia} astrometry for $\sim 400$ stars but also noted a significant zero-point offset in the {\it Gaia} parallaxes.

The calibration of MIR PLRs of RR Lyrae in the GGCs were first provided by \citet{dambis2014} using WISE data. The authors found two significantly different estimates for the zero-points based on statistical and HST trigonometric parallaxes. \citet{neeley2019} calibrated multiband PLZ relations of RR Lyrae using photometry obtained from the {\it Carnegie RR Lyrae Program} and parallaxes from the {\it Gaia} second data release for a sample of 55 Galactic field RR Lyrae stars. They found that the scatter in the PLZ relations is significantly large ($\sim$ 0.2 mag) when compared to theoretical predictions, and is still dominated by uncertainties in the parallaxes from current {\it Gaia} data. 

Despite the metallicity uncertainties, NIR PLRs of RR Lyrae have been used extensively to determine distances to several stellar systems, for example, GGCs M92 \citep{del2005}, M5 \citep{coppola2011}, $\omega$ cen \citep{navarrete2015, braga2018}, M4 \citep{braga2015}, Galactic center \citep{dekany2013}, LMC old cluster Reticulum \citep{dallora2004}, Magellanic Clouds \citep{ripepi2012, moretti2014, muraveva2018}, Carina Dwarf \citep{karczmarek2015}, Fornax \citep{karczmarek2017}, and IC 1613 \citep{hatt2017}. While the application of RR Lyrae PLRs and PWRs to measure distances to individual system is not discussed here, the interested readers are referred to the above mentioned papers.

\section{Type II Cepheids as distance indicators}
\label{sec:sec6}

The discovery of T2Cs played a critical role in the revision of the extragalactic distance scale. In his pioneering work, \citet{baade1944} showed that stellar populations of the galaxies are either similar to those in the solar neighborhood (the slow-moving stars i.e. disk stars) or those 
in the globular clusters. This eventually led to the classification of Population I and Population II stars. In his seminal papers, \citet{baade1958a, baade1958b, baade1958c} introduced a difference in the PLRs of the population II Cepheids in the globular clusters and the classical or population I Cepheids in the spiral arms of the galaxies. The former are T2Cs that represent old, low-mass stellar populations. Before their discovery, both young and old Cepheid populations had been used in the PLRs and distance scale. The distinction of the PLRs for two classes of Cepheids eventually resolved a major issue in the $H_0$ determination at that time, which led to the reduction of the spatial and temporal scales of the universe by a factor of two \citep{baade1956}. While the T2Cs have not been used extensively as distance indicators being fainter than classical Cepheids, they have played crucial roles as excellent tracers of stellar evolution and Galactic structure. I will discuss some empirical properties of T2Cs and focus on recent updates in their PLRs for distance measurements. There are several excellent reviews on T2Cs \citep{harris1985, wallerstein2002, sandage2006, welch2012, feast2010, feast2013, beaton2018} that are recommended to the interested readers. 

\subsection{Pulsation properties of T2Cs}

\begin{figure}[!t]
\begin{center}
\includegraphics[width=0.5\textwidth]{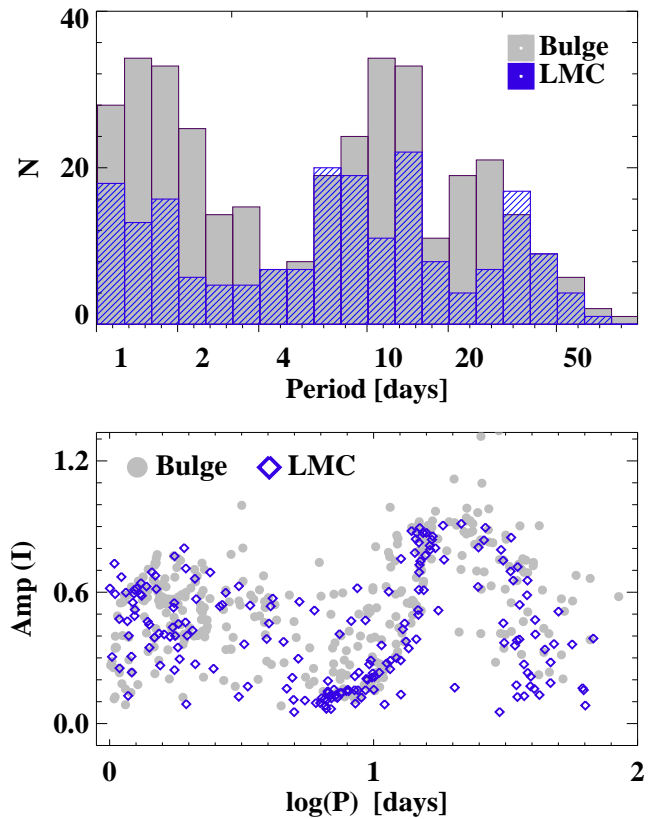}
\caption{{\it Top panel:} The period-distribution of T2Cs in the Galactic bulge and the LMC. {\it Bottom panel}: The $I$-band amplitude as a function of period for T2Cs in the Galactic bulge and the LMC.}
\label{fig:t2c_pul}
\end{center}
\end{figure}

Similar to the evolutionary status, the pulsation properties of T2Cs are distinct among the three subclasses. The classification into BL Her, W Vir and RV Tau is predominantly based on the pulsation period. However, the period range for each group is not universal and depends on the stellar environment. The top panel of Fig.~\ref{fig:t2c_pul} shows the period distribution for T2Cs in the Milky Way bulge and the Magellanic Clouds. It is evident that the minima in the period distributions vary between bulge and LMC due to their significantly different metallicities. \citet{soszynski2011a} found that bulge T2Cs are dominated by short-period BL Her stars which are more luminous than their counterparts in the Magellanic Clouds. The bottom panel of Fig.~\ref{fig:t2c_pul} displays the period-amplitude diagram for T2Cs in different stellar environments and exhibits different structures for each subclass. The amplitudes of BL Her exhibit large scatter at a given period similar to RR Lyrae variables. A sharp rise in the amplitudes for W Vir stars with $ 0.8 < \log(P) < 1.3$ days can be seen while the amplitudes decrease quickly as a function of period for RV Tau stars. The light curves of T2Cs are also quite different from classical Cepheids or RR Lyrae as they exhibit complex variations depending on the subclasses. 

Unlike classical Cepheids and RR Lyrae, theoretical studies on T2C pulsation properties are very limited. Earlier studies were limited to linear models \citep{wallerstein1984} and non-linear pulsation models without accounting for the convective transport \citep{fadeev1985}. Full time-dependent convective pulsation models of BL Her stars were provided by \citet{bono1997c, marconi2007, di2007}. \citet{bono1997c} showed that T2Cs pulsate primarily in the fundamental mode and their masses decrease with increasing period, and also derived metallicity independent period-luminosity-amplitude relations. \citet{marconi2007} presented the topology of the IS, and light and radial velocity curves for BL Her stars. They showed that the first-overtone IS is very narrow and therefore most T2Cs are fundamental pulsators which was also seen empirically for T2Cs in the Magellanic Clouds \citep{soszynski2008, soszynski2018}. Similar to classical Cepheids and RR Lyrae, T2Cs also follow a PLR that can be derived from the pulsation equation \citep[see, Section 3.3 of][]{matsunaga2006, di2007}.

\subsection{Period-Luminosity and Period-Wesenheit relations}

The optical studies of T2Cs in the GGCs provided evidence of a PLRs \citep{harris1985, mcnamara1995} but their investigations as useful distance indicators peaked with modern data from large photometric surveys \citep[][and references therein]{nemec1994, alcock1998, kubiak2003, majaess2009, schmidt2009}. For example, \citet{alcock1998} used MACHO microlensing survey data to discover T2Cs in the Magellanic Clouds and determined PLC relations for W Vir and RV Tau variables. In the past decade, OGLE survey has discovered several Galactic and Magellanic Clouds T2Cs and derived solid, optical-band PLRs \citep{soszynski2008a, soszynski2010,  soszynski2011a, soszynski2017}. Similar to classical Cepheids and RR Lyrae, increased availability of NIR observations have allowed several investigations on T2C PLRs at these wavelengths, where less sensitivity to metallicity and extinction leads to tighter PLRs \citep[][and references within]{matsunaga2006, feast2008, gmat2008, ciech2010, ripepi2015, bhardwaj2017}. 

At NIR wavelengths, \citet{matsunaga2006, matsunaga2009, matsunaga2011} derived NIR PLRs for T2Cs in the GGCs and the Magellanic Clouds. The authors found non-universal slopes of the PLRs in different systems and also noted varying frequency of each subtype. \citet{matsunaga2006} derived PLRs for T2Cs in the GGCs and found a linear relation over entire period range with a typical dispersion of 0.15 mag in $JHK_s$ bands. They obtained distances to individual GGCs using $M_V$-[Fe/H] relation for horizontal branch stars and showed a consistency between RR Lyrae and T2C distance scale. \citet{gmat2008} utilized NIR photometry of T2Cs in the Galactic bulge to estimate a distance to the Galactic center. In the LMC, \citet{ripepi2015} derived T2C PLRs in $JK_s$ using data from VISTA VMC survey with intrinsic dispersion of 0.13 mag in $J$ and 0.09 mag in $K$-band. More recently, \citet{bhardwaj2017a} used data from the LMC NIR synoptic survey to derive PLRs in $JHKs$-bands. Combining with literature data, they presented the largest sample to date of T2Cs with observations and used it to derive PLRs as well as absolute calibration with the known late-type eclipsing binary distance to the LMC. Furthermore, distance estimates to several GGCs from \citet{matsunaga2006} based on the horizontal branch morphology are within $1\sigma$ of the distances obtained by applying the LMC calibrations of T2Cs PLRs \citep{bhardwaj2017b}. 

\begin{figure}[!t]
\begin{center}
\includegraphics[width=0.5\textwidth]{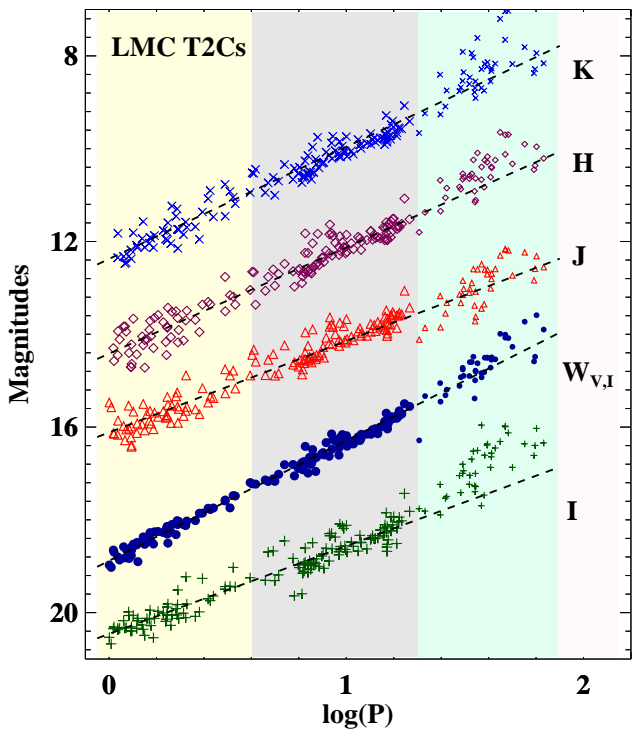}
\caption{Multiband PLRs and optical PWR for T2Cs in the LMC. The shaded regions represent three subclasses of T2Cs. The dashed lines represent a linear regression fitted to BL Her and W Vir subclasses. The magnitudes in different bands are offset by some arbitrary amount for visualization purposes.}
\label{fig:plr_t2cep}
\end{center}
\end{figure}

Figure~\ref{fig:plr_t2cep} shows optical PWR and $IJHK$-band PLRs for T2Cs in the LMC. It can be seen that the PLRs are not linear throughout the period range i.e. for all subclasses as an ensemble. A linear regression over entire period range results in a large dispersion of $\sim 0.6~\&~0.4$ mag in $V$ and $I$-band PLRs for T2Cs in the LMC, which is not useful for precision distance measurements. Even after excluding RV Tau that are distinctly brighter than the BL Her and W Vir stars, resulting PLR fits exhibit large dispersion. This suggests that the contribution to the intrinsic scatter in T2C PLRs at optical bands may have some dependence on metallicity and stellar environment. The presence of peculiar W Vir stars that have distinct light curve shapes also contributes to the scatter in the PLRs as they are systematically brighter than W Vir stars. \citet{soszynski2017} found that a significant fraction of W Vir are in the eclipsing binary systems and thus should be excluded from the PLR fits to obtain a better distance estimates using T2Cs. In Figure~\ref{fig:plr_t2cep}, the PLRs in NIR bands do not show a significant deviation in the slope between BL Her and W Vir subclasses. However, pW Vir and RV Tau are systematically brighter than the PLRs followed by short-period T2Cs. The T2C PLRs and PWRs for a combined sample of BL Her and W Vir stars in different stellar systems are provided here.

\begin{align}
\label{eq:rrl_plr_model}
W_{{V,I}~ LMC} &= 17.32 - 2.49\log(P) ~~~~~(\sigma=0.12), \nonumber\\ 
W_{{V,I}~ SMC} &= 17.59 - 2.54\log(P) ~~~~~(\sigma=0.38), 
\end{align}

\noindent where the optical photometry is taken from \citet{soszynski2018}. A significant reduction in dispersion is evident in the case of the optical PWR when compared to optical PLRs. The $K_s$-band PLRs from \citet{matsunaga2006}, \citet{bhardwaj2017}, and \citet{braga2018} are:

\begin{align}
\label{eq:rrl_plr_model}
\textrm{M}_{K_{{s}~ GGC}} &= -1.10 - 2.41\log(P) ~~~~~(\sigma=0.14), \nonumber\\ 
K_{{s}~ LMC} &= 17.10 - 2.23\log(P) ~~~~~(\sigma=0.18), \nonumber\\ 
K_{{s}~ BLG} &= 13.44 - 2.23\log(P) ~~~~~(\sigma=0.28). 
\end{align}

The slopes of the $K_s$-band PLRs are very similar between the LMC and bulge short-period  B Her + W Vir T2Cs. In $K_s$-band, \citet{bhardwaj2017a} did not find a significant deviation in the slope of PLRs for RV Tau from the B Her + W Vir sample but their photometry for BL Her showed evidence of crowding effects as their target fields were in the central bar of the LMC. 

At present, T2C PLRs are mainly calibrated with zero-point anchored to the LMC thanks to a very precise 1\% late-type eclipsing binary distance. The {\it HST} parallaxes are available for only two T2Cs ($\kappa$ Pav and VY Pyx) and there are two T2Cs with Baade-Wesselink distance \citep{feast2008}. Therefore, a robust Galactic calibration is still lacking but is expected to be delivered with increasingly accurate astrometric data from the {\it Gaia} mission.

Using T2Cs in the Galactic bulge from the VISTA VVV survey, \citet{bhardwaj2017b} derived PLRs in $JHK_s$-bands and estimated a robust distance to the Galactic center. T2Cs are particularly interesting in the extremely crowded regions like the Galactic bulge because their multiband NIR PLR and PWRs can be used to constrain the individual distances and extinction simultaneously without accounting for the metallicity effects. For example, the distance distribution derived using PLRs without accounting metallicity effects is much broader for RR Lyrae in the bulge than T2Cs \citep{bhardwaj2017b} implying a better precision for individual T2C distances. Recently, \citet{braga2018} extended this work with a larger sample of T2Cs in the bulge and derived individual distances to trace the structure and kinematics of old stellar populations \citep[see also,][]{dekany2019}. 

From theoretical point of view,  \citet{di2007} derived PLR and PWRs for T2Cs and estimated distances to several GGCs that were found to be consistent with RR Lyrae based estimates. They also predicted that the slope of the overall PLRs for T2Cs is less steep than that of classical Cepheids, which is also seen in the empirical PLRs. The fact that T2Cs follow similar PLRs as that of RR Lyrae in NIR \citep{matsunaga2006, feast2012, bhardwaj2017a} suggests a continuous transition between evolved RR Lyrae and BL Her evolutionary and pulsational properties. Recent findings have confirmed that BL Her and W Vir show similar PLRs provided pW Vir are excluded from the sample \citep{ripepi2015, bhardwaj2017a, bhardwaj2017b}. However, RV Tau are found to be systematically brighter than the PLRs followed by BL Her and W Vir stars. RV Tau are post-AGB stars and they may have circumstellar envelopes which can make them fainter. However, this can lead to significant variation in luminosities in different pulsation cycles and subsequently contribute to the scatter in the PLRs. 

\subsubsection{Metallicity effects:} 

The theoretical and observational investigations of T2Cs suggest minimal or no dependence of metal-abundances on NIR PLRs unlike RR Lyrae. 
The metallicity effects on NIR PLRs of T2Cs are at the level of $\sim 0.05$ mag/dex according to the theoretical predictions \citep{bono1997c, di2007, marconi2007}. Empirically, \citet{matsunaga2006} found negligible effect of metallicity dependence on NIR PLRs for T2Cs in the GGCs. \citet{bhardwaj2017b} showed that the slope of $K$-band PLR of T2C is statistically similar between GGCs, bulge, LMC and for the Milky Way T2Cs having good parallax measurements. Spectroscopic measurements for T2Cs are very limited \citep[for example,][and references therein]{maas2007, lemasle2015, kovtyukh2018} to investigate metallicity effects but these stars are known to cover a range of metallicities similar to that of RRLs. Without accounting metallicity effects, \citet{ripepi2015, bhardwaj2017a} determined a distance to the LMC based on empirical relations that is in excellent agreement with classical Cepheid and RR Lyrae based estimates. 

Apart from the Magellanic Clouds \citep{soszynski2008}, T2Cs have also been discovered in several other extragalactic stellar systems, for example, IC1613, M31, M33 \citep{majaess2009}, and in dwarf spheroidal galaxy Fornax \citep{bersier2002}. Given that T2Cs are brighter than RR Lyrae, BL Her and W Vir can be used to estimate distances to the galaxies beyond the Local Group up to ($\sim$ 10 Mpc). However, long-term time-domain surveys are critical to identify and classify the T2Cs because of the complex light variations and a broad period range.

\begin{figure*}[!t]
\begin{center}
\includegraphics[width=\textwidth]{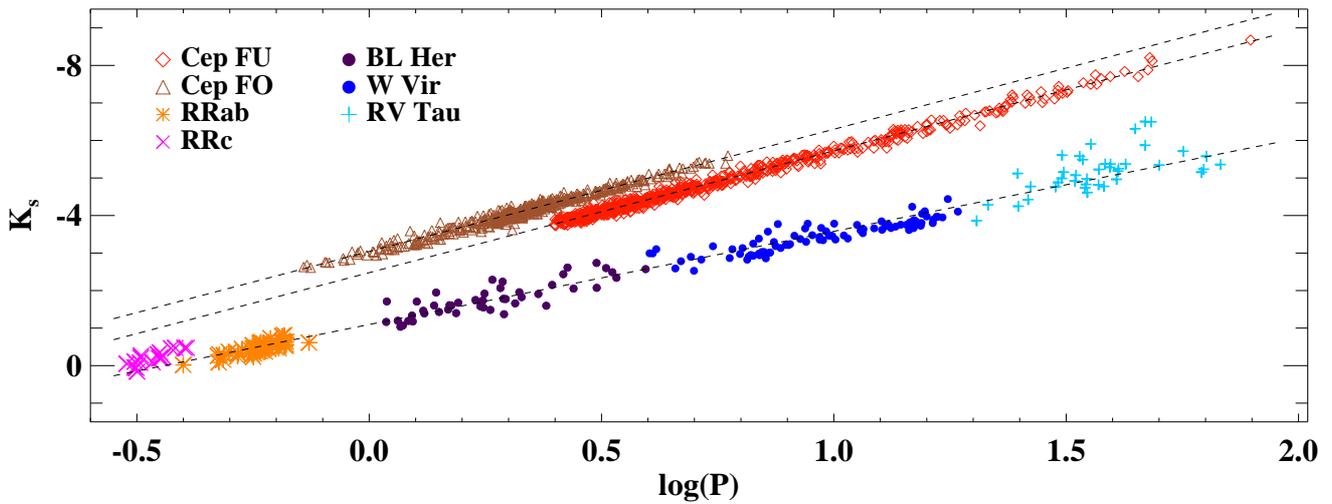}
\caption{The $K_s$-band PLR for classical Cepheids, RR Lyrae and T2Cs in the LMC calibrated with $1\%$ precise late-type eclipsing binary distance from \citet{piet2019}. The dashed lines represent best-fit linear regression over fundamental and first-overtone mode classical Cepheids, and BL Her + W Vir sample of T2Cs.}
\label{fig:all_ks}
\end{center}
\end{figure*}

\subsection{Comparison with classical Cepheids and RR Lyrae}

T2Cs are not as abundant as classical Cepheids and RR Lyrae due to their short evolutionary timescales, which limits a detailed investigation of their pulsation properties. The evolutionary time scales of T2Cs are roughly two orders of magnitude faster than RR Lyrae \citep{marconi2015}. Unlike classical Cepheids, the optical PLRs of T2Cs are non-linear and lack the precision to be useful distance indicators. T2C observations in NIR, where their PLRs do not show significant metallicity dependence (see Sec 6.2), are increasing as there are more NIR variability surveys. They are used both as population tracers and distance indicators complementing RR Lyrae variables. T2Cs are brighter than RR Lyrae and therefore can extend the use of population II standard candles to galaxies beyond 2 Mpc where the application of RR Lyrae is presently limited \citep{dacosta2010}. The light curves of T2Cs are not as distinct as classical Cepheids and RR Lyrae, and a significant overlap with classical Cepheids can be seen on the Fourier plane and color-magnitude diagrams.

Fig.~\ref{fig:all_ks} displays calibrated PLR in $K_s$-band for classical pulsating stars in the LMC. Classical Cepheids are systematically $\sim$1.5-3 magnitude brighter than T2Cs at a fixed period but the exact difference is period-dependent. Depending on the period of T2Cs, these are up to $8$ mag brighter than population II RR Lyrae. The BL Her and W Vir subclasses of T2Cs follow a linear PLRs. Extending their PLR to shorter periods ($<$1 day) clearly suggests that RRab are also located on this relation while the overtone RRc seem to be brighter than the PLR of T2Cs. The RV Tau are typically not included in the PLRs fits for T2Cs. The calibration of PLRs for classical Cepheids, RR Lyrae and T2Cs in the LMC is based on the eclipsing binary distance to the LMC \citep{piet2019}.

\section{Summary and Future Prospects}
\label{sec:sec7}

I discussed observational and theoretical pulsational properties of classical Cepheid, RR Lyrae and T2Cs, and their application to extragalactic distance measurements. The first two sections presented a historical overview and explained the evolutionary and pulsational scenario related to these classical pulsating stars. The section that describes the light curve properties provides an overview of their identification and classification as well as emphasises how their multi-wavelength observations can constrain the stellar evolution and pulsation models. The last three sections are focussed on the use of classical pulsating stars for cosmic distance scale delineating both population I and population II distance indicators. These standard candles have a long history dating back to more than a century and their theoretical and empirical investigations have persistently played significant roles in our understanding of the stellar evolution, Galactic structure and the Universe. 

A discussion on primary calibrations for classical Cepheids, RR Lyrae and T2Cs reveals two major issues that are yet to be addressed properly. First, the lack of robust geometric distances that limits the precision of calibrated PLRs of these standard candles in our own Galaxy. The HST parallaxes are available for a small sample of 16 classical Cepheids \citep{benedict2007, riess2014, riess2018}, 5 RR Lyrae and 2 T2Cs \citep{benedict2011}. With such small statistics even the precise determination of slope and zero-point of PLRs is not possible let alone quantification of other systematics, for example, due to age or metallicity. Therefore, either the theoretical calibrations are adopted or these standard candles in the LMC serve as primary calibrators. In case of the former, theoretical predictions suffer from the lack of observational constraints while in the case of latter, lack of high precision abundances for these stellar populations \citep{mancino2020} in the LMC precludes quantification of other systematic uncertainties. However, {\it Gaia} mission is already providing unprecedently precise astrometry for stellar populations in the solar neighbourhood \citep{lindegren2018}. The final {\it Gaia} parallaxes are predicted to have 10\% precise parallaxes at a distance of 10 kpc and $< 2\%$ within a few kpc distance. Therefore, a robust calibration of nearby Cepheid and RR Lyrae populations will eventually be derived with a percent-level precision. While some of the results from {\it Gaia} second data release are presented in this manuscript, it became apparent that the current data release suffers from a significant zero-point offset in parallaxes for both Cepheid and RR Lyrae \citep[for example,][]{muraveva2018, riess2018a}. Regardless, the plethora of {\it Gaia} astrometric data, spectro-photometry, variability and spectroscopy of bright sources will lead to potential breakthroughs in the studies of classical pulsating variable stars. 

The second outstanding question is related to the impact of composition, metallicity and helium effects in particular, age and evolutionary effects on the primary calibration of the classical pulsating stars. In the course of writing this manuscript, it became evident that the metallicity effects on both the theoretical and empirical PLRs are not well constrained despite decades of efforts. The high-resolution spectroscopic observations for Cepheids and RR Lyrae are very limited and will not be available even with {\it Gaia} data. Therefore, complementary large scale ongoing spectroscopic surveys (e.g. APOGEE \citep{majewski2017}, LAMOST \citep{zhao2012}) and future facilities \citep[e.g. 4MOST,][]{dejong2019} will provide stellar parameters for a statistically significant sample of pulsating stars. In case of the highly extincted and crowded extragalactic systems, a deeper insight in our understanding of the physics and chemistry of the classical pulsating stars will come with higher sensitivity and resolution of JWST in space and 30-m class ground-based extremely large telescopes.

High-precision space-based photometry has revealed interesting additional non-radial modes, period-doubling and amplitude/phase modulations in classical pulsating stars \citep{kolenberg2010, derekas2017, molnar2018}. The photometric revolution with ongoing missions such as {\it Gaia} and {\it TESS}, and future {\it PLATO} mission will continue to explore these phenomena in pulsating variables. The classical pulsators are less explored at UV and X-ray wavelengths where new insights can be gained into evolution and pulsation, and heating and dynamics of their atmospheres \citep{engle2015, neilson2016}. At UV wavelengths, the large amplitudes are particularly interesting not only for the identification but also to provide constraints for the pulsation models, for example, by simultaneous model-fitting of multiband light curves. While the focus of this review is on distance measurements, classical pulsating stars are also used extensively as stellar population tracers for extinction, metallicity, and morphology of their host galaxies. For example, minimum light color of RR Lyrae is an excellent tool for reddening diagnostics \citep[][]{sturch1966, ngeow2017, saha2019}. Classical Cepheid and RR Lyrae (and T2Cs) have been used to trace the spatial distribution and kinematics of young (metal-rich) and old (metal-poor) stellar populations in the Galaxy and the Magellanic Clouds \citep[see][and reference therein for more details]{subramanian2012, dekany2013, deb2014, subramanian2015, piet2015, jacyszyn2016, jacyszyn2017, ripepi2017, muraveva2018a, skowron2019}. Improved absolute calibrations of classical pulsating stars will enable precise individual distance measurements allowing new insights into the structure and kinematics of their underlying stellar populations in the host galaxies.

A most complete census of classical pulsating stars in the Local Group and beyond will be produced by the upcoming revolutionary {\it Vera C. Rubin Observatory Legacy Survey of Space and
Time} previously referred to as the {\it Large Synoptic Survey Telescope}. The optical time-domain observations from LSST will be complemented with infrared observations with JWST and large-aperture ground-based telescopes. In the infrared regime, where extinction and metallicity effects and the intrinsic variations are minimized, standard candles have reached to unprecedented precision and accuracy in the past decade. At these precisions, every possible source of known or unknown systematics becomes relevant and careful investigations are needed, for example, regarding the universality of extinction law, evolutionary effects, differences in the photometric systems (ground versus space-based), possible contributions of additional parameters to the intrinsic dispersion in the PLRs. With predominantly infrared observational facilities in the future, classical pulsating variables will enable potential scientific discoveries related to the structure of the Milky Way to the evolution of stars and our Universe.

\section*{Acknowledgements}

I thank the editorial board of the Journal of Astrophysics and Astronomy for inviting me to write this review article and Marcio Catelan, Marina Rejkuba, Wolfgang Gieren, H. P. Singh, Noriyuki Matsunaga, Richard I. Anderson and Shashi Kanbur for useful comments and suggestions. I also thank the anonymous referee for the quick and constructive report that helped improve the manuscript. 
AB acknowledges research grant $\#11850410434$ awarded by the National Natural Science Foundation of China through the Research Fund for International Young Scientists,
and a China Post-doctoral General Grant, and the Gruber fellowship 2020 grant sponsored by The Gruber Foundation and the International Astronomical Union. This research was supported by the Munich Institute for Astro- and Particle Physics (MIAPP) of the DFG cluster of excellence ``Origin and Structure of the Universe''. 

\vspace{-1em}


\setlength{\bibsep}{0.7pt plus 1.0ex}
\bibliography{../../mybib_final}  

\begin{thebibliography}{337}
\expandafter\ifx\csname natexlab\endcsname\relax\def\natexlab#1{#1}\fi

\bibitem[{{Alcock} {et~al}\mbox{.}(1998){Alcock}, {Allsman}, {Alves},
  {Axelrod}, {Becker}, {Bennett}, {Cook}, {Freeman}, {Griest}, {Lawson},
  {Lehner}, {Marshall}, {Minniti}, {Peterson}, {Pollard}, {Pratt}, {Quinn},
  {Rodgers}, {Sutherland}, {Tomaney}, \& {Welch}}]{alcock1998}
{Alcock} C. {et~al.}, 1998, AJ, 115, 1921

\bibitem[{{Anderson}(2016)}]{anderson2016a}
{Anderson} R.~I., 2016, MNRAS, 463, 1707

\bibitem[{{Anderson}(2019)}]{anderson2019}
{Anderson} R.~I., 2019, A\&A, 631, A165

\bibitem[{{Anderson} {et~al}\mbox{.}(2014){Anderson}, {Ekstr{\"o}m}, {Georgy},
  {Meynet}, {Mowlavi}, \& {Eyer}}]{anderson2014}
{Anderson} R.~I., {Ekstr{\"o}m} S., {Georgy} C., {Meynet} G., {Mowlavi} N.,
  {Eyer} L., 2014, A\&A, 564, A100

\bibitem[{{Anderson} \& {Riess}(2018)}]{anderson2018}
{Anderson} R.~I., {Riess} A.~G., 2018, ApJ, 861, 36

\bibitem[{{Anderson} {et~al}\mbox{.}(2016){Anderson}, {Saio}, {Ekstr{\"o}m},
  {Georgy}, \& {Meynet}}]{anderson2016}
{Anderson} R.~I., {Saio} H., {Ekstr{\"o}m} S., {Georgy} C., {Meynet} G., 2016,
  A\&A, 591, A8

\bibitem[{{Andrievsky} {et~al}\mbox{.}(2002){Andrievsky}, {Kovtyukh}, {Luck},
  {L{\'e}pine}, {Bersier}, {Maciel}, {Barbuy}, {Klochkova}, {Panchuk}, \&
  {Karpischek}}]{andrievsky2002}
{Andrievsky} S.~M., {Kovtyukh} V.~V., {Luck} R.~E., {L{\'e}pine} J.~R.~D.,
  {Bersier} D., {Maciel} W.~J., {Barbuy} B., {Klochkova} V.~G., {Panchuk}
  V.~E., {Karpischek} R.~U., 2002, A\&A, 381, 32

\bibitem[{{Baade}(1944)}]{baade1944}
{Baade} W., 1944, ApJ, 100, 137

\bibitem[{{Baade}(1956)}]{baade1956}
{Baade} W., 1956, PASP, 68, 5

\bibitem[{{Baade}(1958{\natexlab{a}})}]{baade1958a}
{Baade} W., 1958{\natexlab{a}}, Ricerche Astronomiche, 5, 3

\bibitem[{{Baade}(1958{\natexlab{b}})}]{baade1958b}
{Baade} W., 1958{\natexlab{b}}, Ricerche Astronomiche, 5, 165

\bibitem[{{Baade}(1958{\natexlab{c}})}]{baade1958c}
{Baade} W., 1958{\natexlab{c}}, Ricerche Astronomiche, 5, 303

\bibitem[{{Bailey}(1902)}]{bailey1902}
{Bailey} S.~I., 1902, Annals of Harvard College Observatory, 38

\bibitem[{{Bauer} {et~al}\mbox{.}(1999){Bauer}, {Afonso}, {Albert}, {Alard},
  {Andersen}, {Ansari}, {Aubourg}, {Bareyre}, {Beaulieu}, {Bouquet}, {Char},
  {Charlot}, {Couchot}, {Coutures}, {Derue}, {Ferlet}, {Gaucherel},
  {Glicenstein}, {Goldman}, {Gould}, {Graff}, {Gros}, {Haissinski}, {Hamilton},
  {Hardin}, {de Kat}, {Kim}, {Lasserre}, {Lesquoy}, {Loup}, {Magneville},
  {Mansoux}, {Marquette}, {Maurice}, {Milsztajn}, {Moniez},
  {Palanque-Delabrouille}, {Perdereau}, {Pr{\'e}vot}, {Renault}, {Regnault},
  {Rich}, {Spiro}, {Vidal-Madjar}, {Vigroux}, \& {Zylberajch}}]{bauer1999}
{Bauer} F. {et~al.}, 1999, A\&A, 348, 175

\bibitem[{{Beaton} {et~al}\mbox{.}(2018){Beaton}, {Bono}, {Braga}, {Dall'Ora},
  {Fiorentino}, {Jang}, {Mart{\'\i}nez-V{\'a}zquez}, {Matsunaga}, {Monelli},
  {Neeley}, \& {Salaris}}]{beaton2018}
{Beaton} R.~L. {et~al.}, 2018, Space Science Reviews, 214, 113

\bibitem[{{Beaton} {et~al}\mbox{.}(2016){Beaton}, {Freedman}, {Madore}, {Bono},
  {Carlson}, {Clementini}, {Durbin}, {Garofalo}, {Hatt}, {Jang}, {Kollmeier},
  {Lee}, {Monson}, {Rich}, {Scowcroft}, {Seibert}, {Sturch}, \&
  {Yang}}]{beaton2016}
{Beaton} R.~L. {et~al.}, 2016, ApJ, 832, 210

\bibitem[{{Bellinger} {et~al}\mbox{.}(2020){Bellinger}, {Kanbur}, {Bhardwaj},
  \& {Marconi}}]{bellinger2020}
{Bellinger} E.~P., {Kanbur} S.~M., {Bhardwaj} A., {Marconi} M., 2020, MNRAS,
  491, 4752

\bibitem[{{Benedict} {et~al}\mbox{.}(2011){Benedict}, {McArthur}, {Feast},
  {Barnes}, {Harrison}, {Bean}, {Menzies}, {Chaboyer}, {Fossati}, {Nesvacil},
  {Smith}, {Kolenberg}, {Laney}, {Kochukhov}, {Nelan}, {Shulyak}, {Taylor}, \&
  {Freedman}}]{benedict2011}
{Benedict} G.~F. {et~al.}, 2011, AJ, 142, 187

\bibitem[{{Benedict} {et~al}\mbox{.}(2007){Benedict}, {McArthur}, {Feast},
  {Barnes}, {Harrison}, {Patterson}, {Menzies}, {Bean}, \&
  {Freedman}}]{benedict2007}
{Benedict} G.~F., {McArthur} B.~E., {Feast} M.~W., {Barnes} T.~G., {Harrison}
  T.~E., {Patterson} R.~J., {Menzies} J.~W., {Bean} J.~L., {Freedman} W.~L.,
  2007, AJ, 133, 1810

\bibitem[{{Bersier} \& {Wood}(2002)}]{bersier2002}
{Bersier} D., {Wood} P.~R., 2002, AJ, 123, 840

\bibitem[{{Bhardwaj} {et~al}\mbox{.}(2016{\natexlab{a}}){Bhardwaj}, {Kanbur},
  {Macri}, {Singh}, {Ngeow}, \& {Ishida}}]{bhardwaj2016b}
{Bhardwaj} A., {Kanbur} S.~M., {Macri} L.~M., {Singh} H.~P., {Ngeow} C.-C.,
  {Ishida} E.~E.~O., 2016{\natexlab{a}}, MNRAS, 457, 1644

\bibitem[{{Bhardwaj} {et~al}\mbox{.}(2016{\natexlab{b}}){Bhardwaj}, {Kanbur},
  {Macri}, {Singh}, {Ngeow}, {Wagner-Kaiser}, \& {Sarajedini}}]{bhardwaj2016a}
{Bhardwaj} A., {Kanbur} S.~M., {Macri} L.~M., {Singh} H.~P., {Ngeow} C.-C.,
  {Wagner-Kaiser} R., {Sarajedini} A., 2016{\natexlab{b}}, AJ, 151, 88

\bibitem[{{Bhardwaj} {et~al}\mbox{.}(2017{\natexlab{a}}){Bhardwaj}, {Kanbur},
  {Marconi}, {Rejkuba}, {Singh}, \& {Ngeow}}]{bhardwaj2017}
{Bhardwaj} A., {Kanbur} S.~M., {Marconi} M., {Rejkuba} M., {Singh} H.~P.,
  {Ngeow} C.-C., 2017{\natexlab{a}}, MNRAS, 466, 2805

\bibitem[{{Bhardwaj} {et~al}\mbox{.}(2015){Bhardwaj}, {Kanbur}, {Singh},
  {Macri}, \& {Ngeow}}]{bhardwaj2015}
{Bhardwaj} A., {Kanbur} S.~M., {Singh} H.~P., {Macri} L.~M., {Ngeow} C.-C.,
  2015, MNRAS, 447, 3342

\bibitem[{{Bhardwaj} {et~al}\mbox{.}(2014){Bhardwaj}, {Kanbur}, {Singh}, \&
  {Ngeow}}]{bhardwaj2014}
{Bhardwaj} A., {Kanbur} S.~M., {Singh} H.~P., {Ngeow} C.-C., 2014, MNRAS, 445,
  2655

\bibitem[{{Bhardwaj} {et~al}\mbox{.}(2017{\natexlab{b}}){Bhardwaj}, {Macri},
  {Rejkuba}, {Kanbur}, {Ngeow}, \& {Singh}}]{bhardwaj2017a}
{Bhardwaj} A., {Macri} L.~M., {Rejkuba} M., {Kanbur} S.~M., {Ngeow} C.-C.,
  {Singh} H.~P., 2017{\natexlab{b}}, AJ, 153, 154

\bibitem[{{Bhardwaj} {et~al}\mbox{.}(2016{\natexlab{c}}){Bhardwaj}, {Ngeow},
  {Kanbur}, \& {Singh}}]{bhardwaj2016c}
{Bhardwaj} A., {Ngeow} C.-C., {Kanbur} S.~M., {Singh} H.~P.,
  2016{\natexlab{c}}, MNRAS, 458, 3705

\bibitem[{{Bhardwaj} {et~al}\mbox{.}(2017{\natexlab{c}}){Bhardwaj}, {Rejkuba},
  {Minniti}, {Surot}, {Valenti}, {Zoccali}, {Gonzalez}, {Romaniello}, {Kanbur},
  \& {Singh}}]{bhardwaj2017b}
{Bhardwaj} A., {Rejkuba} M., {Minniti} D., {Surot} F., {Valenti} E., {Zoccali}
  M., {Gonzalez} O.~A., {Romaniello} M., {Kanbur} S.~M., {Singh} H.~P.,
  2017{\natexlab{c}}, A\&A, 605, A100

\bibitem[{{Bla{\v{z}}ko}(1907)}]{blazhko1907}
{Bla{\v{z}}ko} S., 1907, Astronomische Nachrichten, 175, 325

\bibitem[{{Bono} {et~al}\mbox{.}(1997){Bono}, {Caputo}, {Cassisi},
  {Castellani}, \& {Marconi}}]{bono1997a}
{Bono} G., {Caputo} F., {Cassisi} S., {Castellani} V., {Marconi} M., 1997, ApJ,
  479, 279

\bibitem[{{Bono} {et~al}\mbox{.}(2000){Bono}, {Caputo}, {Cassisi}, {Marconi},
  {Piersanti}, \& {Tornamb{\`e}}}]{bono2000}
{Bono} G., {Caputo} F., {Cassisi} S., {Marconi} M., {Piersanti} L.,
  {Tornamb{\`e}} A., 2000, ApJ, 543, 955

\bibitem[{{Bono} {et~al}\mbox{.}(1999){Bono}, {Caputo}, {Castellani}, \&
  {Marconi}}]{bono1999b}
{Bono} G., {Caputo} F., {Castellani} V., {Marconi} M., 1999, ApJ, 512, 711

\bibitem[{{Bono} {et~al}\mbox{.}(2003){Bono}, {Caputo}, {Castellani},
  {Marconi}, {Storm}, \& {Degl'Innocenti}}]{bono2003}
{Bono} G., {Caputo} F., {Castellani} V., {Marconi} M., {Storm} J.,
  {Degl'Innocenti} S., 2003, MNRAS, 344, 1097

\bibitem[{{Bono}, {Caputo} \& {Di Criscienzo}(2007){Bono}, {Caputo}, \& {Di
  Criscienzo}}]{bono2007}
{Bono} G., {Caputo} F., {Di Criscienzo} M., 2007, A\&A, 476, 779

\bibitem[{{Bono} {et~al}\mbox{.}(2010){Bono}, {Caputo}, {Marconi}, \&
  {Musella}}]{bono2010}
{Bono} G., {Caputo} F., {Marconi} M., {Musella} I., 2010, ApJ, 715, 277

\bibitem[{{Bono}, {Caputo} \& {Santolamazza}(1997){Bono}, {Caputo}, \&
  {Santolamazza}}]{bono1997c}
{Bono} G., {Caputo} F., {Santolamazza} P., 1997, A\&A, 317, 171

\bibitem[{{Bono} {et~al}\mbox{.}(1995){Bono}, {Castellani}, {degl'Innocenti},
  \& {Pulone}}]{bono1995}
{Bono} G., {Castellani} V., {degl'Innocenti} S., {Pulone} L., 1995, A\&A, 297,
  115

\bibitem[{{Bono} {et~al}\mbox{.}(2001){Bono}, {Gieren}, {Marconi},
  {Fouqu{\'e}}, \& {Caputo}}]{bono2001}
{Bono} G., {Gieren} W.~P., {Marconi} M., {Fouqu{\'e}} P., {Caputo} F., 2001,
  ApJ, 563, 319

\bibitem[{{Bono} {et~al}\mbox{.}(2002){Bono}, {Groenewegen}, {Marconi}, \&
  {Caputo}}]{bono2002a}
{Bono} G., {Groenewegen} M.~A.~T., {Marconi} M., {Caputo} F., 2002, ApJL, 574,
  L33

\bibitem[{{Bono}, {Incerpi} \& {Marconi}(1996){Bono}, {Incerpi}, \&
  {Marconi}}]{bono1996}
{Bono} G., {Incerpi} R., {Marconi} M., 1996, ApJL, 467, L97

\bibitem[{{Bono}, {Marconi} \& {Stellingwerf}(1999){Bono}, {Marconi}, \&
  {Stellingwerf}}]{bono1999a}
{Bono} G., {Marconi} M., {Stellingwerf} R.~F., 1999, ApJS, 122, 167

\bibitem[{{Bono}, {Marconi} \& {Stellingwerf}(2000){Bono}, {Marconi}, \&
  {Stellingwerf}}]{bono2000d}
{Bono} G., {Marconi} M., {Stellingwerf} R.~F., 2000, A\&A, 360, 245

\bibitem[{{Bono} {et~al}\mbox{.}(2016){Bono}, {Pietrinferni}, {Marconi},
  {Braga}, {Fiorentino}, {Stetson}, {Buonanno}, {Castellani}, {Dall'Ora},
  {Fabrizio}, {Ferraro}, {Giuffrida}, {Iannicola}, {Marengo}, {Magurno},
  {Martinez-Vazquez}, {Matsunaga}, {Monelli}, {Neeley}, {Rastello}, {Salaris},
  {Short}, \& {Stellingwerf}}]{bono2016}
{Bono} G. {et~al.}, 2016, Commmunications of the Konkoly Observatory Hungary,
  105, 149

\bibitem[{{Bono} \& {Stellingwerf}(1994)}]{bono1994}
{Bono} G., {Stellingwerf} R.~F., 1994, ApJS, 93, 233

\bibitem[{{Borissova} {et~al}\mbox{.}(2006){Borissova}, {Minniti}, {Rejkuba},
  \& {Alves}}]{borissova2006}
{Borissova} J., {Minniti} D., {Rejkuba} M., {Alves} D., 2006, A\&A, 460, 459

\bibitem[{{Borissova} {et~al}\mbox{.}(2004){Borissova}, {Minniti}, {Rejkuba},
  {Alves}, {Cook}, \& {Freeman}}]{borissova2004}
{Borissova} J., {Minniti} D., {Rejkuba} M., {Alves} D., {Cook} K.~H., {Freeman}
  K.~C., 2004, A\&A, 423, 97

\bibitem[{{Borissova} {et~al}\mbox{.}(2009){Borissova}, {Rejkuba}, {Minniti},
  {Catelan}, \& {Ivanov}}]{borissova2009}
{Borissova} J., {Rejkuba} M., {Minniti} D., {Catelan} M., {Ivanov} V.~D., 2009,
  A\&A, 502, 505

\bibitem[{{Braga} {et~al}\mbox{.}(2015){Braga}, {Dall'Ora}, {Bono}, {Stetson},
  {Ferraro}, {Iannicola}, {Marengo}, {Neeley}, {Persson}, {Buonanno},
  {Coppola}, {Freedman}, {Madore}, {Marconi}, {Matsunaga}, {Monson}, {Rich},
  {Scowcroft}, \& {Seibert}}]{braga2015}
{Braga} V.~F. {et~al.}, 2015, ApJ, 799, 165

\bibitem[{{Braga} {et~al}\mbox{.}(2016){Braga}, {Stetson}, {Bono}, {Dall'Ora},
  {Ferraro}, {Fiorentino}, {Freyhammer}, {Iannicola}, {Marengo}, {Neeley},
  {Valenti}, {Buonanno}, {Calamida}, {Castellani}, {da Silva},
  {Degl'Innocenti}, {Di Cecco}, {Fabrizio}, {Freedman}, {Giuffrida}, {Lub},
  {Madore}, {Marconi}, {Marinoni}, {Matsunaga}, {Monelli}, {Persson},
  {Piersimoni}, {Pietrinferni}, {Prada-Moroni}, {Pulone}, {Stellingwerf},
  {Tognelli}, \& {Walker}}]{braga2016}
{Braga} V.~F. {et~al.}, 2016, AJ, 152, 170

\bibitem[{{Braga} {et~al}\mbox{.}(2019){Braga}, {Stetson}, {Bono}, {Dall'Ora},
  {Ferraro}, {Fiorentino}, {Iannicola}, {Inno}, {Marengo}, {Neeley}, {Beaton},
  {Buonanno}, {Calamida}, {Contreras Ramos}, {Chaboyer}, {Fabrizio},
  {Freedman}, {Gilligan}, {Johnston}, {Lub}, {Madore}, {Magurno}, {Marconi},
  {Marinoni}, {Marrese}, {Mateo}, {Matsunaga}, {Minniti}, {Monson}, {Monelli},
  {Nonino}, {Persson}, {Pietrinferni}, {Sneden}, {Storm}, {Walker}, {Valenti},
  \& {Zoccali}}]{braga2019}
{Braga} V.~F. {et~al.}, 2019, A\&A, 625, A1

\bibitem[{{Braga} {et~al}\mbox{.}(2018){Braga}, {Stetson}, {Bono}, {Dall'Ora},
  {Ferraro}, {Fiorentino}, {Iannicola}, {Marconi}, {Marengo}, {Monson},
  {Neeley}, {Persson}, {Beaton}, {Buonanno}, {Calamida}, {Castellani}, {Di
  Carlo}, {Fabrizio}, {Freedman}, {Inno}, {Madore}, {Magurno}, {Marchetti},
  {Marinoni}, {Marrese}, {Matsunaga}, {Minniti}, {Monelli}, {Nonino},
  {Piersimoni}, {Pietrinferni}, {Prada-Moroni}, {Pulone}, {Stellingwerf},
  {Tognelli}, {Walker}, {Valenti}, \& {Zoccali}}]{braga2018}
{Braga} V.~F. {et~al.}, 2018, AJ, 155, 137

\bibitem[{{Buchler} \& {Koll{\'a}th}(2011)}]{buchler2011}
{Buchler} J.~R., {Koll{\'a}th} Z., 2011, ApJ, 731, 24

\bibitem[{{Butler}(2003)}]{butler2003}
{Butler} D.~J., 2003, A\&A, 405, 981

\bibitem[{{Caputo}(1997)}]{caputo1997}
{Caputo} F., 1997, MNRAS, 284, 994

\bibitem[{{Caputo} {et~al}\mbox{.}(2005){Caputo}, {Bono}, {Fiorentino},
  {Marconi}, \& {Musella}}]{caputo2005d}
{Caputo} F., {Bono} G., {Fiorentino} G., {Marconi} M., {Musella} I., 2005, ApJ,
  629, 1021

\bibitem[{{Caputo} {et~al}\mbox{.}(2000){Caputo}, {Castellani}, {Marconi}, \&
  {Ripepi}}]{caputo2000}
{Caputo} F., {Castellani} V., {Marconi} M., {Ripepi} V., 2000, MNRAS, 316, 819

\bibitem[{{Caputo}, {Marconi} \& {Musella}(2000){Caputo}, {Marconi}, \&
  {Musella}}]{caputo2000b}
{Caputo} F., {Marconi} M., {Musella} I., 2000, A\&A, 354, 610

\bibitem[{{Cardelli}, {Clayton} \& {Mathis}(1989){Cardelli}, {Clayton}, \&
  {Mathis}}]{card1989}
{Cardelli} J.~A., {Clayton} G.~C., {Mathis} J.~S., 1989, ApJ, 345, 245

\bibitem[{{Carini} {et~al}\mbox{.}(2017){Carini}, {Brocato}, {Raimondo}, \&
  {Marconi}}]{carini2017}
{Carini} R., {Brocato} E., {Raimondo} G., {Marconi} M., 2017, MNRAS, 469, 1532

\bibitem[{{Catelan}(2009)}]{catelan2009}
{Catelan} M., 2009, AP\&SS, 320, 261

\bibitem[{{Catelan}, {Pritzl} \& {Smith}(2004){Catelan}, {Pritzl}, \&
  {Smith}}]{catelan2004}
{Catelan} M., {Pritzl} B.~J., {Smith} H.~A., 2004, ApJS, 154, 633

\bibitem[{{Catelan} \& {Smith}(2015)}]{catelan2015}
{Catelan} M., {Smith} H.~A., 2015, {Pulsating Stars}. Wiley-VCH

\bibitem[{{Ceraski}(1905)}]{ceraski1905}
{Ceraski} W., 1905, Astronomische Nachrichten, 168, 29

\bibitem[{{Chiosi} {et~al}\mbox{.}(1992){Chiosi}, {Wood}, {Bertelli}, \&
  {Bressan}}]{chiosi1992}
{Chiosi} C., {Wood} P., {Bertelli} G., {Bressan} A., 1992, ApJ, 387, 320

\bibitem[{{Christy}(1966)}]{christy1966}
{Christy} R.~F., 1966, ApJ, 144, 108

\bibitem[{{Christy}(1968)}]{christy1968}
{Christy} R.~F., 1968, Quarterly Journal of the Royal Astronomical Society, 9,
  13

\bibitem[{{Ciechanowska} {et~al}\mbox{.}(2010){Ciechanowska},
  {Pietrzy{\'n}ski}, {Szewczyk}, {Gieren}, \& {Soszy{\'n}ski}}]{ciech2010}
{Ciechanowska} A., {Pietrzy{\'n}ski} G., {Szewczyk} O., {Gieren} W.,
  {Soszy{\'n}ski} I., 2010, Acta Astron., 60, 233

\bibitem[{{Cioni} {et~al}\mbox{.}(2011){Cioni}, {Clementini}, {Girardi},
  {Guandalini}, {Gullieuszik}, {Miszalski}, {Moretti}, {Ripepi}, {Rubele},
  {Bagheri}, {Bekki}, {Cross}, {de Blok}, {de Grijs}, {Emerson}, {Evans},
  {Gibson}, {Gonzales-Solares}, {Groenewegen}, {Irwin}, {Ivanov}, {Lewis},
  {Marconi}, {Marquette}, {Mastropietro}, {Moore}, {Napiwotzki}, {Naylor},
  {Oliveira}, {Read}, {Sutorius}, {van Loon}, {Wilkinson}, \&
  {Wood}}]{cioni2011}
{Cioni} M.-R.~L. {et~al.}, 2011, A\&A, 527, A116

\bibitem[{{Clement} {et~al}\mbox{.}(2001){Clement}, {Muzzin}, {Dufton},
  {Ponnampalam}, {Wang}, {Burford}, {Richardson}, {Rosebery}, {Rowe}, \&
  {Hogg}}]{clement2001}
{Clement} C.~M., {Muzzin} A., {Dufton} Q., {Ponnampalam} T., {Wang} J.,
  {Burford} J., {Richardson} A., {Rosebery} T., {Rowe} J., {Hogg} H.~S., 2001,
  AJ, 122, 2587

\bibitem[{{Clementini} {et~al}\mbox{.}(1995){Clementini}, {Carretta},
  {Gratton}, {Merighi}, {Mould}, \& {McCarthy}}]{clementini1995}
{Clementini} G., {Carretta} E., {Gratton} R., {Merighi} R., {Mould} J.~R.,
  {McCarthy} J.~K., 1995, AJ, 110, 2319

\bibitem[{{Clementini} {et~al}\mbox{.}(2017){Clementini}, {Eyer}, {Muraveva},
  {Garofalo}, {Ripepi}, {Marconi}, {Sarro}, {Palmer}, {Luri}, {Molinaro},
  {Rimoldini}, {Szabados}, {Anderson}, \& {Musella}}]{clementini2017}
{Clementini} G. {et~al.}, 2017, in European Physical Journal Web of
  Conferences, Vol. 152, European Physical Journal Web of Conferences, p. 02003

\bibitem[{{Clementini} {et~al}\mbox{.}(2003){Clementini}, {Gratton},
  {Bragaglia}, {Carretta}, {Di Fabrizio}, \& {Maio}}]{clementini2003}
{Clementini} G., {Gratton} R., {Bragaglia} A., {Carretta} E., {Di Fabrizio} L.,
  {Maio} M., 2003, AJ, 125, 1309

\bibitem[{{Coppola} {et~al}\mbox{.}(2011){Coppola}, {Dall'Ora}, {Ripepi},
  {Marconi}, {Musella}, {Bono}, {Piersimoni}, {Stetson}, \&
  {Storm}}]{coppola2011}
{Coppola} G., {Dall'Ora} M., {Ripepi} V., {Marconi} M., {Musella} I., {Bono}
  G., {Piersimoni} A.~M., {Stetson} P.~B., {Storm} J., 2011, MNRAS, 416, 1056

\bibitem[{{Cox}(1980{\natexlab{a}})}]{cox1980b}
{Cox} A.~N., 1980{\natexlab{a}}, ARA\&A, 18, 15

\bibitem[{{Cox}(1980{\natexlab{b}})}]{cox1980a}
{Cox} J.~P., 1980{\natexlab{b}}, {Theory of stellar pulsation}. Princeton
  University Press, NJ

\bibitem[{{Da Costa} {et~al}\mbox{.}(2010){Da Costa}, {Rejkuba}, {Jerjen}, \&
  {Grebel}}]{dacosta2010}
{Da Costa} G.~S., {Rejkuba} M., {Jerjen} H., {Grebel} E.~K., 2010, ApJL, 708,
  L121

\bibitem[{{Dall'Ora} {et~al}\mbox{.}(2004){Dall'Ora}, {Storm}, {Bono},
  {Ripepi}, {Monelli}, {Testa}, {Andreuzzi}, {Buonanno}, {Caputo},
  {Castellani}, {Corsi}, {Marconi}, {Marconi}, {Pulone}, \&
  {Stetson}}]{dallora2004}
{Dall'Ora} M. {et~al.}, 2004, ApJ, 610, 269

\bibitem[{{Dambis}, {Rastorguev} \& {Zabolotskikh}(2014){Dambis}, {Rastorguev},
  \& {Zabolotskikh}}]{dambis2014}
{Dambis} A.~K., {Rastorguev} A.~S., {Zabolotskikh} M.~V., 2014, MNRAS, 439,
  3765

\bibitem[{{Das} {et~al}\mbox{.}(2018){Das}, {Bhardwaj}, {Kanbur}, {Singh}, \&
  {Marconi}}]{das2018}
{Das} S., {Bhardwaj} A., {Kanbur} S.~M., {Singh} H.~P., {Marconi} M., 2018,
  MNRAS

\bibitem[{{de Grijs}(2011)}]{grijs2011}
{de Grijs} R., 2011, An Introduction to Distance Measurement in Astronomy,
  ISBN: 978-1-119-97817-6

\bibitem[{{de Jong}(2019)}]{dejong2019}
{de Jong} R. e.~a., 2019, The Messenger, 175, 3

\bibitem[{{Deb} \& {Singh}(2009)}]{deb2009}
{Deb} S., {Singh} H.~P., 2009, A\&A, 507, 1729

\bibitem[{{Deb} \& {Singh}(2014)}]{deb2014}
{Deb} S., {Singh} H.~P., 2014, MNRAS, 438, 2440

\bibitem[{{D{\'e}k{\'a}ny} {et~al}\mbox{.}(2019){D{\'e}k{\'a}ny}, {Hajdu},
  {Grebel}, \& {Catelan}}]{dekany2019}
{D{\'e}k{\'a}ny} I., {Hajdu} G., {Grebel} E.~K., {Catelan} M., 2019, ApJ, 883,
  58

\bibitem[{{D{\'e}k{\'a}ny} {et~al}\mbox{.}(2013){D{\'e}k{\'a}ny}, {Minniti},
  {Catelan}, {Zoccali}, {Saito}, {Hempel}, \& {Gonzalez}}]{dekany2013}
{D{\'e}k{\'a}ny} I., {Minniti} D., {Catelan} M., {Zoccali} M., {Saito} R.~K.,
  {Hempel} M., {Gonzalez} O.~A., 2013, ApJL, 776, L19

\bibitem[{{D{\'e}k{\'a}ny} {et~al}\mbox{.}(2015){D{\'e}k{\'a}ny}, {Minniti},
  {Majaess}, {Zoccali}, {Hajdu}, {Alonso-Garc{\'\i}a}, {Catelan}, {Gieren}, \&
  {Borissova}}]{dekany2015}
{D{\'e}k{\'a}ny} I., {Minniti} D., {Majaess} D., {Zoccali} M., {Hajdu} G.,
  {Alonso-Garc{\'\i}a} J., {Catelan} M., {Gieren} W., {Borissova} J., 2015,
  ApJL, 812, L29

\bibitem[{{Del Principe} {et~al}\mbox{.}(2005){Del Principe}, {Piersimoni},
  {Bono}, {Di Paola}, {Dolci}, \& {Marconi}}]{del2005}
{Del Principe} M., {Piersimoni} A.~M., {Bono} G., {Di Paola} A., {Dolci} M.,
  {Marconi} M., 2005, AJ, 129, 2714

\bibitem[{{Derekas} {et~al}\mbox{.}(2017){Derekas}, {Plachy}, {Moln{\'a}r},
  {S{\'o}dor}, {Benk{\H{o}}}, {Szabados}, {Bogn{\'a}r}, {Cs{\'a}k},
  {Szab{\'o}}, {Szab{\'o}}, \& {P{\'a}l}}]{derekas2017}
{Derekas} A. {et~al.}, 2017, MNRAS, 464, 1553

\bibitem[{{Di Criscienzo} {et~al}\mbox{.}(2007){Di Criscienzo}, {Caputo},
  {Marconi}, \& {Cassisi}}]{di2007}
{Di Criscienzo} M., {Caputo} F., {Marconi} M., {Cassisi} S., 2007, A\&A, 471,
  893

\bibitem[{{Downes} {et~al}\mbox{.}(2004){Downes}, {Margon}, {Homer}, \&
  {Anderson}}]{downes2004}
{Downes} R.~A., {Margon} B., {Homer} L., {Anderson} S.~F., 2004, AJ, 128, 2288

\bibitem[{{Dziembowski} \& {Mizerski}(2004)}]{dziembowski2004}
{Dziembowski} W.~A., {Mizerski} T., 2004, Acta Astron., 54, 363

\bibitem[{{Eddington}(1918)}]{eddington1918}
{Eddington} A.~S., 1918, MNRAS, 79, 2

\bibitem[{{Eddington}(1919)}]{eddington1919}
{Eddington} A.~S., 1919, MNRAS, 79, 171

\bibitem[{{Eddington}(1926)}]{eddington1926}
{Eddington} A.~S., 1926, {The Internal Constitution of the Stars}

\bibitem[{{Engle}(2015)}]{engle2015}
{Engle} S., 2015, PhD thesis, James Cook University

\bibitem[{{Fadeev} \& {Fokin}(1985)}]{fadeev1985}
{Fadeev} I.~A., {Fokin} A.~B., 1985, APSS, 111, 355

\bibitem[{{Fazio} {et~al}\mbox{.}(2004){Fazio}, {Hora}, {Allen}, {Ashby},
  {Barmby}, {Deutsch}, {Huang}, {Kleiner}, {Marengo}, {Megeath}, {Melnick},
  {Pahre}, {Patten}, {Polizotti}, {Smith}, {Taylor}, {Wang}, {Willner},
  {Hoffmann}, {Pipher}, {Forrest}, {McMurty}, {McCreight}, {McKelvey},
  {McMurray}, {Koch}, {Moseley}, {Arendt}, {Mentzell}, {Marx}, {Losch},
  {Mayman}, {Eichhorn}, {Krebs}, {Jhabvala}, {Gezari}, {Fixsen}, {Flores},
  {Shakoorzadeh}, {Jungo}, {Hakun}, {Workman}, {Karpati}, {Kichak}, {Whitley},
  {Mann}, {Tollestrup}, {Eisenhardt}, {Stern}, {Gorjian}, {Bhattacharya},
  {Carey}, {Nelson}, {Glaccum}, {Lacy}, {Lowrance}, {Laine}, {Reach},
  {Stauffer}, {Surace}, {Wilson}, {Wright}, {Hoffman}, {Domingo}, \&
  {Cohen}}]{fazio2004}
{Fazio} G.~G. {et~al.}, 2004, ApJS, 154, 10

\bibitem[{{Feast}(1999)}]{feast1999}
{Feast} M., 1999, PASP, 111, 775

\bibitem[{{Feast}(2010)}]{feast2010}
{Feast} M.~W., 2010, in Variable Stars, the Galactic halo and Galaxy Formation,
  p.~45

\bibitem[{{Feast}(2013)}]{feast2013}
{Feast} M.~W., 2013, {Galactic Distance Scales}, Vol.~5, p. 829

\bibitem[{{Feast} {et~al}\mbox{.}(2008){Feast}, {Laney}, {Kinman}, {van
  Leeuwen}, \& {Whitelock}}]{feast2008}
{Feast} M.~W., {Laney} C.~D., {Kinman} T.~D., {van Leeuwen} F., {Whitelock}
  P.~A., 2008, MNRAS, 386, 2115

\bibitem[{{Feast} {et~al}\mbox{.}(2012){Feast}, {Whitelock}, {Menzies}, \&
  {Matsunaga}}]{feast2012}
{Feast} M.~W., {Whitelock} P.~A., {Menzies} J.~W., {Matsunaga} N., 2012, MNRAS,
  421, 2998

\bibitem[{{Fernley} {et~al}\mbox{.}(1998){Fernley}, {Barnes}, {Skillen},
  {Hawley}, {Hanley}, {Evans}, {Solano}, \& {Garrido}}]{fernley1998}
{Fernley} J., {Barnes} T.~G., {Skillen} I., {Hawley} S.~L., {Hanley} C.~J.,
  {Evans} D.~W., {Solano} E., {Garrido} R., 1998, A\&A, 330, 515

\bibitem[{{Fiorentino} {et~al}\mbox{.}(2002){Fiorentino}, {Caputo}, {Marconi},
  \& {Musella}}]{fiorentino2002}
{Fiorentino} G., {Caputo} F., {Marconi} M., {Musella} I., 2002, ApJ, 576, 402

\bibitem[{{Fiorentino} {et~al}\mbox{.}(2006){Fiorentino}, {Limongi}, {Caputo},
  \& {Marconi}}]{fiorentino2006}
{Fiorentino} G., {Limongi} M., {Caputo} F., {Marconi} M., 2006, A\&A, 460, 155

\bibitem[{{Fiorentino} {et~al}\mbox{.}(2007){Fiorentino}, {Marconi}, {Musella},
  \& {Caputo}}]{fiorentino2007}
{Fiorentino} G., {Marconi} M., {Musella} I., {Caputo} F., 2007, A\&A, 476, 863

\bibitem[{{For}, {Sneden} \& {Preston}(2011){For}, {Sneden}, \&
  {Preston}}]{for2011}
{For} B.-Q., {Sneden} C., {Preston} G.~W., 2011, ApJS, 197, 29

\bibitem[{{Fouqu{\'e}} {et~al}\mbox{.}(2007){Fouqu{\'e}}, {Arriagada}, {Storm},
  {Barnes}, {Nardetto}, {M{\'e}rand}, {Kervella}, {Gieren}, {Bersier},
  {Benedict}, \& {McArthur}}]{fouque2007}
{Fouqu{\'e}} P. {et~al.}, 2007, A\&A, 476, 73

\bibitem[{{Freedman} \& {Madore}(2010)}]{freedman2010}
{Freedman} W.~L., {Madore} B.~F., 2010, ARA\&A, 48, 673

\bibitem[{{Freedman} {et~al}\mbox{.}(2001){Freedman}, {Madore}, {Gibson},
  {Ferrarese}, {Kelson}, {Sakai}, {Mould}, {Kennicutt}, {Ford}, {Graham},
  {Huchra}, {Hughes}, {Illingworth}, {Macri}, \& {Stetson}}]{freedman2001}
{Freedman} W.~L. {et~al.}, 2001, ApJ, 553, 47

\bibitem[{{Freedman} {et~al}\mbox{.}(2019){Freedman}, {Madore}, {Hatt}, {Hoyt},
  {Jang}, {Beaton}, {Burns}, {Lee}, {Monson}, {Neeley}, {Phillips}, {Rich}, \&
  {Seibert}}]{freedman2019}
{Freedman} W.~L. {et~al.}, 2019, ApJ, 882, 34

\bibitem[{{Freedman} {et~al}\mbox{.}(2008){Freedman}, {Madore}, {Rigby},
  {Persson}, \& {Sturch}}]{freedman2008}
{Freedman} W.~L., {Madore} B.~F., {Rigby} J., {Persson} S.~E., {Sturch} L.,
  2008, ApJ, 679, 71

\bibitem[{{Freedman} {et~al}\mbox{.}(2012){Freedman}, {Madore}, {Scowcroft},
  {Burns}, {Monson}, {Persson}, {Seibert}, \& {Rigby}}]{freedman2012}
{Freedman} W.~L., {Madore} B.~F., {Scowcroft} V., {Burns} C., {Monson} A.,
  {Persson} S.~E., {Seibert} M., {Rigby} J., 2012, ApJ, 758, 24

\bibitem[{{Freedman} {et~al}\mbox{.}(2011){Freedman}, {Madore}, {Scowcroft},
  {Monson}, {Persson}, {Seibert}, {Rigby}, {Sturch}, \&
  {Stetson}}]{freedman2011}
{Freedman} W.~L., {Madore} B.~F., {Scowcroft} V., {Monson} A., {Persson} S.~E.,
  {Seibert} M., {Rigby} J.~R., {Sturch} L., {Stetson} P., 2011, AJ, 142, 192

\bibitem[{{Freedman} {et~al}\mbox{.}(2009){Freedman}, {Rigby}, {Madore},
  {Persson}, {Sturch}, \& {Mager}}]{freedman2009}
{Freedman} W.~L., {Rigby} J., {Madore} B.~F., {Persson} S.~E., {Sturch} L.,
  {Mager} V., 2009, ApJ, 695, 996

\bibitem[{{Garc{\'{\i}}a-Varela}, {Sabogal} \&
  {Ram{\'{\i}}rez-Tannus}(2013){Garc{\'{\i}}a-Varela}, {Sabogal}, \&
  {Ram{\'{\i}}rez-Tannus}}]{varela2013}
{Garc{\'{\i}}a-Varela} A., {Sabogal} B.~E., {Ram{\'{\i}}rez-Tannus} M.~C.,
  2013, MNRAS, 431, 2278

\bibitem[{{Genovali} {et~al}\mbox{.}(2014){Genovali}, {Lemasle}, {Bono},
  {Romaniello}, {Fabrizio}, {Ferraro}, {Iannicola}, {Laney}, {Nonino},
  {Bergemann}, {Buonanno}, {Fran{\c{c}}ois}, {Inno}, {Kudritzki}, {Matsunaga},
  {Pedicelli}, {Primas}, \& {Th{\'e}venin}}]{genovali2014}
{Genovali} K. {et~al.}, 2014, A\&A, 566, A37

\bibitem[{{Genovali} {et~al}\mbox{.}(2013){Genovali}, {Lemasle}, {Bono},
  {Romaniello}, {Primas}, {Fabrizio}, {Buonanno}, {Fran{\c{c}}ois}, {Inno},
  {Laney}, {Matsunaga}, {Pedicelli}, \& {Th{\'e}venin}}]{genovali2013}
{Genovali} K. {et~al.}, 2013, A\&A, 554, A132

\bibitem[{{Genovali} {et~al}\mbox{.}(2015){Genovali}, {Lemasle}, {da Silva},
  {Bono}, {Fabrizio}, {Bergemann}, {Buonanno}, {Ferraro}, {Fran{\c{c}}ois},
  {Iannicola}, {Inno}, {Laney}, {Kudritzki}, {Matsunaga}, {Nonino}, {Primas},
  {Romaniello}, {Urbaneja}, \& {Th{\'e}venin}}]{genovali2015}
{Genovali} K. {et~al.}, 2015, A\&A, 580, A17

\bibitem[{{Gieren} {et~al}\mbox{.}(2013){Gieren}, {G{\'o}rski},
  {Pietrzy{\'n}ski}, {Konorski}, {Suchomska}, {Graczyk}, {Pilecki}, {Bresolin},
  {Kudritzki}, {Storm}, {Karczmarek}, {Gallenne}, {Calder{\'o}n}, \&
  {Geisler}}]{gieren2013}
{Gieren} W. {et~al.}, 2013, ApJ, 773, 69

\bibitem[{{Gieren} {et~al}\mbox{.}(2005){Gieren}, {Pietrzy{\'n}ski},
  {Soszy{\'n}ski}, {Bresolin}, {Kudritzki}, {Minniti}, \& {Storm}}]{gieren2005}
{Gieren} W., {Pietrzy{\'n}ski} G., {Soszy{\'n}ski} I., {Bresolin} F.,
  {Kudritzki} R.-P., {Minniti} D., {Storm} J., 2005, ApJ, 628, 695

\bibitem[{{Gieren} {et~al}\mbox{.}(2009){Gieren}, {Pietrzy{\'n}ski},
  {Soszy{\'n}ski}, {Szewczyk}, {Bresolin}, {Kudritzki}, {Urbaneja}, {Storm},
  {Minniti}, \& {Garc{\'{\i}}a-Varela}}]{gieren2009}
{Gieren} W., {Pietrzy{\'n}ski} G., {Soszy{\'n}ski} I., {Szewczyk} O.,
  {Bresolin} F., {Kudritzki} R.-P., {Urbaneja} M.~A., {Storm} J., {Minniti} D.,
  {Garc{\'{\i}}a-Varela} A., 2009, ApJ, 700, 1141

\bibitem[{{Gieren} {et~al}\mbox{.}(2018){Gieren}, {Storm}, {Konorski},
  {G{\'o}rski}, {Pilecki}, {Thompson}, {Pietrzy{\'n}ski}, {Graczyk}, {Barnes},
  {Fouqu{\'e}}, {Nardetto}, {Gallenne}, {Karczmarek}, {Suchomska},
  {Wielg{\'o}rski}, {Taormina}, \& {Zgirski}}]{gieren2018}
{Gieren} W. {et~al.}, 2018, A\&A, 620, A99

\bibitem[{{Gieren}, {Fouque} \& {Gomez}(1998){Gieren}, {Fouque}, \&
  {Gomez}}]{gieren1998}
{Gieren} W.~P., {Fouque} P., {Gomez} M., 1998, ApJ, 496, 17

\bibitem[{{Gingold}(1976)}]{gingold1976}
{Gingold} R.~A., 1976, ApJ, 204, 116

\bibitem[{{Goodricke}(1783)}]{goodricke1783}
{Goodricke} J., 1783, Philosophical Transactions of the Royal Society of London
  Series I, 73, 474

\bibitem[{{Goodricke}(1786)}]{goodricke1786}
{Goodricke} J., 1786, Philosophical Transactions of the Royal Society of London
  Series I, 76, 48

\bibitem[{{Gratton} {et~al}\mbox{.}(2004){Gratton}, {Bragaglia}, {Clementini},
  {Carretta}, {Di Fabrizio}, {Maio}, \& {Taribello}}]{gratton2004}
{Gratton} R.~G., {Bragaglia} A., {Clementini} G., {Carretta} E., {Di Fabrizio}
  L., {Maio} M., {Taribello} E., 2004, A\&A, 421, 937

\bibitem[{{Groenewegen}(2013)}]{gmat2013}
{Groenewegen} M.~A.~T., 2013, A\&A, 550, A70

\bibitem[{{Groenewegen}(2018)}]{groenewegen2018}
{Groenewegen} M.~A.~T., 2018, A\&A, 619, A8

\bibitem[{{Groenewegen} \& {Jurkovic}(2017{\natexlab{a}})}]{groenewegen2017a}
{Groenewegen} M.~A.~T., {Jurkovic} M.~I., 2017{\natexlab{a}}, A\&A, 603, A70

\bibitem[{{Groenewegen} \& {Jurkovic}(2017{\natexlab{b}})}]{groenewegen2017}
{Groenewegen} M.~A.~T., {Jurkovic} M.~I., 2017{\natexlab{b}}, A\&A, 604, A29

\bibitem[{{Groenewegen} {et~al}\mbox{.}(2004){Groenewegen}, {Romaniello},
  {Primas}, \& {Mottini}}]{groenewegen2004}
{Groenewegen} M.~A.~T., {Romaniello} M., {Primas} F., {Mottini} M., 2004, A\&A,
  420, 655

\bibitem[{{Groenewegen}, {Udalski} \& {Bono}(2008){Groenewegen}, {Udalski}, \&
  {Bono}}]{gmat2008}
{Groenewegen} M.~A.~T., {Udalski} A., {Bono} G., 2008, A\&A, 481, 441

\bibitem[{{Hajdu} {et~al}\mbox{.}(2018){Hajdu}, {D{\'e}k{\'a}ny}, {Catelan},
  {Grebel}, \& {Jurcsik}}]{hajdu2018}
{Hajdu} G., {D{\'e}k{\'a}ny} I., {Catelan} M., {Grebel} E.~K., {Jurcsik} J.,
  2018, ApJ, 857, 55

\bibitem[{{Harris}(1985)}]{harris1985}
{Harris} H.~C., 1985, in {Population II Cepheids}, {Madore} B.~F., ed., pp.
  232--245

\bibitem[{{Haschke} {et~al}\mbox{.}(2012){Haschke}, {Grebel}, {Frebel},
  {Duffau}, {Hansen}, \& {Koch}}]{haschke2012}
{Haschke} R., {Grebel} E.~K., {Frebel} A., {Duffau} S., {Hansen} C.~J., {Koch}
  A., 2012, AJ, 144, 88

\bibitem[{{Hatt} {et~al}\mbox{.}(2017){Hatt}, {Beaton}, {Freedman}, {Madore},
  {Jang}, {Hoyt}, {Lee}, {Monson}, {Rich}, {Scowcroft}, \&
  {Seibert}}]{hatt2017}
{Hatt} D. {et~al.}, 2017, ApJ, 845, 146

\bibitem[{{Hertzsprung}(1926)}]{hertzsprung1926}
{Hertzsprung} E., 1926, BAIN, 3, 115

\bibitem[{{Hoffmeister}(1929)}]{hoffmeister1929}
{Hoffmeister} C., 1929, Astronomische Nachrichten, 236, 233

\bibitem[{{Hubble}(1929)}]{hubble1929}
{Hubble} E., 1929, Proceedings of the National Academy of Science, 15, 168

\bibitem[{{Hubble}(1926)}]{hubble1926}
{Hubble} E.~P., 1926, ApJ, 64, 321

\bibitem[{{Inno} {et~al}\mbox{.}(2013){Inno}, {Matsunaga}, {Bono}, {Caputo},
  {Buonanno}, {Genovali}, {Laney}, {Marconi}, {Piersimoni}, {Primas}, \&
  {Romaniello}}]{inno2013}
{Inno} L. {et~al.}, 2013, ApJ, 764, 84

\bibitem[{{Inno} {et~al}\mbox{.}(2015){Inno}, {Matsunaga}, {Romaniello},
  {Bono}, {Monson}, {Ferraro}, {Iannicola}, {Persson}, {Buonanno}, {Freedman},
  {Gieren}, {Groenewegen}, {Ita}, {Laney}, {Lemasle}, {Madore}, {Nagayama},
  {Nakada}, {Nonino}, {Pietrzy{\'n}ski}, {Primas}, {Scowcroft},
  {Soszy{\'n}ski}, {Tanab{\'e}}, \& {Udalski}}]{inno2015}
{Inno} L. {et~al.}, 2015, A\&A, 576, A30

\bibitem[{{Ita} {et~al}\mbox{.}(2004{\natexlab{a}}){Ita}, {Tanab{\'e}},
  {Matsunaga}, {Nakajima}, {Nagashima}, {Nagayama}, {Kato}, {Kurita}, {Nagata},
  {Sato}, {Tamura}, {Nakaya}, \& {Nakada}}]{ita2004b}
{Ita} Y. {et~al.}, 2004{\natexlab{a}}, MNRAS, 353, 705

\bibitem[{{Ita} {et~al}\mbox{.}(2004{\natexlab{b}}){Ita}, {Tanab{\'e}},
  {Matsunaga}, {Nakajima}, {Nagashima}, {Nagayama}, {Kato}, {Kurita}, {Nagata},
  {Sato}, {Tamura}, {Nakaya}, \& {Nakada}}]{ita2004a}
{Ita} Y. {et~al.}, 2004{\natexlab{b}}, MNRAS, 347, 720

\bibitem[{{Jacyszyn-Dobrzeniecka} {et~al}\mbox{.}(2016){Jacyszyn-Dobrzeniecka},
  {Skowron}, {Mr{\'o}z}, {Skowron}, {Soszy{\'n}ski}, {Udalski}, {Pietrukowicz},
  {Koz{\l}owski}, {Wyrzykowski}, {Poleski}, {Pawlak}, {Szyma{\'n}ski}, \&
  {Ulaczyk}}]{jacyszyn2016}
{Jacyszyn-Dobrzeniecka} A.~M. {et~al.}, 2016, Acta Astron., 66, 149

\bibitem[{{Jacyszyn-Dobrzeniecka} {et~al}\mbox{.}(2017){Jacyszyn-Dobrzeniecka},
  {Skowron}, {Mr{\'o}z}, {Soszy{\'n}ski}, {Udalski}, {Pietrukowicz}, {Skowron},
  {Poleski}, {Koz{\l}owski}, {Wyrzykowski}, {Pawlak}, {Szyma{\'n}ski}, \&
  {Ulaczyk}}]{jacyszyn2017}
{Jacyszyn-Dobrzeniecka} A.~M. {et~al.}, 2017, Acta Astron., 67, 1

\bibitem[{{Jeffery} \& {Saio}(2016)}]{jeffery2016}
{Jeffery} C.~S., {Saio} H., 2016, MNRAS, 458, 1352

\bibitem[{{Jurcsik} \& {Kovacs}(1996)}]{jk1996}
{Jurcsik} J., {Kovacs} G., 1996, A\&A, 312, 111

\bibitem[{{Jurcsik} {et~al}\mbox{.}(2009){Jurcsik}, {S{\'o}dor}, {Szeidl},
  {Hurta}, {V{\'a}radi}, {Posztob{\'a}nyi}, {Vida}, {Hajdu},
  {K{\H{o}}v{\'a}ri}, {Nagy}, {Moln{\'a}r}, \& {Belucz}}]{jurcsik2009}
{Jurcsik} J. {et~al.}, 2009, MNRAS, 400, 1006

\bibitem[{{Jurkovic}(2018)}]{jurkovic2018}
{Jurkovic} M.~I., 2018, Serbian Astronomical Journal, 197, 13

\bibitem[{{Kains} {et~al}\mbox{.}(2019){Kains}, {Calamida}, {Rejkuba},
  {Bhardwaj}, {Inno}, {Sahu}, {Zoccali}, {Bono}, {Surot}, {Anderson}, \&
  {Casertano}}]{kains2019}
{Kains} N. {et~al.}, 2019, MNRAS, 482, 3058

\bibitem[{{Kanbur} \& {Ngeow}(2005)}]{kanbur2005a}
{Kanbur} S., {Ngeow} C., 2005, in ESA Special Publication, Vol. 576, The
  Three-Dimensional Universe with Gaia, {Turon} C., {O'Flaherty} K.~S.,
  {Perryman} M.~A.~C., eds., p. 691

\bibitem[{{Kanbur} {et~al}\mbox{.}(2010){Kanbur}, {Marconi}, {Ngeow},
  {Musella}, {Turner}, {James}, {Magin}, \& {Halsey}}]{kanbur2010}
{Kanbur} S.~M., {Marconi} M., {Ngeow} C., {Musella} I., {Turner} M., {James}
  A., {Magin} S., {Halsey} J., 2010, MNRAS, 408, 695

\bibitem[{{Kanbur} {et~al}\mbox{.}(2007){Kanbur}, {Ngeow}, {Nanthakumar}, \&
  {Stevens}}]{kanbur2007}
{Kanbur} S.~M., {Ngeow} C., {Nanthakumar} A., {Stevens} R., 2007, PASP, 119,
  512

\bibitem[{{Karczmarek} {et~al}\mbox{.}(2015){Karczmarek}, {Pietrzy{\'n}ski},
  {Gieren}, {Suchomska}, {Konorski}, {G{\'o}rski}, {Pilecki}, {Graczyk}, \&
  {Wielg{\'o}rski}}]{karczmarek2015}
{Karczmarek} P., {Pietrzy{\'n}ski} G., {Gieren} W., {Suchomska} K., {Konorski}
  P., {G{\'o}rski} M., {Pilecki} B., {Graczyk} D., {Wielg{\'o}rski} P., 2015,
  AJ, 150, 90

\bibitem[{{Karczmarek} {et~al}\mbox{.}(2017){Karczmarek}, {Pietrzy{\'n}ski},
  {G{\'o}rski}, {Gieren}, \& {Bersier}}]{karczmarek2017}
{Karczmarek} P., {Pietrzy{\'n}ski} G., {G{\'o}rski} M., {Gieren} W., {Bersier}
  D., 2017, AJ, 154, 263

\bibitem[{{Kennicutt} {et~al}\mbox{.}(1998){Kennicutt}, {Stetson}, {Saha},
  {Kelson}, {Rawson}, {Sakai}, {Madore}, {Mould}, {Freedman}, {Bresolin},
  {Ferrarese}, {Ford}, {Gibson}, {Graham}, {Han}, {Harding}, {Hoessel},
  {Huchra}, {Hughes}, {Illingworth}, {Macri}, {Phelps}, {Silbermann}, {Turner},
  \& {Wood}}]{kennicutt1998}
{Kennicutt}, Robert~C. J. {et~al.}, 1998, ApJ, 498, 181

\bibitem[{{Kervella} {et~al}\mbox{.}(2004){Kervella}, {Nardetto}, {Bersier},
  {Mourard}, \& {Coud{\'e} du Foresto}}]{kervella2004}
{Kervella} P., {Nardetto} N., {Bersier} D., {Mourard} D., {Coud{\'e} du
  Foresto} V., 2004, A\&A, 416, 941

\bibitem[{{Kinemuchi} {et~al}\mbox{.}(2006){Kinemuchi}, {Smith}, {Wo{\'z}niak},
  {McKay}, \& {ROTSE Collaboration}}]{kinemuchi2006}
{Kinemuchi} K., {Smith} H.~A., {Wo{\'z}niak} P.~R., {McKay} T.~A., {ROTSE
  Collaboration}, 2006, AJ, 132, 1202

\bibitem[{{Kinman} \& {Brown}(2014)}]{kinman2014}
{Kinman} T.~D., {Brown} W.~R., 2014, AJ, 148, 121

\bibitem[{{Kippenhahn} \& {Weigert}(1991)}]{kippenhahn1991}
{Kippenhahn} R., {Weigert} A., 1991, SSR, 58, 190

\bibitem[{{Klein} {et~al}\mbox{.}(2011){Klein}, {Richards}, {Butler}, \&
  {Bloom}}]{klein2011}
{Klein} C.~R., {Richards} J.~W., {Butler} N.~R., {Bloom} J.~S., 2011, ApJ, 738,
  185

\bibitem[{{Klein} {et~al}\mbox{.}(2014){Klein}, {Richards}, {Butler}, \&
  {Bloom}}]{klein2014}
{Klein} C.~R., {Richards} J.~W., {Butler} N.~R., {Bloom} J.~S., 2014, MNRAS,
  440, L96

\bibitem[{{Kodric} {et~al}\mbox{.}(2018){Kodric}, {Riffeser}, {Hopp}, {Goessl},
  {Seitz}, {Bender}, {Koppenhoefer}, {Obermeier}, {Snigula}, {Lee}, {Burgett},
  {Draper}, {Hodapp}, {Kaiser}, {Kudritzki}, {Metcalfe}, {Tonry}, \&
  {Wainscoat}}]{kodric2018}
{Kodric} M. {et~al.}, 2018, AJ, 156, 130

\bibitem[{{Kodric} {et~al}\mbox{.}(2015){Kodric}, {Riffeser}, {Seitz},
  {Snigula}, {Hopp}, {Lee}, {Goessl}, {Koppenhoefer}, {Bender}, \&
  {Gieren}}]{kodric2015}
{Kodric} M., {Riffeser} A., {Seitz} S., {Snigula} J., {Hopp} U., {Lee} C.-H.,
  {Goessl} C., {Koppenhoefer} J., {Bender} R., {Gieren} W., 2015, ApJ, 799, 144

\bibitem[{{Kolenberg} {et~al}\mbox{.}(2010){Kolenberg}, {Szab{\'o}}, {Kurtz},
  {Gilliland }, {Christensen-Dalsgaard}, {Kjeldsen}, {Brown}, {Benk{\H{o}}},
  {Chadid}, {Derekas}, {Di Criscienzo}, {Guggenberger}, {Kinemuchi}, {Kunder},
  {Koll{\'a}th}, {Kopacki}, {Moskalik}, {Nemec}, {Nuspl}, {Silvotti}, {Suran},
  {Borucki}, {Koch}, \& {Jenkins}}]{kolenberg2010}
{Kolenberg} K. {et~al.}, 2010, ApJL, 713, L198

\bibitem[{{Kovtyukh} {et~al}\mbox{.}(2018){Kovtyukh}, {Yegorova}, {Andrievsky},
  {Korotin}, {Saviane}, {Lemasle}, {Chekhonadskikh}, \& {Belik}}]{kovtyukh2018}
{Kovtyukh} V., {Yegorova} I., {Andrievsky} S., {Korotin} S., {Saviane} I.,
  {Lemasle} B., {Chekhonadskikh} F., {Belik} S., 2018, MNRAS, 477, 2276

\bibitem[{{Kubiak} \& {Udalski}(2003)}]{kubiak2003}
{Kubiak} M., {Udalski} A., 2003, Acta Astron., 53, 117

\bibitem[{{Kunder} \& {Chaboyer}(2009)}]{kunder2009}
{Kunder} A., {Chaboyer} B., 2009, AJ, 138, 1284

\bibitem[{{Kunder} {et~al}\mbox{.}(2018){Kunder}, {Valenti}, {Dall'Ora},
  {Pietrukowicz}, {Sneden}, {Bono}, {Braga}, {Ferraro}, {Fiorentino},
  {Iannicola}, {Marconi}, {Mart{\'\i}nez-V{\'a}zquez}, {Monelli}, {Musella},
  {Ripepi}, {Salaris}, \& {Stetson}}]{kunder2018}
{Kunder} A. {et~al.}, 2018, Space Science Reviews, 214, 90

\bibitem[{{Leavitt}(1908)}]{leavitt1908}
{Leavitt} H.~S., 1908, Annals of Harvard College Observatory, 60, 87

\bibitem[{{Leavitt} \& {Pickering}(1912)}]{leavitt1912}
{Leavitt} H.~S., {Pickering} E.~C., 1912, Harvard College Observatory Circular,
  173, 1

\bibitem[{{Lemasle} {et~al}\mbox{.}(2013){Lemasle}, {Fran{\c{c}}ois},
  {Genovali}, {Kovtyukh}, {Bono}, {Inno}, {Laney}, {Kaper}, {Bergemann},
  {Fabrizio}, {Matsunaga}, {Pedicelli}, {Primas}, \&
  {Romaniello}}]{lemasle2013}
{Lemasle} B. {et~al.}, 2013, A\&A, 558, A31

\bibitem[{{Lemasle} {et~al}\mbox{.}(2017){Lemasle}, {Groenewegen}, {Grebel},
  {Bono}, {Fiorentino}, {Fran{\c{c}}ois}, {Inno}, {Kovtyukh}, {Matsunaga},
  {Pedicelli}, {Primas}, {Pritchard}, {Romaniello}, \& {da
  Silva}}]{lemasle2017}
{Lemasle} B. {et~al.}, 2017, A\&A, 608, A85

\bibitem[{{Lemasle} {et~al}\mbox{.}(2015){Lemasle}, {Kovtyukh}, {Bono},
  {Fran{\c c}ois}, {Saviane}, {Yegorova}, {Genovali}, {Inno}, {Galazutdinov},
  \& {da Silva}}]{lemasle2015}
{Lemasle} B., {Kovtyukh} V., {Bono} G., {Fran{\c c}ois} P., {Saviane} I.,
  {Yegorova} I., {Genovali} K., {Inno} L., {Galazutdinov} G., {da Silva} R.,
  2015, A\&A, 579, A47

\bibitem[{{Lindegren} {et~al}\mbox{.}(2018){Lindegren}, {Hern{\'a}ndez},
  {Bombrun}, {Klioner}, {Bastian}, {Ramos-Lerate}, {de Torres},
  {Steidelm{\"u}ller}, {Stephenson}, {Hobbs}, {Lammers}, {Biermann}, {Geyer},
  {Hilger}, {Michalik}, {Stampa}, {McMillan}, {Casta{\~n}eda}, {Clotet},
  {Comoretto}, {Davidson}, {Fabricius}, {Gracia}, {Hambly}, {Hutton}, {Mora},
  {Portell}, {van Leeuwen}, {Abbas}, {Abreu}, {Altmann}, {Andrei}, {Anglada},
  {Balaguer-N{\'u}{\~n}ez}, {Barache}, {Becciani}, {Bertone}, {Bianchi},
  {Bouquillon}, {Bourda}, {Br{\"u}semeister}, {Bucciarelli}, {Busonero},
  {Buzzi}, {Cancelliere}, {Carlucci}, {Charlot}, {Cheek}, {Crosta}, {Crowley},
  {de Bruijne}, {de Felice}, {Drimmel}, {Esquej}, {Fienga}, {Fraile}, {Gai},
  {Garralda}, {Gonz{\'a}lez-Vidal}, {Guerra}, {Hauser}, {Hofmann}, {Holl},
  {Jordan}, {Lattanzi}, {Lenhardt}, {Liao}, {Licata}, {Lister}, {L{\"o}ffler},
  {Marchant}, {Martin-Fleitas}, {Messineo}, {Mignard}, {Morbidelli}, {Poggio},
  {Riva}, {Rowell}, {Salguero}, {Sarasso}, {Sciacca}, {Siddiqui}, {Smart},
  {Spagna}, {Steele}, {Taris}, {Torra}, {van Elteren}, {van Reeven}, \&
  {Vecchiato}}]{lindegren2018}
{Lindegren} L. {et~al.}, 2018, A\&A, 616, A2

\bibitem[{{Lindegren} {et~al}\mbox{.}(2016){Lindegren}, {Lammers}, {Bastian},
  {Hern{\'a}ndez}, {Klioner}, {Hobbs}, {Bombrun}, {Michalik}, {Ramos-Lerate},
  {Butkevich}, {Comoretto}, {Joliet}, {Holl}, {Hutton}, {Parsons},
  {Steidelm{\"u}ller}, {Abbas}, {Altmann}, {Andrei}, {Anton}, {Bach},
  {Barache}, {Becciani}, {Berthier}, {Bianchi}, {Biermann}, {Bouquillon},
  {Bourda}, {Br{\"u}semeister}, {Bucciarelli}, {Busonero}, {Carlucci},
  {Casta{\~n}eda}, {Charlot}, {Clotet}, {Crosta}, {Davidson}, {de Felice},
  {Drimmel}, {Fabricius}, {Fienga}, {Figueras}, {Fraile}, {Gai}, {Garralda},
  {Geyer}, {Gonz{\'a}lez-Vidal}, {Guerra}, {Hambly}, {Hauser}, {Jordan},
  {Lattanzi}, {Lenhardt}, {Liao}, {L{\"o}ffler}, {McMillan}, {Mignard}, {Mora},
  {Morbidelli}, {Portell}, {Riva}, {Sarasso}, {Serraller}, {Siddiqui}, {Smart},
  {Spagna}, {Stampa}, {Steele}, {Taris}, {Torra}, {van Reeven}, {Vecchiato},
  {Zschocke}, {de Bruijne}, {Gracia}, {Raison}, {Lister}, {Marchant},
  {Messineo}, {Soffel}, {Osorio}, {de Torres}, \& {O'Mullane}}]{lindegren2016}
{Lindegren} L. {et~al.}, 2016, A\&A, 595, A4

\bibitem[{{Longmore}, {Fernley} \& {Jameson}(1986){Longmore}, {Fernley}, \&
  {Jameson}}]{longmore1986}
{Longmore} A.~J., {Fernley} J.~A., {Jameson} R.~F., 1986, MNRAS, 220, 279

\bibitem[{{Maas}, {Giridhar} \& {Lambert}(2007){Maas}, {Giridhar}, \&
  {Lambert}}]{maas2007}
{Maas} T., {Giridhar} S., {Lambert} D.~L., 2007, ApJ, 666, 378

\bibitem[{{Macri} {et~al}\mbox{.}(2015){Macri}, {Ngeow}, {Kanbur}, {Mahzooni},
  \& {Smitka}}]{macri2015}
{Macri} L.~M., {Ngeow} C.-C., {Kanbur} S.~M., {Mahzooni} S., {Smitka} M.~T.,
  2015, AJ, 149, 117

\bibitem[{{Macri} {et~al}\mbox{.}(2006){Macri}, {Stanek}, {Bersier},
  {Greenhill}, \& {Reid}}]{macri2006}
{Macri} L.~M., {Stanek} K.~Z., {Bersier} D., {Greenhill} L.~J., {Reid} M.~J.,
  2006, ApJ, 652, 1133

\bibitem[{{Madore}(1982)}]{madore1982}
{Madore} B.~F., 1982, ApJ, 253, 575

\bibitem[{{Madore} \& {Freedman}(1991)}]{madore1991}
{Madore} B.~F., {Freedman} W.~L., 1991, PASP, 103, 933

\bibitem[{{Madore} \& {Freedman}(2009)}]{madore2009}
{Madore} B.~F., {Freedman} W.~L., 2009, ApJ, 696, 1498

\bibitem[{{Madore} {et~al}\mbox{.}(2009){Madore}, {Freedman}, {Rigby},
  {Persson}, {Sturch}, \& {Mager}}]{madore2009a}
{Madore} B.~F., {Freedman} W.~L., {Rigby} J., {Persson} S.~E., {Sturch} L.,
  {Mager} V., 2009, ApJ, 695, 988

\bibitem[{{Madore} {et~al}\mbox{.}(2013){Madore}, {Hoffman}, {Freedman},
  {Kollmeier}, {Monson}, {Persson}, {Rich}, {Scowcroft}, \&
  {Seibert}}]{madore2013}
{Madore} B.~F., {Hoffman} D., {Freedman} W.~L., {Kollmeier} J.~A., {Monson} A.,
  {Persson} S.~E., {Rich}, Jeff~A. J., {Scowcroft} V., {Seibert} M., 2013, ApJ,
  776, 135

\bibitem[{{Magurno} {et~al}\mbox{.}(2019){Magurno}, {Sneden}, {Bono}, {Braga},
  {Mateo}, {Persson}, {Preston}, {Th{\'e}venin}, {da Silva}, {Dall'Ora},
  {Fabrizio}, {Ferraro}, {Fiorentino}, {Iannicola}, {Inno}, {Marengo},
  {Marinoni}, {Marrese}, {Mart{\'\i}nez-V{\'a}zquez}, {Matsunaga}, {Monelli},
  {Neeley}, {Nonino}, \& {Walker}}]{magurno2019}
{Magurno} D. {et~al.}, 2019, ApJ, 881, 104

\bibitem[{{Magurno} {et~al}\mbox{.}(2018){Magurno}, {Sneden}, {Braga}, {Bono},
  {Mateo}, {Persson}, {DallOra}, {Marengo}, {Monelli}, \&
  {Neeley}}]{magurno2018}
{Magurno} D., {Sneden} C., {Braga} V.~F., {Bono} G., {Mateo} M., {Persson}
  S.~E., {DallOra} M., {Marengo} M., {Monelli} M., {Neeley} J.~R., 2018, ApJ,
  864, 57

\bibitem[{{Majaess}, {Turner} \& {Lane}(2009){Majaess}, {Turner}, \&
  {Lane}}]{majaess2009}
{Majaess} D., {Turner} D., {Lane} D., 2009, Acta Astron., 59, 403

\bibitem[{{Majewski} {et~al}\mbox{.}(2017){Majewski}, {Schiavon}, {Frinchaboy},
  {Allende Prieto}, {Barkhouser}, {Bizyaev}, {Blank}, {Brunner}, {Burton},
  {Carrera}, {Chojnowski}, {Cunha}, {Epstein}, {Fitzgerald}, {Garc{\'\i}a
  P{\'e}rez}, {Hearty}, {Henderson}, {Holtzman}, {Johnson}, {Lam}, {Lawler},
  {Maseman}, {M{\'e}sz{\'a}ros}, {Nelson}, {Nguyen}, {Nidever}, {Pinsonneault},
  {Shetrone}, {Smee}, {Smith}, {Stolberg}, {Skrutskie}, {Walker}, {Wilson},
  {Zasowski}, {Anders}, {Basu}, {Beland}, {Blanton}, {Bovy}, {Brownstein},
  {Carlberg}, {Chaplin}, {Chiappini}, {Eisenstein}, {Elsworth}, {Feuillet},
  {Fleming}, {Galbraith-Frew}, {Garc{\'\i}a}, {Garc{\'\i}a-Hern{\'a}ndez},
  {Gillespie}, {Girardi}, {Gunn}, {Hasselquist}, {Hayden}, {Hekker}, {Ivans},
  {Kinemuchi}, {Klaene}, {Mahadevan}, {Mathur}, {Mosser}, {Muna}, {Munn},
  {Nichol}, {O'Connell}, {Parejko}, {Robin}, {Rocha-Pinto}, {Schultheis},
  {Serenelli}, {Shane}, {Silva Aguirre}, {Sobeck}, {Thompson}, {Troup},
  {Weinberg}, \& {Zamora}}]{majewski2017}
{Majewski} S.~R. {et~al.}, 2017, AJ, 154, 94

\bibitem[{{Mancino} {et~al}\mbox{.}(2020){Mancino}, {Romaniello}, {Anderson},
  \& {Kudritzki}}]{mancino2020}
{Mancino} S., {Romaniello} M., {Anderson} R.~I., {Kudritzki} R.-P., 2020, arXiv
  e-prints, arXiv:2001.05881

\bibitem[{{Manick} {et~al}\mbox{.}(2018){Manick}, {Van Winckel}, {Kamath},
  {Sekaran}, \& {Kolenberg}}]{manick2018}
{Manick} R., {Van Winckel} H., {Kamath} D., {Sekaran} S., {Kolenberg} K., 2018,
  A\&A, 618, A21

\bibitem[{{Marconi} {et~al}\mbox{.}(2018){Marconi}, {Bono}, {Pietrinferni},
  {Braga}, {Castellani}, \& {Stellingwerf}}]{marconi2018a}
{Marconi} M., {Bono} G., {Pietrinferni} A., {Braga} V.~F., {Castellani} M.,
  {Stellingwerf} R.~F., 2018, ApJL, 864, L13

\bibitem[{{Marconi} {et~al}\mbox{.}(2015){Marconi}, {Coppola}, {Bono}, {Braga},
  {Pietrinferni}, {Buonanno}, {Castellani}, {Musella}, {Ripepi}, \&
  {Stellingwerf}}]{marconi2015}
{Marconi} M., {Coppola} G., {Bono} G., {Braga} V., {Pietrinferni} A.,
  {Buonanno} R., {Castellani} M., {Musella} I., {Ripepi} V., {Stellingwerf}
  R.~F., 2015, ApJ, 808, 50

\bibitem[{{Marconi} \& {Di Criscienzo}(2007)}]{marconi2007}
{Marconi} M., {Di Criscienzo} M., 2007, A\&A, 467, 223

\bibitem[{{Marconi} \& {Minniti}(2018)}]{marconi2018}
{Marconi} M., {Minniti} D., 2018, ApJL, 853, L20

\bibitem[{{Marconi} {et~al}\mbox{.}(2013{\natexlab{a}}){Marconi}, {Molinaro},
  {Bono}, {Pietrzy{\'n}ski}, {Gieren}, {Pilecki}, {Stellingwerf}, {Graczyk},
  {Smolec}, {Konorski}, {Suchomska}, {G{\'o}rski}, \&
  {Karczmarek}}]{marconi2013a}
{Marconi} M. {et~al.}, 2013{\natexlab{a}}, ApJ, 768, L6

\bibitem[{{Marconi} {et~al}\mbox{.}(2017){Marconi}, {Molinaro}, {Ripepi},
  {Cioni}, {Clementini}, {Moretti}, {Ragosta}, {de Grijs}, {Groenewegen}, \&
  {Ivanov}}]{marconi2017}
{Marconi} M., {Molinaro} R., {Ripepi} V., {Cioni} M.-R.~L., {Clementini} G.,
  {Moretti} M.~I., {Ragosta} F., {de Grijs} R., {Groenewegen} M.~A.~T.,
  {Ivanov} V.~D., 2017, MNRAS, 466, 3206

\bibitem[{{Marconi} {et~al}\mbox{.}(2013{\natexlab{b}}){Marconi}, {Molinaro},
  {Ripepi}, {Musella}, \& {Brocato}}]{marconi2013}
{Marconi} M., {Molinaro} R., {Ripepi} V., {Musella} I., {Brocato} E.,
  2013{\natexlab{b}}, MNRAS, 428, 2185

\bibitem[{{Marconi}, {Musella} \& {Fiorentino}(2005){Marconi}, {Musella}, \&
  {Fiorentino}}]{marconi2005}
{Marconi} M., {Musella} I., {Fiorentino} G., 2005, ApJ, 632, 590

\bibitem[{{Marengo} {et~al}\mbox{.}(2010){Marengo}, {Evans}, {Barmby}, {Bono},
  {Welch}, \& {Romaniello}}]{marengo2010}
{Marengo} M., {Evans} N.~R., {Barmby} P., {Bono} G., {Welch} D.~L.,
  {Romaniello} M., 2010, ApJ, 709, 120

\bibitem[{{Martin} \& {Plummer}(1915)}]{martin1915}
{Martin} C., {Plummer} H.~C., 1915, MNRAS, 75, 566

\bibitem[{{Matsunaga} {et~al}\mbox{.}(2018){Matsunaga}, {Bono}, {Chen}, {de
  Grijs}, {Inno}, \& {Nishiyama}}]{matsunaga2018}
{Matsunaga} N., {Bono} G., {Chen} X., {de Grijs} R., {Inno} L., {Nishiyama} S.,
  2018, Space Science Reviews, 214, 74

\bibitem[{{Matsunaga} {et~al}\mbox{.}(2016){Matsunaga}, {Feast}, {Bono},
  {Kobayashi}, {Inno}, {Nagayama}, {Nishiyama}, {Matsuoka}, \&
  {Nagata}}]{matsunaga2016}
{Matsunaga} N., {Feast} M.~W., {Bono} G., {Kobayashi} N., {Inno} L., {Nagayama}
  T., {Nishiyama} S., {Matsuoka} Y., {Nagata} T., 2016, MNRAS, 462, 414

\bibitem[{{Matsunaga}, {Feast} \& {Menzies}(2009){Matsunaga}, {Feast}, \&
  {Menzies}}]{matsunaga2009}
{Matsunaga} N., {Feast} M.~W., {Menzies} J.~W., 2009, MNRAS, 397, 933

\bibitem[{{Matsunaga}, {Feast} \& {Soszy{\'n}ski}(2011){Matsunaga}, {Feast}, \&
  {Soszy{\'n}ski}}]{matsunaga2011}
{Matsunaga} N., {Feast} M.~W., {Soszy{\'n}ski} I., 2011, MNRAS, 413, 223

\bibitem[{{Matsunaga} {et~al}\mbox{.}(2006){Matsunaga}, {Fukushi}, {Nakada},
  {Tanab{\'e}}, {Feast}, {Menzies}, {Ita}, {Nishiyama}, {Baba}, {Naoi},
  {Nakaya}, {Kawadu}, {Ishihara}, \& {Kato}}]{matsunaga2006}
{Matsunaga} N. {et~al.}, 2006, MNRAS, 370, 1979

\bibitem[{{Matsunaga} {et~al}\mbox{.}(2011){Matsunaga}, {Kawadu}, {Nishiyama},
  {Nagayama}, {Kobayashi}, {Tamura}, {Bono}, {Feast}, \&
  {Nagata}}]{matsunaga2011a}
{Matsunaga} N., {Kawadu} T., {Nishiyama} S., {Nagayama} T., {Kobayashi} N.,
  {Tamura} M., {Bono} G., {Feast} M.~W., {Nagata} T., 2011, Nature, 477, 188

\bibitem[{{McGonegal} {et~al}\mbox{.}(1982){McGonegal}, {McAlary}, {Madore}, \&
  {McLaren}}]{mcgonegal1982}
{McGonegal} R., {McAlary} C.~W., {Madore} B.~F., {McLaren} R.~A., 1982, ApJL,
  257, L33

\bibitem[{{McNamara}(1995)}]{mcnamara1995}
{McNamara} D.~H., 1995, AJ, 109, 2134

\bibitem[{{McWilliam}(2011)}]{mcwilliam2011}
{McWilliam} A., 2011, RR Lyrae Stars, Metal-Poor Stars, and the Galaxy
  conference proceedings, 5

\bibitem[{{M{\'e}rand} {et~al}\mbox{.}(2015){M{\'e}rand}, {Kervella},
  {Breitfelder}, {Gallenne}, {Coud{\'e} du Foresto}, {ten Brummelaar},
  {McAlister}, {Ridgway}, {Sturmann}, {Sturmann}, \& {Turner}}]{merand2015}
{M{\'e}rand} A. {et~al.}, 2015, A\&A, 584, A80

\bibitem[{{Minniti} {et~al}\mbox{.}(2010){Minniti}, {Lucas}, {Emerson},
  {Saito}, {Hempel}, {Pietrukowicz}, {Ahumada}, {Alonso}, {Alonso-Garcia},
  {Arias}, {Bandyopadhyay}, {Barb{\'a}}, {Barbuy}, {Bedin}, {Bica},
  {Borissova}, {Bronfman}, {Carraro}, {Catelan}, {Clari{\'a}}, {Cross}, {de
  Grijs}, {D{\'e}k{\'a}ny}, {Drew}, {Fari{\~n}a}, {Feinstein}, {Fern{\'a}ndez
  Laj{\'u}s}, {Gamen}, {Geisler}, {Gieren}, {Goldman}, {Gonzalez}, {Gunthardt},
  {Gurovich}, {Hambly}, {Irwin}, {Ivanov}, {Jord{\'a}n}, {Kerins}, {Kinemuchi},
  {Kurtev}, {L{\'o}pez-Corredoira}, {Maccarone}, {Masetti}, {Merlo},
  {Messineo}, {Mirabel}, {Monaco}, {Morelli}, {Padilla}, {Palma}, {Parisi},
  {Pignata}, {Rejkuba}, {Roman-Lopes}, {Sale}, {Schreiber}, {Schr{\"o}der},
  {Smith}, {}, {Soto}, {Tamura}, {Tappert}, {Thompson}, {Toledo}, {Zoccali}, \&
  {Pietrzynski}}]{minnitivvv2010}
{Minniti} D. {et~al.}, 2010, New Astronomy, 15, 433

\bibitem[{{Moln{\'a}r}(2018)}]{molnar2018}
{Moln{\'a}r} L., 2018, in The RR Lyrae 2017 Conference. Revival of the
  Classical Pulsators: from Galactic Structure to Stellar Interior Diagnostics,
  {Smolec} R., {Kinemuchi} K., {Anderson} R.~I., eds., Vol.~6, pp. 106--113

\bibitem[{{Monson} {et~al}\mbox{.}(2012){Monson}, {Freedman}, {Madore},
  {Persson}, {Scowcroft}, {Seibert}, \& {Rigby}}]{monson2012}
{Monson} A.~J., {Freedman} W.~L., {Madore} B.~F., {Persson} S.~E., {Scowcroft}
  V., {Seibert} M., {Rigby} J.~R., 2012, ApJ, 759, 146

\bibitem[{{Moretti} {et~al}\mbox{.}(2014{\natexlab{a}}){Moretti}, {Clementini},
  {Muraveva}, {Ripepi}, {Marquette}, {Cioni}, {Marconi}, {Girardi}, {Rubele},
  {Tisserand}, {de Grijs}, {Groenewegen}, {Guandalini}, {Ivanov}, \& {van
  Loon}}]{morretti2014}
{Moretti} M.~I. {et~al.}, 2014{\natexlab{a}}, MNRAS, 437, 2702

\bibitem[{{Moretti} {et~al}\mbox{.}(2014{\natexlab{b}}){Moretti}, {Clementini},
  {Muraveva}, {Ripepi}, {Marquette}, {Cioni}, {Marconi}, {Girardi}, {Rubele},
  {Tisserand}, {de Grijs}, {Groenewegen}, {Guandalini}, {Ivanov}, \& {van
  Loon}}]{moretti2014}
{Moretti} M.~I. {et~al.}, 2014{\natexlab{b}}, MNRAS, 437, 2702

\bibitem[{{Moskalik} {et~al}\mbox{.}(2015){Moskalik}, {Smolec}, {Kolenberg},
  {Moln{\'a}r}, {Kurtz}, {Szab{\'o}}, {Benk{\H{o}}}, {Nemec}, {Chadid},
  {Guggenberger}, {Ngeow}, {Jeon}, {Kopacki}, \& {Kanbur}}]{moskalik2015}
{Moskalik} P. {et~al.}, 2015, MNRAS, 447, 2348

\bibitem[{{Muraveva} {et~al}\mbox{.}(2018{\natexlab{a}}){Muraveva}, {Delgado},
  {Clementini}, {Sarro}, \& {Garofalo}}]{muraveva2018}
{Muraveva} T., {Delgado} H.~E., {Clementini} G., {Sarro} L.~M., {Garofalo} A.,
  2018{\natexlab{a}}, MNRAS, 481, 1195

\bibitem[{{Muraveva} {et~al}\mbox{.}(2015){Muraveva}, {Palmer}, {Clementini},
  {Luri}, {Cioni}, {Moretti}, {Marconi}, {Ripepi}, \& {Rubele}}]{muraveva2015}
{Muraveva} T., {Palmer} M., {Clementini} G., {Luri} X., {Cioni} M.-R.~L.,
  {Moretti} M.~I., {Marconi} M., {Ripepi} V., {Rubele} S., 2015, ApJ, 807, 127

\bibitem[{{Muraveva} {et~al}\mbox{.}(2018{\natexlab{b}}){Muraveva},
  {Subramanian}, {Clementini}, {Cioni}, {Palmer}, {van Loon}, {Moretti}, {de
  Grijs}, {Molinaro}, \& {Ripepi}}]{muraveva2018a}
{Muraveva} T., {Subramanian} S., {Clementini} G., {Cioni} M. R.~L., {Palmer}
  M., {van Loon} J.~T., {Moretti} M.~I., {de Grijs} R., {Molinaro} R., {Ripepi}
  V., 2018{\natexlab{b}}, MNRAS, 473, 3131

\bibitem[{{Nataf} {et~al}\mbox{.}(2016){Nataf}, {Gonzalez}, {Casagrande},
  {Zasowski}, {Wegg}, {Wolf}, {Kunder}, {Alonso-Garcia}, {Minniti}, {Rejkuba},
  {Saito}, {Valenti}, {Zoccali}, {Poleski}, {Pietrzy{\'n}ski}, {Skowron},
  {Soszy{\'n}ski}, {Szyma{\'n}ski}, {Udalski}, {Ulaczyk}, \&
  {Wyrzykowski}}]{nataf2016}
{Nataf} D.~M. {et~al.}, 2016, MNRAS, 456, 2692

\bibitem[{{Navarrete} {et~al}\mbox{.}(2015){Navarrete}, {Contreras Ramos},
  {Catelan}, {Clement}, {Gran}, {Alonso-Garc{\'{\i}}a}, {Angeloni}, {Hempel},
  {D{\'e}k{\'a}ny}, \& {Minniti}}]{navarrete2015}
{Navarrete} C., {Contreras Ramos} R., {Catelan} M., {Clement} C.~M., {Gran} F.,
  {Alonso-Garc{\'{\i}}a} J., {Angeloni} R., {Hempel} M., {D{\'e}k{\'a}ny} I.,
  {Minniti} D., 2015, A\&A, 577, A99

\bibitem[{{Neeley} {et~al}\mbox{.}(2017){Neeley}, {Marengo}, {Bono}, {Braga},
  {Dall'Ora}, {Magurno}, {Marconi}, {Trueba}, {Tognelli}, {Prada Moroni},
  {Beaton}, {Freedman}, {Madore}, {Monson}, {Scowcroft}, {Seibert}, \&
  {Stetson}}]{neeley2017}
{Neeley} J.~R. {et~al.}, 2017, ApJ, 841, 84

\bibitem[{{Neeley} {et~al}\mbox{.}(2019){Neeley}, {Marengo}, {Freedman},
  {Madore}, {Beaton}, {Hatt}, {Hoyt}, {Monson}, {Rich}, {Sarajedini},
  {Seibert}, \& {Scowcroft}}]{neeley2019}
{Neeley} J.~R. {et~al.}, 2019, MNRAS, 490, 4254

\bibitem[{{Neilson}, {Cantiello} \& {Langer}(2011){Neilson}, {Cantiello}, \&
  {Langer}}]{neilson2011}
{Neilson} H.~R., {Cantiello} M., {Langer} N., 2011, A\&A, 529, L9

\bibitem[{{Neilson} {et~al}\mbox{.}(2016){Neilson}, {Engle}, {Guinan}, {Bisol},
  \& {Butterworth}}]{neilson2016}
{Neilson} H.~R., {Engle} S.~G., {Guinan} E.~F., {Bisol} A.~C., {Butterworth}
  N., 2016, ApJ, 824, 1

\bibitem[{{Nemec} {et~al}\mbox{.}(2013){Nemec}, {Cohen}, {Ripepi}, {Derekas},
  {Moskalik}, {Sesar}, {Chadid}, \& {Bruntt}}]{nemec2013}
{Nemec} J.~M., {Cohen} J.~G., {Ripepi} V., {Derekas} A., {Moskalik} P., {Sesar}
  B., {Chadid} M., {Bruntt} H., 2013, ApJ, 773, 181

\bibitem[{{Nemec}, {Nemec} \& {Lutz}(1994){Nemec}, {Nemec}, \&
  {Lutz}}]{nemec1994}
{Nemec} J.~M., {Nemec} A.~F.~L., {Lutz} T.~E., 1994, AJ, 108, 222

\bibitem[{{Netzel}, {Smolec} \& {Moskalik}(2015){Netzel}, {Smolec}, \&
  {Moskalik}}]{netzel2015}
{Netzel} H., {Smolec} R., {Moskalik} P., 2015, MNRAS, 447, 1173

\bibitem[{{Ngeow} \& {Kanbur}(2006{\natexlab{a}})}]{ngeow2006a}
{Ngeow} C., {Kanbur} S.~M., 2006{\natexlab{a}}, ApJ, 650, 180

\bibitem[{{Ngeow} \& {Kanbur}(2006{\natexlab{b}})}]{ngeow2006}
{Ngeow} C., {Kanbur} S.~M., 2006{\natexlab{b}}, ApJL, 642, L29

\bibitem[{{Ngeow}, {Kanbur} \& {Nanthakumar}(2008){Ngeow}, {Kanbur}, \&
  {Nanthakumar}}]{ngeow2008}
{Ngeow} C., {Kanbur} S.~M., {Nanthakumar} A., 2008, A\&A, 477, 621

\bibitem[{{Ngeow}(2012)}]{ngeow2012}
{Ngeow} C.-C., 2012, ApJ, 747, 50

\bibitem[{{Ngeow} {et~al}\mbox{.}(2017){Ngeow}, {Kanbur}, {Bhardwaj},
  {Schrecengost}, \& {Singh}}]{ngeow2017}
{Ngeow} C.-C., {Kanbur} S.~M., {Bhardwaj} A., {Schrecengost} Z., {Singh} H.~P.,
  2017, ApJ, 834, 160

\bibitem[{{Ngeow} {et~al}\mbox{.}(2015){Ngeow}, {Kanbur}, {Bhardwaj}, \&
  {Singh}}]{ngeow2015}
{Ngeow} C.-C., {Kanbur} S.~M., {Bhardwaj} A., {Singh} H.~P., 2015, ApJ, 808, 67

\bibitem[{{Ngeow} {et~al}\mbox{.}(2009){Ngeow}, {Kanbur}, {Neilson},
  {Nanthakumar}, \& {Buonaccorsi}}]{ngeow2009}
{Ngeow} C.-C., {Kanbur} S.~M., {Neilson} H.~R., {Nanthakumar} A., {Buonaccorsi}
  J., 2009, ApJ, 693, 691

\bibitem[{{Ngeow} {et~al}\mbox{.}(2005){Ngeow}, {Kanbur}, {Nikolaev},
  {Buonaccorsi}, {Cook}, \& {Welch}}]{ngeow2005}
{Ngeow} C.-C., {Kanbur} S.~M., {Nikolaev} S., {Buonaccorsi} J., {Cook} K.~H.,
  {Welch} D.~L., 2005, MNRAS, 363, 831

\bibitem[{{Nishiyama} {et~al}\mbox{.}(2006){Nishiyama}, {Nagata}, {Kusakabe},
  {Matsunaga}, {Naoi}, {Kato}, {Nagashima}, {Sugitani}, {Tamura}, {Tanab{\'e}},
  \& {Sato}}]{nishiyama2006}
{Nishiyama} S. {et~al.}, 2006, ApJ, 638, 839

\bibitem[{{Nishiyama} {et~al}\mbox{.}(2009){Nishiyama}, {Tamura}, {Hatano},
  {Kato}, {Tanab{\'e}}, {Sugitani}, \& {Nagata}}]{nishiyama2009}
{Nishiyama} S., {Tamura} M., {Hatano} H., {Kato} D., {Tanab{\'e}} T.,
  {Sugitani} K., {Nagata} T., 2009, ApJ, 696, 1407

\bibitem[{{Oosterhoff}(1939)}]{oosterhoff1939}
{Oosterhoff} P.~T., 1939, The Observatory, 62, 104

\bibitem[{{Pancino} {et~al}\mbox{.}(2015){Pancino}, {Britavskiy}, {Romano},
  {Cacciari}, {Mucciarelli}, \& {Clementini}}]{pancino2015}
{Pancino} E., {Britavskiy} N., {Romano} D., {Cacciari} C., {Mucciarelli} A.,
  {Clementini} G., 2015, MNRAS, 447, 2404

\bibitem[{{Persson} {et~al}\mbox{.}(2004){Persson}, {Madore}, {Krzemi{\'n}ski},
  {Freedman}, {Roth}, \& {Murphy}}]{persson2004}
{Persson} S.~E., {Madore} B.~F., {Krzemi{\'n}ski} W., {Freedman} W.~L., {Roth}
  M., {Murphy} D.~C., 2004, AJ, 128, 2239

\bibitem[{{Petersen} \& {Diethelm}(1986)}]{petersen1986}
{Petersen} J.~O., {Diethelm} R., 1986, A\&A, 156, 337

\bibitem[{{Pickering}(1889)}]{pickering1889}
{Pickering} E.~C., 1889, Astronomische Nachrichten, 123, 207

\bibitem[{{Pickering} {et~al}\mbox{.}(1901){Pickering}, {Colson}, {Fleming}, \&
  {Wells}}]{pickering1901}
{Pickering} E.~C., {Colson} H.~R., {Fleming} W.~P., {Wells} L.~D., 1901, ApJ,
  13, 226

\bibitem[{{Pietrukowicz} {et~al}\mbox{.}(2015){Pietrukowicz}, {Koz{\l}owski},
  {Skowron}, {Soszy{\'n}ski}, {Udalski}, {Poleski}, {Wyrzykowski},
  {Szyma{\'n}ski}, {Pietrzy{\'n}ski}, {Ulaczyk}, {Mr{\'o}z}, {Skowron}, \&
  {Kubiak}}]{piet2015}
{Pietrukowicz} P. {et~al.}, 2015, ApJ, 811, 113

\bibitem[{{Pietrzy{\'n}ski} \& {Gieren}(2006)}]{piet2006}
{Pietrzy{\'n}ski} G., {Gieren} W., 2006, MmSAI, 77, 239

\bibitem[{{Pietrzy{\'n}ski} {et~al}\mbox{.}(2007){Pietrzy{\'n}ski}, {Gieren},
  {Udalski}, {Soszy{\'n}ski}, {Bresolin}, {Kudritzki}, {Garcia}, {Minniti},
  {Mennickent}, {Szewczyk}, {Szyma{\'n}ski}, {Kubiak}, \&
  {Wyrzykowski}}]{piet2007}
{Pietrzy{\'n}ski} G. {et~al.}, 2007, AJ, 134, 594

\bibitem[{{Pietrzy{\'n}ski} {et~al}\mbox{.}(2019){Pietrzy{\'n}ski}, {Graczyk},
  {Gallenne}, {Gieren}, {Thompson}, {Pilecki}, {Karczmarek}, {G{\'o}rski},
  {Suchomska}, \& {Taormina}}]{piet2019}
{Pietrzy{\'n}ski} G., {Graczyk} D., {Gallenne} A., {Gieren} W., {Thompson}
  I.~B., {Pilecki} B., {Karczmarek} P., {G{\'o}rski} M., {Suchomska} K.,
  {Taormina} M., 2019, Nature, 567, 200

\bibitem[{{Pietrzy{\'n}ski} {et~al}\mbox{.}(2010){Pietrzy{\'n}ski}, {Thompson},
  {Gieren}, {Graczyk}, {Bono}, {Udalski}, {Soszy{\'n}ski}, {Minniti}, \&
  {Pilecki}}]{piet2010}
{Pietrzy{\'n}ski} G., {Thompson} I.~B., {Gieren} W., {Graczyk} D., {Bono} G.,
  {Udalski} A., {Soszy{\'n}ski} I., {Minniti} D., {Pilecki} B., 2010, Nature,
  468, 542

\bibitem[{{Pilecki} {et~al}\mbox{.}(2018){Pilecki}, {Gieren},
  {Pietrzy{\'n}ski}, {Thompson}, {Smolec}, {Graczyk}, {Taormina}, {Udalski},
  {Storm}, {Nardetto}, {Gallenne}, {Kervella}, {Soszy{\'n}ski}, {G{\'o}rski},
  {Wielg{\'o}rski}, {Suchomska}, {Karczmarek}, \& {Zgirski}}]{pilecki2018}
{Pilecki} B. {et~al.}, 2018, ApJ, 862, 43

\bibitem[{{Planck Collaboration} {et~al}\mbox{.}(2018){Planck Collaboration},
  {Aghanim}, {Akrami}, {Ashdown}, {Aumont}, {Baccigalupi}, {Ballardini},
  {Banday}, {Barreiro}, {Bartolo}, {Basak}, {Battye}, {Benabed}, {Bernard},
  {Bersanelli}, {Bielewicz}, {Bock}, {Bond}, {Borrill}, {Bouchet}, {Boulanger},
  {Bucher}, {Burigana}, {Butler}, {Calabrese}, {Cardoso}, {Carron},
  {Challinor}, {Chiang}, {Chluba}, {Colombo}, {Combet}, {Contreras}, {Crill},
  {Cuttaia}, {de Bernardis}, {de Zotti}, {Delabrouille}, {Delouis}, {Di
  Valentino}, {Diego}, {Dor{\'e}}, {Douspis}, {Ducout}, {Dupac}, {Dusini},
  {Efstathiou}, {Elsner}, {En{\ss}lin}, {Eriksen}, {Fantaye}, {Farhang},
  {Fergusson}, {Fernandez-Cobos}, {Finelli}, {Forastieri}, {Frailis},
  {Franceschi}, {Frolov}, {Galeotta}, {Galli}, {Ganga}, {G{\'e}nova-Santos},
  {Gerbino}, {Ghosh}, {Gonz{\'a}lez-Nuevo}, {G{\'o}rski}, {Gratton},
  {Gruppuso}, {Gudmundsson}, {Hamann}, {Handley}, {Herranz}, {Hivon}, {Huang},
  {Jaffe}, {Jones}, {Karakci}, {Keih{\"a}nen}, {Keskitalo}, {Kiiveri}, {Kim},
  {Kisner}, {Knox}, {Krachmalnicoff}, {Kunz}, {Kurki-Suonio}, {Lagache},
  {Lamarre}, {Lasenby}, {Lattanzi}, {Lawrence}, {Le Jeune}, {Lemos},
  {Lesgourgues}, {Levrier}, {Lewis}, {Liguori}, {Lilje}, {Lilley}, {Lindholm},
  {L{\'o}pez-Caniego}, {Lubin}, {Ma}, {Mac{\'{\i}}as-P{\'e}rez}, {Maggio},
  {Maino}, {Mandolesi}, {Mangilli}, {Marcos-Caballero}, {Maris}, {Martin},
  {Martinelli}, {Mart{\'{\i}}nez-Gonz{\'a}lez}, {Matarrese}, {Mauri}, {McEwen},
  {Meinhold}, {Melchiorri}, {Mennella}, {Migliaccio}, {Millea}, {Mitra},
  {Miville-Desch{\^e}nes}, {Molinari}, {Montier}, {Morgante}, {Moss}, {Natoli},
  {N{\o}rgaard-Nielsen}, {Pagano}, {Paoletti}, {Partridge}, {Patanchon},
  {Peiris}, {Perrotta}, {Pettorino}, {Piacentini}, {Polastri}, {Polenta},
  {Puget}, {Rachen}, {Reinecke}, {Remazeilles}, {Renzi}, {Rocha}, {Rosset},
  {Roudier}, {Rubi{\~n}o-Mart{\'{\i}}n}, {Ruiz-Granados}, {Salvati}, {Sandri},
  {Savelainen}, {Scott}, {Shellard}, {Sirignano}, {Sirri}, {Spencer},
  {Sunyaev}, {Suur-Uski}, {Tauber}, {Tavagnacco}, {Tenti}, {Toffolatti},
  {Tomasi}, {Trombetti}, {Valenziano}, {Valiviita}, {Van Tent}, {Vibert},
  {Vielva}, {Villa}, {Vittorio}, {Wandelt}, {Wehus}, {White}, {White},
  {Zacchei}, \& {Zonca}}]{planck2018}
{Planck Collaboration} {et~al.}, 2018, ArXiv, 1807.06209

\bibitem[{{Prada Moroni} {et~al}\mbox{.}(2012){Prada Moroni}, {Gennaro},
  {Bono}, {Pietrzy{\'n}ski}, {Gieren}, {Pilecki}, {Graczyk}, \&
  {Thompson}}]{prada2012}
{Prada Moroni} P.~G., {Gennaro} M., {Bono} G., {Pietrzy{\'n}ski} G., {Gieren}
  W., {Pilecki} B., {Graczyk} D., {Thompson} I.~B., 2012, ApJ, 749, 108

\bibitem[{{Proxauf} {et~al}\mbox{.}(2018){Proxauf}, {da Silva}, {Kovtyukh},
  {Bono}, {Inno}, {Lemasle}, {Pritchard}, {Przybilla}, {Storm}, {Urbaneja},
  {Valenti}, {Bergemann}, {Buonanno}, {D'Orazi}, {Fabrizio}, {Ferraro},
  {Fiorentino}, {Fran{\c{c}}ois}, {Iannicola}, {Laney}, {Kudritzki},
  {Matsunaga}, {Nonino}, {Primas}, {Romaniello}, \&
  {Th{\'e}venin}}]{proxauf2018}
{Proxauf} B. {et~al.}, 2018, A\&A, 616, A82

\bibitem[{{Riess} {et~al}\mbox{.}(2014){Riess}, {Casertano}, {Anderson},
  {MacKenty}, \& {Filippenko}}]{riess2014}
{Riess} A.~G., {Casertano} S., {Anderson} J., {MacKenty} J., {Filippenko}
  A.~V., 2014, ApJ, 785, 161

\bibitem[{{Riess} {et~al}\mbox{.}(2018{\natexlab{a}}){Riess}, {Casertano},
  {Yuan}, {Macri}, {Anderson}, {MacKenty}, {Bowers}, {Clubb}, {Filippenko},
  {Jones}, \& {Tucker}}]{riess2018}
{Riess} A.~G. {et~al.}, 2018{\natexlab{a}}, ApJ, 855, 136

\bibitem[{{Riess} {et~al}\mbox{.}(2018{\natexlab{b}}){Riess}, {Casertano},
  {Yuan}, {Macri}, {Bucciarelli}, {Lattanzi}, {MacKenty}, {Bowers}, {Zheng},
  {Filippenko}, {Huang}, \& {Anderson}}]{riess2018a}
{Riess} A.~G. {et~al.}, 2018{\natexlab{b}}, ApJ, 861, 126

\bibitem[{{Riess} {et~al}\mbox{.}(2019){Riess}, {Casertano}, {Yuan}, {Macri},
  \& {Scolnic}}]{riess2019}
{Riess} A.~G., {Casertano} S., {Yuan} W., {Macri} L.~M., {Scolnic} D., 2019,
  ApJ, 876, 85

\bibitem[{{Riess} {et~al}\mbox{.}(2011){Riess}, {Macri}, {Casertano},
  {Lampeitl}, {Ferguson}, {Filippenko}, {Jha}, {Li}, \& {Chornock}}]{riess2011}
{Riess} A.~G., {Macri} L., {Casertano} S., {Lampeitl} H., {Ferguson} H.~C.,
  {Filippenko} A.~V., {Jha} S.~W., {Li} W., {Chornock} R., 2011, ApJ, 730, 119

\bibitem[{{Riess} {et~al}\mbox{.}(2016){Riess}, {Macri}, {Hoffmann}, {Scolnic},
  {Casertano}, {Filippenko}, {Tucker}, {Reid}, {Jones}, {Silverman},
  {Chornock}, {Challis}, {Yuan}, {Brown}, \& {Foley}}]{riess2016}
{Riess} A.~G. {et~al.}, 2016, ApJ, 826, 56

\bibitem[{{Ripepi} {et~al}\mbox{.}(2017){Ripepi}, {Cioni}, {Moretti},
  {Marconi}, {Bekki}, {Clementini}, {de Grijs}, {Emerson}, {Groenewegen},
  {Ivanov}, {Molinaro}, {Muraveva}, {Oliveira}, {Piatti}, {Subramanian}, \&
  {van Loon}}]{ripepi2017}
{Ripepi} V. {et~al.}, 2017, MNRAS, 472, 808

\bibitem[{{Ripepi} {et~al}\mbox{.}(2018){Ripepi}, {Molinaro}, {Musella},
  {Marconi}, {Leccia}, \& {Eyer}}]{ripepi2018}
{Ripepi} V., {Molinaro} R., {Musella} I., {Marconi} M., {Leccia} S., {Eyer} L.,
  2018, arXiv e-prints, arXiv:1810.10486

\bibitem[{{Ripepi} {et~al}\mbox{.}(2015){Ripepi}, {Moretti}, {Marconi},
  {Clementini}, {Cioni}, {de Grijs}, {Emerson}, {Groenewegen}, {Ivanov},
  {Muraveva}, {Piatti}, \& {Subramanian}}]{ripepi2015}
{Ripepi} V. {et~al.}, 2015, MNRAS, 446, 3034

\bibitem[{{Ripepi} {et~al}\mbox{.}(2012){Ripepi}, {Moretti}, {Marconi},
  {Clementini}, {Cioni}, {Marquette}, {Girardi}, {Rubele}, {Groenewegen}, {de
  Grijs}, {Gibson}, {Oliveira}, {van Loon}, \& {Emerson}}]{ripepi2012}
{Ripepi} V. {et~al.}, 2012, MNRAS, 424, 1807

\bibitem[{{Ritter}(1879)}]{ritter1879}
{Ritter} A., 1879, Annalen der Physik, 244, 157

\bibitem[{{Romaniello} {et~al}\mbox{.}(2005){Romaniello}, {Primas}, {Mottini},
  {Groenewegen}, {Bono}, \& {Fran{\c{c}}ois}}]{romaniello2005}
{Romaniello} M., {Primas} F., {Mottini} M., {Groenewegen} M., {Bono} G.,
  {Fran{\c{c}}ois} P., 2005, A\&A, 429, L37

\bibitem[{{Romaniello} {et~al}\mbox{.}(2008){Romaniello}, {Primas}, {Mottini},
  {Pedicelli}, {Lemasle}, {Bono}, {Fran{\c{c}}ois}, {Groenewegen}, \&
  {Laney}}]{romaniello2008}
{Romaniello} M., {Primas} F., {Mottini} M., {Pedicelli} S., {Lemasle} B.,
  {Bono} G., {Fran{\c{c}}ois} P., {Groenewegen} M.~A.~T., {Laney} C.~D., 2008,
  A\&A, 488, 731

\bibitem[{{Rood}(1973)}]{rood1973}
{Rood} R.~T., 1973, ApJ, 184, 815

\bibitem[{{Sachkov}, {Bertone} \& {Chavez}(2018){Sachkov}, {Bertone}, \&
  {Chavez}}]{sachkov2018}
{Sachkov} M.~E., {Bertone} E., {Chavez} M., 2018, in A.A. Boyarchuk Memorial
  Conference, {Bisikalo} D.~V., {Wiebe} D.~S., eds., pp. 100--104

\bibitem[{{Saha} {et~al}\mbox{.}(2019){Saha}, {Vivas}, {Olszewski}, {Smith},
  {Olsen}, {Blum}, {Valdes}, {Claver}, {Calamida}, {Walker}, {Matheson},
  {Narayan}, {Soraisam}, {Cunha}, {Axelrod}, {Bloom}, {Cenko}, {Frye}, {Juric},
  {Kaleida}, {Kunder}, {Miller}, {Nidever}, \& {Ridgway}}]{saha2019}
{Saha} A. {et~al.}, 2019, ApJ, 874, 30

\bibitem[{{Salaris} \& {Cassisi}(2005)}]{salaris2005}
{Salaris} M., {Cassisi} S., 2005, {Evolution of Stars and Stellar Populations},
  Vol. pp. 400. ISBN 0-470-09220-3. Wiley-VCH. Evolution of Stars and Stellar
  Populations, by Maurizio Salaris, Santi Cassisi

\bibitem[{{Sandage}(1958)}]{sandage1958a}
{Sandage} A., 1958, ApJ, 128, 150

\bibitem[{{Sandage}(1982)}]{sandage1982}
{Sandage} A., 1982, ApJ, 252, 553

\bibitem[{{Sandage} \& {Tammann}(2006)}]{sandage2006}
{Sandage} A., {Tammann} G.~A., 2006, ARA\&A, 44, 93

\bibitem[{{Sandage}, {Tammann} \& {Reindl}(2004){Sandage}, {Tammann}, \&
  {Reindl}}]{sandage2004}
{Sandage} A., {Tammann} G.~A., {Reindl} B., 2004, A\&A, 424, 43

\bibitem[{{Schmidt} {et~al}\mbox{.}(2009){Schmidt}, {Hemen}, {Rogalla}, \&
  {Thacker-Lynn}}]{schmidt2009}
{Schmidt} E.~G., {Hemen} B., {Rogalla} D., {Thacker-Lynn} L., 2009, AJ, 137,
  4598

\bibitem[{{Sch\"onfeld}(1866)}]{schonfeld1866}
{Sch\"onfeld} J., 1866, Jahresberichte of the Mannheim Physical Society, 32

\bibitem[{{Scowcroft} {et~al}\mbox{.}(2016{\natexlab{a}}){Scowcroft},
  {Freedman}, {Madore}, {Monson}, {Persson}, {Rich}, {Seibert}, \&
  {Rigby}}]{scowcroft2016}
{Scowcroft} V., {Freedman} W.~L., {Madore} B.~F., {Monson} A., {Persson} S.~E.,
  {Rich} J., {Seibert} M., {Rigby} J.~R., 2016{\natexlab{a}}, ApJ, 816, 49

\bibitem[{{Scowcroft} {et~al}\mbox{.}(2011){Scowcroft}, {Freedman}, {Madore},
  {Monson}, {Persson}, {Seibert}, {Rigby}, \& {Sturch}}]{scowcroft2011}
{Scowcroft} V., {Freedman} W.~L., {Madore} B.~F., {Monson} A.~J., {Persson}
  S.~E., {Seibert} M., {Rigby} J.~R., {Sturch} L., 2011, ApJ, 743, 76

\bibitem[{{Scowcroft} {et~al}\mbox{.}(2016{\natexlab{b}}){Scowcroft},
  {Seibert}, {Freedman}, {Beaton}, {Madore}, {Monson}, {Rich}, \&
  {Rigby}}]{scowcroft2016a}
{Scowcroft} V., {Seibert} M., {Freedman} W.~L., {Beaton} R.~L., {Madore} B.~F.,
  {Monson} A.~J., {Rich} J.~A., {Rigby} J.~R., 2016{\natexlab{b}}, MNRAS, 459,
  1170

\bibitem[{{Shapley}(1914)}]{shapley1914}
{Shapley} H., 1914, ApJ, 40, 448

\bibitem[{{Siegel} {et~al}\mbox{.}(2015){Siegel}, {Porterfield}, {Balzer}, \&
  {Hagen}}]{siegel2015}
{Siegel} M.~H., {Porterfield} B.~L., {Balzer} B.~G., {Hagen} L. M.~Z., 2015,
  AJ, 150, 129

\bibitem[{{Simon} \& {Davis}(1983)}]{simon1983}
{Simon} N.~R., {Davis} C.~G., 1983, ApJ, 266, 787

\bibitem[{{Simon} \& {Lee}(1981)}]{slee1981}
{Simon} N.~R., {Lee} A.~S., 1981, ApJ, 248, 291

\bibitem[{{Simon} \& {Moffett}(1985)}]{simon1985}
{Simon} N.~R., {Moffett} T.~J., 1985, PSAP, 97, 1078

\bibitem[{{Simon} \& {Schmidt}(1976)}]{simon1976}
{Simon} N.~R., {Schmidt} E.~G., 1976, ApJ, 205, 162

\bibitem[{{Simon} \& {Teays}(1982)}]{simon1982}
{Simon} N.~R., {Teays} T.~J., 1982, ApJ, 261, 586

\bibitem[{{Skarka}, {Prudil} \& {Jurcsik}(2020){Skarka}, {Prudil}, \&
  {Jurcsik}}]{skara2020}
{Skarka} M., {Prudil} Z., {Jurcsik} J., 2020, MNRAS, 494, 1237

\bibitem[{{Skowron} {et~al}\mbox{.}(2019){Skowron}, {Skowron}, {Mr{\'o}z},
  {Udalski}, {Pietrukowicz}, {Soszy{\'n}ski}, {Szyma{\'n}ski}, {Poleski},
  {Koz{\l}owski}, {Ulaczyk}, {Rybicki}, \& {Iwanek}}]{skowron2019}
{Skowron} D.~M. {et~al.}, 2019, Science, 365, 478

\bibitem[{{Smith}(1995)}]{smith1995}
{Smith} H.~A., 1995, RR Lyrae stars, Cambridge Astrophysics Series 27

\bibitem[{{Smolec}(2016)}]{smolec2016}
{Smolec} R., 2016, MNRAS, 456, 3475

\bibitem[{{Smolec} \& {{\'S}niegowska}(2016)}]{smolec2016a}
{Smolec} R., {{\'S}niegowska} M., 2016, MNRAS, 458, 3561

\bibitem[{{Sollima} {et~al}\mbox{.}(2006){Sollima}, {Borissova}, {Catelan},
  {Smith}, {Minniti}, {Cacciari}, \& {Ferraro}}]{sollima2006a}
{Sollima} A., {Borissova} J., {Catelan} M., {Smith} H.~A., {Minniti} D.,
  {Cacciari} C., {Ferraro} F.~R., 2006, ApJL, 640, L43

\bibitem[{{Sollima}, {Cacciari} \& {Valenti}(2006){Sollima}, {Cacciari}, \&
  {Valenti}}]{sollima2006}
{Sollima} A., {Cacciari} C., {Valenti} E., 2006, MNRAS, 372, 1675

\bibitem[{{Soszy{\'n}ski}, {Gieren} \& {Pietrzy{\'n}ski}(2005){Soszy{\'n}ski},
  {Gieren}, \& {Pietrzy{\'n}ski}}]{soszynski2005}
{Soszy{\'n}ski} I., {Gieren} W., {Pietrzy{\'n}ski} G., 2005, PASP, 117, 823

\bibitem[{{Soszynski} {et~al}\mbox{.}(2008){Soszynski}, {Poleski}, {Udalski},
  {Szymanski}, {Kubiak}, {Pietrzynski}, {Wyrzykowski}, {Szewczyk}, \&
  {Ulaczyk}}]{soszynski2008}
{Soszynski} I., {Poleski} R., {Udalski} A., {Szymanski} M.~K., {Kubiak} M.,
  {Pietrzynski} G., {Wyrzykowski} L., {Szewczyk} O., {Ulaczyk} K., 2008, Acta
  Astron., 58, 163

\bibitem[{{Soszy{\'n}ski} {et~al}\mbox{.}(2019){Soszy{\'n}ski}, {Smolec},
  {Udalski}, \& {Pietrukowicz}}]{soszynski2019}
{Soszy{\'n}ski} I., {Smolec} R., {Udalski} A., {Pietrukowicz} P., 2019, ApJ,
  873, 43

\bibitem[{{Soszy{\'n}ski} {et~al}\mbox{.}(2011){Soszy{\'n}ski}, {Udalski},
  {Pietrukowicz}, {Szyma{\'n}ski}, {Kubiak}, {Pietrzy{\'n}ski}, {Wyrzykowski},
  {Ulaczyk}, {Poleski}, \& {Koz{\l}owski}}]{soszynski2011a}
{Soszy{\'n}ski} I., {Udalski} A., {Pietrukowicz} P., {Szyma{\'n}ski} M.~K.,
  {Kubiak} M., {Pietrzy{\'n}ski} G., {Wyrzykowski} {\L}., {Ulaczyk} K.,
  {Poleski} R., {Koz{\l}owski} S., 2011, Acta Astron., 61, 285

\bibitem[{{Soszy{\'n}ski} {et~al}\mbox{.}(2008){Soszy{\'n}ski}, {Udalski},
  {Szyma{\'n}ski}, {Kubiak}, {Pietrzy{\'n}ski}, {Wyrzykowski}, {Szewczyk},
  {Ulaczyk}, \& {Poleski}}]{soszynski2008a}
{Soszy{\'n}ski} I., {Udalski} A., {Szyma{\'n}ski} M.~K., {Kubiak} M.,
  {Pietrzy{\'n}ski} G., {Wyrzykowski} {\L}., {Szewczyk} O., {Ulaczyk} K.,
  {Poleski} R., 2008, Acta Astron., 58, 293

\bibitem[{{Soszy{\'n}ski} {et~al}\mbox{.}(2010){Soszy{\'n}ski}, {Udalski},
  {Szyma{\'n}ski}, {Kubiak}, {Pietrzy{\'n}ski}, {Wyrzykowski}, {Ulaczyk}, \&
  {Poleski}}]{soszynski2010}
{Soszy{\'n}ski} I., {Udalski} A., {Szyma{\'n}ski} M.~K., {Kubiak} M.,
  {Pietrzy{\'n}ski} G., {Wyrzykowski} {\L}., {Ulaczyk} K., {Poleski} R., 2010,
  Acta Astron., 60, 91

\bibitem[{{Soszy{\'n}ski} {et~al}\mbox{.}(2015){Soszy{\'n}ski}, {Udalski},
  {Szyma{\'n}ski}, {Skowron}, {Pietrzy{\'n}ski}, {Poleski}, {Pietrukowicz},
  {Skowron}, {Mr{\'o}z}, {Koz{\l}owski}, {Wyrzykowski}, {Ulaczyk}, \&
  {Pawlak}}]{soszynski2015}
{Soszy{\'n}ski} I. {et~al.}, 2015, Acta Astron., 65, 297

\bibitem[{{Soszy{\'n}ski} {et~al}\mbox{.}(2018){Soszy{\'n}ski}, {Udalski},
  {Szyma{\'n}ski}, {Wyrzykowski}, {Ulaczyk}, {Poleski}, {Pietrukowicz},
  {Koz{\l}owski}, {Skowron}, {Skowron}, {Mr{\'o}z}, {Rybicki}, \&
  {Iwanek}}]{soszynski2018}
{Soszy{\'n}ski} I. {et~al.}, 2018, Acta Astron., 68, 89

\bibitem[{{Soszy{\'n}ski} {et~al}\mbox{.}(2016){Soszy{\'n}ski}, {Udalski},
  {Szyma{\'n}ski}, {Wyrzykowski}, {Ulaczyk}, {Poleski}, {Pietrukowicz},
  {Koz{\l}owski}, {Skowron}, {Skowron}, {Mr{\'o}z}, \&
  {Pawlak}}]{soszynski2016}
{Soszy{\'n}ski} I. {et~al.}, 2016, Acta Astron., 66, 131

\bibitem[{{Soszy{\'n}ski} {et~al}\mbox{.}(2017{\natexlab{a}}){Soszy{\'n}ski},
  {Udalski}, {Szyma{\'n}ski}, {Wyrzykowski}, {Ulaczyk}, {Poleski},
  {Pietrukowicz}, {Koz{\l}owski}, {Skowron}, {Skowron}, {Mr{\'o}z}, \&
  {Pawlak}}]{soszynski2017a}
{Soszy{\'n}ski} I. {et~al.}, 2017{\natexlab{a}}, Acta Astron., 67, 103

\bibitem[{{Soszy{\'n}ski} {et~al}\mbox{.}(2017{\natexlab{b}}){Soszy{\'n}ski},
  {Udalski}, {Szyma{\'n}ski}, {Wyrzykowski}, {Ulaczyk}, {Poleski},
  {Pietrukowicz}, {Koz{\l}owski}, {Skowron}, {Skowron}, {Mr{\'o}z}, {Pawlak},
  {Rybicki}, \& {Jacyszyn-Dobrzeniecka}}]{soszynski2017}
{Soszy{\'n}ski} I. {et~al.}, 2017{\natexlab{b}}, Acta Astron., 67, 297

\bibitem[{{Stellingwerf}(1982)}]{stellingwerf1982}
{Stellingwerf} R.~F., 1982, ApJ, 262, 330

\bibitem[{{Stellingwerf}(1984)}]{stellingwerf1984}
{Stellingwerf} R.~F., 1984, ApJ, 277, 327

\bibitem[{{Stellingwerf} \& {Donohoe}(1986)}]{stelling1986}
{Stellingwerf} R.~F., {Donohoe} M., 1986, ApJ, 306, 183

\bibitem[{{Stobie}(1969{\natexlab{a}})}]{stobie1969a}
{Stobie} R.~S., 1969{\natexlab{a}}, MNRAS, 144, 461

\bibitem[{{Stobie}(1969{\natexlab{b}})}]{stobie1969b}
{Stobie} R.~S., 1969{\natexlab{b}}, MNRAS, 144, 485

\bibitem[{{Storm} {et~al}\mbox{.}(2004){Storm}, {Carney}, {Gieren},
  {Fouqu{\'e}}, {Latham}, \& {Fry}}]{storm2004}
{Storm} J., {Carney} B.~W., {Gieren} W.~P., {Fouqu{\'e}} P., {Latham} D.~W.,
  {Fry} A.~M., 2004, A\&A, 415, 531

\bibitem[{{Storm} {et~al}\mbox{.}(2011){Storm}, {Gieren}, {Fouqu{\'e}},
  {Barnes}, {Soszy{\'n}ski}, {Pietrzy{\'n}ski}, {Nardetto}, \&
  {Queloz}}]{storm2011}
{Storm} J., {Gieren} W., {Fouqu{\'e}} P., {Barnes} T.~G., {Soszy{\'n}ski} I.,
  {Pietrzy{\'n}ski} G., {Nardetto} N., {Queloz} D., 2011, A\&A, 534, A95

\bibitem[{{Sturch}(1966)}]{sturch1966}
{Sturch} C., 1966, ApJ, 143, 774

\bibitem[{{Subramanian} {et~al}\mbox{.}(2017){Subramanian}, {Marengo},
  {Bhardwaj}, {Huang}, {Inno}, {Nakagawa}, \& {Storm}}]{subramanian2017}
{Subramanian} S., {Marengo} M., {Bhardwaj} A., {Huang} Y., {Inno} L.,
  {Nakagawa} A., {Storm} J., 2017, Space Science Reviews, 212, 1817

\bibitem[{{Subramanian} \& {Subramaniam}(2012)}]{subramanian2012}
{Subramanian} S., {Subramaniam} A., 2012, ApJ, 744, 128

\bibitem[{{Subramanian} \& {Subramaniam}(2015)}]{subramanian2015}
{Subramanian} S., {Subramaniam} A., 2015, A\&A, 573, A135

\bibitem[{{Sweigart} \& {Catelan}(1998)}]{sweigart1998}
{Sweigart} A.~V., {Catelan} M., 1998, ApJL, 501, L63

\bibitem[{{Szab{\'o}} {et~al}\mbox{.}(2010){Szab{\'o}}, {Koll{\'a}th},
  {Moln{\'a}r}, {Kolenberg}, {Kurtz}, {Bryson}, {Benk{\H{o}}},
  {Christensen-Dalsgaard}, {Kjeldsen}, {Borucki}, {Koch}, {Twicken}, {Chadid},
  {di Criscienzo}, {Jeon}, {Moskalik}, {Nemec}, \& {Nuspl}}]{szabo2010}
{Szab{\'o}} R. {et~al.}, 2010, MNRAS, 409, 1244

\bibitem[{{Tammann}, {Sandage} \& {Reindl}(2003){Tammann}, {Sandage}, \&
  {Reindl}}]{tammann2003}
{Tammann} G.~A., {Sandage} A., {Reindl} B., 2003, A\&A, 404, 423

\bibitem[{{Turner}(2010)}]{turner2010}
{Turner} D.~G., 2010, AP \& SS, 326, 219

\bibitem[{{Udalski} {et~al}\mbox{.}(1993){Udalski}, {Szymanski}, {Kaluzny},
  {Kubiak}, {Krzeminski}, {Mateo}, {Preston}, \& {Paczynski}}]{udalski1993}
{Udalski} A., {Szymanski} M., {Kaluzny} J., {Kubiak} M., {Krzeminski} W.,
  {Mateo} M., {Preston} G.~W., {Paczynski} B., 1993, Acta Astron., 43, 289

\bibitem[{{Ulaczyk} {et~al}\mbox{.}(2013){Ulaczyk}, {Szyma{\'n}ski}, {Udalski},
  {Kubiak}, {Pietrzy{\'n}ski}, {Soszy{\'n}ski}, {Wyrzykowski}, {Poleski},
  {Gieren}, {Walker}, \& {Garcia-Varela}}]{ulac2013}
{Ulaczyk} K. {et~al.}, 2013, Acta Astronomica, 63, 159

\bibitem[{{van den Bergh}(1975)}]{vanden1975}
{van den Bergh} S., 1975, {The Extragalactic Distance Scale}, The Extragalactic
  Distance Scale, {Sandage}, Allan and {Sandage}, Mary and {Kristian}, Jerome,
  p. 509

\bibitem[{{van Leeuwen}(2007)}]{van2007}
{van Leeuwen} F., 2007, A\&A, 474, 653

\bibitem[{{Verde}, {Treu} \& {Riess}(2019){Verde}, {Treu}, \&
  {Riess}}]{verde2019}
{Verde} L., {Treu} T., {Riess} A.~G., 2019, arXiv:1907.10625, arXiv:1907.10625

\bibitem[{{Wallerstein}(2002)}]{wallerstein2002}
{Wallerstein} G., 2002, PASP, 114, 689

\bibitem[{{Wallerstein} \& {Cox}(1984)}]{wallerstein1984}
{Wallerstein} G., {Cox} A.~N., 1984, PASP, 96, 677

\bibitem[{{Welch}(2012)}]{welch2012}
{Welch} D.~L., 2012, Journal of the American Association of Variable Star
  Observers (JAAVSO), 40, 492

\bibitem[{{Wheatley}, {Welsh} \& {Browne}(2012){Wheatley}, {Welsh}, \&
  {Browne}}]{wheatley2012}
{Wheatley} J., {Welsh} B.~Y., {Browne} S.~E., 2012, PASP, 124, 552

\bibitem[{{Wielg{\'o}rski} {et~al}\mbox{.}(2017){Wielg{\'o}rski},
  {Pietrzy{\'n}ski}, {Gieren}, {G{\'o}rski}, {Kudritzki}, {Zgirski},
  {Bresolin}, {Storm}, {Matsunaga}, {Graczyk}, \&
  {Soszy{\'n}ski}}]{wielgorski2017}
{Wielg{\'o}rski} P. {et~al.}, 2017, ApJ, 842, 116

\bibitem[{{Wood}, {Arnold} \& {Sebo}(1997){Wood}, {Arnold}, \&
  {Sebo}}]{wood1997}
{Wood} P.~R., {Arnold} A.~S., {Sebo} K.~M., 1997, ApJL, 485, L25

\bibitem[{{Zaritsky} {et~al}\mbox{.}(2004){Zaritsky}, {Harris}, {Thompson}, \&
  {Grebel}}]{zaritsky2004}
{Zaritsky} D., {Harris} J., {Thompson} I.~B., {Grebel} E.~K., 2004, AJ, 128,
  1606

\bibitem[{{Zgirski} {et~al}\mbox{.}(2017){Zgirski}, {Gieren},
  {Pietrzy{\'n}ski}, {Karczmarek}, {Gorski}, {Wielgorski}, {Narloch},
  {Graczyk}, {Kudritzki}, \& {Bresolin}}]{zgirski2017}
{Zgirski} B., {Gieren} W., {Pietrzy{\'n}ski} G., {Karczmarek} P., {Gorski} M.,
  {Wielgorski} P., {Narloch} W., {Graczyk} D., {Kudritzki} R.-P., {Bresolin}
  F., 2017, ApJ, 847, 88

\bibitem[{{Zhao} {et~al}\mbox{.}(2012){Zhao}, {Zhao}, {Chu}, {Jing}, \&
  {Deng}}]{zhao2012}
{Zhao} G., {Zhao} Y.-H., {Chu} Y.-Q., {Jing} Y.-P., {Deng} L.-C., 2012, RAA,
  12, 723

\end{thebibliography}



\end{document}